\documentclass[conference]{IEEEtran}
\usepackage{amssymb}
\usepackage{amsthm}
\usepackage{booktabs}
\usepackage{caption}
\usepackage{cite}
\usepackage{color}
\usepackage{colortbl}
\usepackage{cuted}
\usepackage{dblfloatfix} 
\usepackage{enumerate}
\let\labelindent\relax 
\usepackage{enumitem}
\usepackage{floatrow}
\usepackage{graphicx}
\usepackage{hyperref}
\hypersetup{
	colorlinks   = true,
	citecolor    = blue
}
\usepackage{makecell}
\usepackage{multirow}
\usepackage{multicol}
\usepackage{pgf}
\usepackage{pgfplots}
\usepackage{pgfplotstable}
\usepackage{ragged2e}
\usepackage{stmaryrd}
\usepackage{subcaption}
\usepackage{tabu}
\usepackage{tabularx}
\usepackage{threeparttable}
\usepackage{tikz}
\usepackage{todonotes}
\usepackage{url}
\usepackage{verbatim}
\usepackage{wrapfig}
\usepackage{xcolor}
\usepackage[framemethod=tikz]{mdframed} 
\usepackage{algpseudocode}
\usepackage{nccmath}

\newtheorem{theorem}{Theorem}[section]

\newtheorem{lemma}[theorem]{Lemma}

\theoremstyle{definition}

\newcommand{\xor}{\oplus}

\newcommand{\band}{\odot}

\newcommand{\BigO}[1]{\ensuremath{\operatorname{O}\left(#1\right)}}

\newcommand{\Z}[1]{\ensuremath{\mathbb{Z}}_{2^{#1}}}
\newcommand{\F}{\ensuremath{\mathbb{F}}}

\setlist[description]{style=unboxed,leftmargin=0cm}

\newenvironment{myitemize}{
	\begin{list}{{$\bullet$}}{
			\setlength\partopsep{0pt}
			\setlength\parskip{0pt}
			\setlength\parsep{0pt}
			\setlength\topsep{2pt}
			\setlength\itemsep{4pt}
			\setlength{\itemindent}{0pt}
			\setlength{\leftmargin}{8pt}
		}
	}{
	\vspace{1mm}
	\end{list}
}

\newenvironment{mylist}{
	\begin{list}{{$\bullet$}}{
			\setlength\partopsep{0pt}
			\setlength\parskip{0pt}
			\setlength\parsep{0pt}
			\setlength\topsep{2pt}
			\setlength\itemsep{1pt}
			\setlength{\itemindent}{0pt}
			\setlength{\leftmargin}{9pt}
		}
	}{
	\end{list}
}

\newenvironment{myenumerate}
{\setcounter{itemcount}{0}\begin{list}
{\arabic{itemcount}.}{\usecounter{itemcount} \itemindent=-0.2cm
\itemsep=0.0in
\parsep=0.0in
\topsep=5pt
\partopsep=0.0in}}{\end{list}}

\newcounter{itemcount}

\newcommand{\tabref}[1]{Table~\protect\ref{tab:#1}}
\newcommand{\secref}[1]{Section~\protect\ref{sec:#1}}

\newcommand{\figref}[1]{Fig.~\ref{fig:#1}}
\newcommand{\figlab}[1]{\label{fig:#1}}

\newenvironment{boxfig*}[2]{
	\begin{figure*}[h!]		
		\fontsize{5}{5}\selectfont
		\newcommand{\FigCaption}{#1}
		\newcommand{\FigLabel}{#2}
		\vspace{-.05cm}
		\begin{center}
			\begin{small}			 
				\begin{adjustbox}{max width=\textwidth}
					\begin{tabular}{@{}|@{~~}l@{~~}|@{}}
						\hline
						\rule[-1ex]{0pt}{1ex}\begin{minipage}[b]{.95\linewidth}
							\vspace{1ex}	
						}{%
						\end{minipage}\\
						\hline
					\end{tabular}	
				\end{adjustbox}		
			\end{small}
			\vspace{-0.1cm}
			\caption{\FigCaption}
			\figlab{\FigLabel}
		\end{center}
		\vspace{-.38cm}
	\end{figure*}
}

\newenvironment{myboxfig*}[2]{
	\begin{figure*}[!htb]		
		\fontsize{5}{5}\selectfont
		\newcommand{\FigCaption}{#1}
		\newcommand{\FigLabel}{#2}
		\vspace{-.10cm}
		\begin{center}
			\caption{\FigCaption}
			\begin{small}			 
				\begin{adjustbox}{max width=\textwidth}
					\begin{tabular}{@{}|@{~~}l@{~~}|@{}}
						\hline
						\rule[-1ex]{0pt}{1ex}\begin{minipage}[b]{.95\linewidth}
							\vspace{1ex}	
						}{%
						\end{minipage}\\
						\hline
					\end{tabular}	
				\end{adjustbox}		
			\end{small}
			\vspace{-0.25cm}
			\figlab{\FigLabel}
		\end{center}
		\vspace{-.38cm}
	\end{figure*}
}

\newcommand{\boxref}[1]{Fig.~\ref{#1}}

\RequirePackage{mdframed}

\newenvironment{titlebox}[5]
{\mdfsetup{
		style=#2,
		innertopmargin=1.1\baselineskip,
		skipabove={\dimexpr0.2\baselineskip+\topskip\relax},
		skipbelow={1em},needspace=3\baselineskip,
		singleextra={\node[#3,right=10pt,overlay] at (P-|O){~{\sffamily\bfseries #1 }};},%
		firstextra={\node[#3,right=10pt,overlay] at (P-|O) {~{\sffamily\bfseries #1 }};},
		frametitleaboveskip=9em,
		innerrightmargin=5pt
	}
	\newcommand{\TitleCaption}{#4}
	\newcommand{\TitleLabel}{#5}
	\begin{mdframed}[font=\small]
		\setlist[itemize]{leftmargin=13pt}\setlist[enumerate]{leftmargin=13pt}\raggedright%
	}
	{\end{mdframed}
	\vspace{-1.5em}
	{\captionof{figure}{\small \TitleCaption}\label{\TitleLabel}}
	\medskip
}

\tikzstyle{normal} = [thick, fill=white, text=black, draw, rounded corners, rectangle, minimum height=.7cm, inner sep=3pt]
\tikzstyle{gray} = [thick, fill=gray!90, text=white, rounded corners, rectangle, minimum height=.7cm, inner sep=3pt]

\mdfdefinestyle{commonbox}{%
	align=center, middlelinewidth=1.1pt,userdefinedwidth=\linewidth
	innerrightmargin=0pt,innerleftmargin=2pt,innertopmargin=5pt,
	splittopskip=7pt,splitbottomskip=3pt
}

\mdfdefinestyle{roundbox}{style=commonbox,roundcorner=5pt,userdefinedwidth=\linewidth}

\newenvironment{systembox}[3]
{\vspace{\baselineskip}\begin{titlebox}{Functionality \normalfont #1}{roundbox}{normal}{#2}{#3}}
	{\end{titlebox}}

\newenvironment{protocolbox}[3]
{\begin{titlebox}{Protocol \normalfont #1}{commonbox}{normal}{#2}{#3}}
	{\end{titlebox}}

\newenvironment{simulatorbox}[3]
{\begin{titlebox}{Simulator \normalfont #1}{commonbox}{normal}{#2}{#3}}
	{\end{titlebox}}

\newenvironment{splittitlebox}[5]
{\mdfsetup{
		style=#2,
		innertopmargin=1.1\baselineskip,
		skipabove={\dimexpr0.2\baselineskip+\topskip\relax},
		skipbelow={1em},needspace=3\baselineskip,
		singleextra={\node[#3,right=10pt,overlay] at (P-|O){~{\sffamily\bfseries #1 }};},%
		firstextra={\node[#3,right=10pt,overlay] at (P-|O) {~{\sffamily\bfseries #1 }};},
		frametitleaboveskip=9em,
		innerrightmargin=5pt
	}
	\newcommand{\TitleCaption}{#4}
	\newcommand{\TitleLabel}{#5}
	\begin{mdframed}[font=\small]
		\setlist[itemize]{leftmargin=13pt}\setlist[enumerate]{leftmargin=13pt}\raggedright%
	}
	{\end{mdframed}
	\vspace{-1.5em}
	{\captionof{figure}{\small \TitleCaption}\label{\TitleLabel}}
	\medskip
}

\mdfdefinestyle{commonsplitbox}{%
	align=center, middlelinewidth=1.1pt,userdefinedwidth=\linewidth
	innerrightmargin=0pt,innerleftmargin=2pt,innertopmargin=5pt,
	splittopskip=7pt,splitbottomskip=3pt
}

\newenvironment{protocolsplitbox}[3]
{\begin{splittitlebox}{Protocol \normalfont #1}{commonsplitbox}{normal}{#2}{#3}}
	{\end{splittitlebox}}

\newenvironment{systembox*}[3]
{\begin{strip}
\vspace{\baselineskip}\begin{titlebox}{Functionality \normalfont #1}{roundbox}{normal}{#2}{#3}}
	{\end{titlebox}
\end{strip}}

\newenvironment{gsystembox*}[3]
{\begin{strip}
\vspace{\baselineskip}\begin{titlebox}{Global Functionality \normalfont #1}{roundbox}{normal}{#2}{#3}}
	{\end{titlebox}
\end{strip}}

\newenvironment{protocolbox*}[3]
{\begin{strip}
\begin{titlebox}{Protocol \normalfont #1}{commonbox}{normal}{#2}{#3}}
	{\end{titlebox}
\end{strip}}

\newenvironment{algobox*}[3]
{\begin{strip}
\begin{titlebox}{Algorithm \normalfont #1}{commonbox}{normal}{#2}{#3}}
	{\end{titlebox}
\end{strip}}

\newenvironment{reductionbox*}[3]
{\begin{strip}
\begin{titlebox}{Reduction \normalfont #1}{commonbox}{normal}{#2}{#3}}
	{\end{titlebox}
\end{strip}}

\newenvironment{gamebox*}[3]
{\begin{strip}
\begin{titlebox}{Game \normalfont #1}{commonbox}{gray}{#2}{#3}}
	{\end{titlebox}
\end{strip}}

\newenvironment{simulatorbox*}[3]
{\begin{strip}
\begin{titlebox}{Simulator \normalfont #1}{commonbox}{normal}{#2}{#3}}
	{\end{titlebox}
\end{strip}}

\mdfdefinestyle{mycommonbox}{%
	align=center, middlelinewidth=1.1pt,userdefinedwidth=\linewidth
	innerrightmargin=0pt,innerleftmargin=2pt,innertopmargin=3pt,
	splittopskip=7pt,splitbottomskip=3pt
}
\mdfdefinestyle{myroundbox}{style=mycommonbox,roundcorner=5pt,userdefinedwidth=\linewidth}

\newenvironment{titlebox*}[5]
{\mdfsetup{
		style=#2,
		innertopmargin=0.3\baselineskip,
		skipabove={0.4em},
		skipbelow={1em},needspace=3\baselineskip,
		frametitleaboveskip=5em,
		innerrightmargin=5pt
	}
	\newcommand{\TitleCaption}{#4}
	\newcommand{\TitleLabel}{#5}
	\begin{mdframed}[font=\small]
		\setlist[itemize]{leftmargin=13pt}\setlist[enumerate]{leftmargin=13pt}\raggedright%
	}
	{\end{mdframed}
	\vspace{-2em}
	{\captionof{figure}{\normalfont \TitleCaption}\label{\TitleLabel}}
	\medskip
}

\newenvironment{mysystembox*}[3]
{\begin{strip}
		\vspace{\baselineskip}\begin{titlebox*}{Functionality \normalfont #1}{myroundbox}{normal}{#2}{#3}}
		{\end{titlebox*}
\end{strip}}

\newenvironment{mygsystembox*}[3]
{\begin{strip}
		\vspace{\baselineskip}\begin{titlebox*}{Global Functionality \normalfont #1}{myroundbox}{normal}{#2}{#3}}
		{\end{titlebox*}
\end{strip}}

\newenvironment{myprotocolbox*}[3]
{\begin{strip}
		\begin{titlebox*}{Protocol \normalfont #1}{mycommonbox}{normal}{#2}{#3}}
		{\end{titlebox*}
\end{strip}}

\newenvironment{myalgobox*}[3]
{\begin{strip}
		\begin{titlebox*}{Algorithm \normalfont #1}{mycommonbox}{normal}{#2}{#3}}
		{\end{titlebox*}
\end{strip}}

\newenvironment{myreductionbox*}[3]
{\begin{strip}
		\begin{titlebox*}{Reduction \normalfont #1}{mycommonbox}{normal}{#2}{#3}}
		{\end{titlebox*}
\end{strip}}

\newenvironment{mygamebox*}[3]
{\begin{strip}
		\begin{titlebox*}{Game \normalfont #1}{mycommonbox}{gray}{#2}{#3}}
		{\end{titlebox*}
\end{strip}}

\newenvironment{mysimulatorbox*}[3]
{\begin{strip}
		\begin{titlebox*}{Simulator \normalfont #1}{mycommonbox}{normal}{#2}{#3}}
		{\end{titlebox*}
\end{strip}}

\newenvironment{mytbox}[5]
{\mdfsetup{
		style=#2,
		innertopmargin=1.8\baselineskip,
		skipabove={\dimexpr0.2\baselineskip+\topskip\relax},
		skipbelow={1em},needspace=3\baselineskip,
		singleextra={\node[#3,right=10pt,overlay] at (P-|O){~{\sffamily\bfseries #1 }};},%
		firstextra={\node[#3,right=10pt,overlay] at (P-|O) {~{\sffamily\bfseries #1 }};},
		frametitleaboveskip=9em,
		innerrightmargin=5pt
	}
	\newcommand{\TitleCaption}{#4}
	\newcommand{\TitleLabel}{#5}
	\begin{mdframed}[font=\small]
		\setlist[itemize]{leftmargin=13pt}\setlist[enumerate]{leftmargin=13pt}\raggedright%
	}
	{\end{mdframed}
	\vspace{-1.5em}
	{\captionof{figure}{\small \TitleCaption}\label{\TitleLabel}}
	\medskip
}

\newenvironment{mytsplitbox}[5]
{\mdfsetup{
		style=#2,
		innertopmargin=1.8\baselineskip,
		skipabove={\dimexpr0.2\baselineskip+\topskip\relax},
		skipbelow={1em},needspace=3\baselineskip,
		singleextra={\node[#3,right=10pt,overlay] at (P-|O){~{\sffamily\bfseries #1 }};},%
		firstextra={\node[#3,right=10pt,overlay] at (P-|O) {~{\sffamily\bfseries #1 }};},
		frametitleaboveskip=9em,
		innerrightmargin=5pt
	}
	\newcommand{\TitleCaption}{#4}
	\newcommand{\TitleLabel}{#5}
	\begin{mdframed}[font=\small]
		\setlist[itemize]{leftmargin=13pt}\setlist[enumerate]{leftmargin=13pt}\raggedright%
	}
	{\end{mdframed}
	\vspace{-0.8em}
	{\captionof{figure}{\small \TitleCaption}\label{\TitleLabel}}
	\medskip
}

\newenvironment{mypbox}[3]
{\begin{mytbox}{Protocol \normalfont #1}{commonbox}{normal}{#2}{#3}}
	{\end{mytbox}}

\newcommand{\algoHead}[1]{\vspace{0.2em} \underline{\textbf{#1}} \vspace{0.3em}}

\makeatletter
\algnewcommand{\ExtendedState}[1]{\State
	\parbox[t]{\dimexpr\linewidth-\ALG@thistlm}{\hangindent=\algorithmicindent\strut\hangafter=3#1\strut}}
\makeatother

\algnewcommand\algorithmicinput{\textbf{Input:}}
\algnewcommand\Input{\item[\algorithmicinput]}

\algrenewcommand{\algorithmiccomment}[1]{{\color{gray}// #1}}

\newcommand{\ckt}{\ensuremath{\mathsf{ckt}}}

\newcommand{\wx}{\mathsf{x}}
\newcommand{\wy}{\mathsf{y}}
\newcommand{\wz}{\mathsf{z}}

\newcommand{\negl}{\ensuremath{\mathsf{negl}}}
\newcommand{\csec}{\kappa}

\newcommand{\abort}{\ensuremath{\mathtt{abort}}}
\newcommand{\continue}{\ensuremath{\mathtt{continue}}}
\newcommand{\flag}{\ensuremath{\mathsf{flag}}}

\newcommand{\Partyset}{\ensuremath{\mathcal{P}}}

\newcommand{\Adv}{\ensuremath{\mathcal{A}}}
\newcommand{\Sim}{\ensuremath{\mathcal{S}}}

\newcommand{\Hash}{\ensuremath{\mathsf{H}}}

\newcommand{\rtt}{\ensuremath{\mathsf{rtt}}}

\newcommand{\SELECT}{\ensuremath{\mathsf{select}}}

\newcommand{\OUTPUT}{\ensuremath{\mathsf{Output}}}

\newcommand{\FSETUP}{\ensuremath{\mathcal{F}_{\mathsf{setup}}}} 

\newcommand{\sqr}[1]{\ensuremath{\left[#1\right]}}  
\newcommand{\sqrA}[1]{\ensuremath{\left[#1\right]_{1}}}
\newcommand{\sqrB}[1]{\ensuremath{\left[#1\right]_{2}}}

\newcommand{\sgr}[1]{\ensuremath{\langle #1 \rangle}}

\newcommand{\shr}[1]{\ensuremath{\llbracket #1 \rrbracket}}

\newcommand{\shrB}[1]{\ensuremath{\llbracket #1 \rrbracket}^{\bf B}} 
\newcommand{\arval}[1]{\ensuremath{(#1)^{\bf A}}}  

\newcommand{\av}[1]{\ensuremath{\alpha_{#1}}}   
\newcommand{\bv}[1]{\ensuremath{\beta_{#1}}}    
\newcommand{\gv}[1]{\ensuremath{\gamma_{#1}}}   
\newcommand{\lv}[1]{\ensuremath{\lambda_{#1}}}  

\newcommand{\val}{\ensuremath{\mathsf{v}}}

\newcommand{\Sh}{\ensuremath{\mathsf{sh}}}
\newcommand{\JSh}{\ensuremath{\mathsf{jsh}}} 
\newcommand{\Rec}{\ensuremath{\mathsf{rec}}}
\newcommand{\fRec}{\ensuremath{\mathsf{frec}}} 

\newcommand{\Mult}{\ensuremath{\mathsf{mult}}}
\newcommand{\ZKPC}{\ensuremath{\mathsf{mulZK}}}
\newcommand{\BitExt}{\ensuremath{\mathsf{bitext}}}
\newcommand{\BitA}{\mathsf{bit2A}}
\newcommand{\DotP}{\ensuremath{\mathsf{dotp}}}
\newcommand{\Trunc}{\ensuremath{\mathsf{trgen}}}
\newcommand{\DotPTr}{\ensuremath{\mathsf{dotpt}}}
\newcommand{\ReLU}{\ensuremath{\mathsf{relu}}}
\newcommand{\Sig}{\ensuremath{\mathsf{sig}}}
\newcommand{\MSB}{\ensuremath{\mathsf{msb}}}

\newcommand{\piSh}{\ensuremath{\Pi_{\Sh}}}
\newcommand{\piJSh}{\ensuremath{\Pi_{\JSh}}} 
\newcommand{\piRec}{\ensuremath{\Pi_{\Rec}}}
\newcommand{\pifRec}{\ensuremath{\Pi_{\fRec}}} 
\newcommand{\piMult}{\ensuremath{\Pi_{\Mult}}}
\newcommand{\piZKPC}{\ensuremath{\Pi_{\ZKPC}}}
\newcommand{\piBitExt}{\ensuremath{\Pi_{\BitExt}}}
\newcommand{\PiBitA}{\ensuremath{\Pi_{\BitA}}}
\newcommand{\piDotP}{\ensuremath{\Pi_{\DotP}}}
\newcommand{\piTrunc}{\ensuremath{\Pi_{\Trunc}}}
\newcommand{\piDotPTr}{\ensuremath{\Pi_{\DotPTr}}}
\newcommand{\piReLU}{\ensuremath{\Pi_{\ReLU}}}
\newcommand{\piSig}{\ensuremath{\Pi_{\Sig}}}

\newcommand{\FSh}{\ensuremath{\mathcal{F}_{\Sh}}} 
\newcommand{\FJSh}{\ensuremath{\mathcal{F}_{\JSh}}} 
\newcommand{\FRec}{\ensuremath{\mathcal{F}_{\Rec}}} 
\newcommand{\FMult}{\ensuremath{\mathcal{F}_{\Mult}}} 
\newcommand{\FZKPC}{\ensuremath{\mathcal{F}_{\ZKPC}}} 
\newcommand{\FBitExt}{\ensuremath{\mathcal{F}_{\BitExt}}} 
\newcommand{\FBitA}{\ensuremath{\mathcal{F}_{\BitA}}} 
\newcommand{\FDotP}{\ensuremath{\mathcal{F}_{\DotP}}} 
\newcommand{\FTrunc}{\ensuremath{\mathcal{F}_{\Trunc}}}

\newcommand{\ESet}{\ensuremath{P_1, P_2}}		
\newcommand{\EInSet}{\ensuremath{ \{1, 2\}}}	
\newcommand{\PInSet}{\ensuremath{\{0, 1, 2\}}}	

\newcommand{\md}{\ensuremath{\mathsf{d}}} 
\newcommand{\me}{\ensuremath{\mathsf{e}}} 
\newcommand{\mf}{\ensuremath{\mathsf{f}}} 

\newcommand{\bitb}{\ensuremath{\mathsf{b}}}

\newcommand{\sqd}{\ensuremath{\left[\cdot\right]}}

\newcommand{\sqrV}[2]{\ensuremath{\left[#1\right]_{#2}}}

\newcommand{\sgrd}{\ensuremath{\langle \cdot \rangle}}

\newcommand{\shrd}{\ensuremath{\llbracket \cdot \rrbracket}}

\newcommand{\Key}[1]{\ensuremath{k_{#1}}}

\newcommand{\AVal}[1]{\ensuremath{\alpha_{#1}}}

\newcommand{\AValV}[2]{\ensuremath{\sqrV{\alpha_{#1}}{#2}}}

\newcommand{\BVal}[1]{\ensuremath{\beta_{#1}}}

\newcommand{\GVal}[1]{\ensuremath{\gamma_{#1}}}

\newcommand{\LVal}[1]{\ensuremath{\lambda_{#1}}}
\newcommand{\LValA}[1]{\ensuremath{\sqrV{\lambda_{#1}}{1}}}
\newcommand{\LValB}[1]{\ensuremath{\sqrV{\lambda_{#1}}{2}}}
\newcommand{\LValV}[2]{\ensuremath{\sqrV{\lambda_{#1}}{#2}}}

\newcommand{\GammaV}[1]{\ensuremath{\Gamma_{#1}}}

\newcommand{\Gammaxy}{\ensuremath{\Gamma_{\wx \wy}}}

\newcommand{\GammaxyV}[1]{\ensuremath{\sqrV{\Gamma_{\wx \wy}}{#1}}}

\newcommand{\GammaxyiV}[1]{\ensuremath{\sqrV{\Gamma_{{\wx}_i {\wy}_i}}{#1}}}

\newcommand{\Chi}{\ensuremath{\chi}}

\newcommand{\ChiA}[1]{\ensuremath{\sqrV{\chi}{1}}}
\newcommand{\ChiB}[1]{\ensuremath{\sqrV{\chi}{2}}}

\newcommand{\starbeta}[1]{\ensuremath{\mathsf{\beta}_{#1}^{\star}}}

\newcommand{\nf}{\ensuremath{\mathsf{n}}} 
\newcommand{\TP}{\ensuremath{\mathsf{TP}}}

\newcommand{\gainR}[2]{\makecell[r]{\ensuremath{\mathbf{#1}\times\\\text{to}~\mathbf{#2}\times}}} 

\newcommand{\hrulesep}{\unskip\ \hrule\ }

\newcommand{\shareA}[1]{\ensuremath{\llbracket #1 \rrbracket}^{\bf A}}
\newcommand{\shareB}[1]{\ensuremath{\llbracket #1 \rrbracket}^{\bf B}}

\newcommand{\commit}{\ensuremath{\mathsf{Com}}}

\newcommand{\va}{\ensuremath{\mathsf{a}}}
\newcommand{\vp}{\ensuremath{\mathsf{p}}}
\newcommand{\vq}{\ensuremath{\mathsf{q}}}
\newcommand{\vr}{\ensuremath{\mathsf{r}}}
\newcommand{\vu}{\ensuremath{\mathsf{u}}}
\newcommand{\vv}{\ensuremath{\mathsf{v}}}
\newcommand{\vx}{\ensuremath{\mathsf{x}}}
\newcommand{\vy}{\ensuremath{\mathsf{y}}}

\newcommand{\vecW}{\ensuremath{\vec{\mathbf{w}}}}

\newcommand{\Mat}[1]{\ensuremath{\mathbf{#1}}}

\newcommand{\trunc}[1]{\ensuremath{{#1}^{d}}}
\newcommand{\vrt}{\ensuremath{\mathsf{r}^d}}
\newcommand{\vrd}{\ensuremath{\mathsf{r}_{d}}}

\newcommand{\vecX}{\ensuremath{\vec{\mathbf{x}}}}
\newcommand{\vecY}{\ensuremath{\vec{\mathbf{y}}}}

\newcommand{\maxv}{\ensuremath{\mathsf{max}}}

\newcommand{\Prover}{\ensuremath{\mathsf{P}}}
\newcommand{\Verifier}{\ensuremath{\mathsf{V}}}

\definecolor{UniBlau}{cmyk}{1,0.7,0,0}
\definecolor{UniGruen}{cmyk}{0.6,0,1,0}
\definecolor{UniOrange}{cmyk}{0,0.3,1,0}
\definecolor{UniRot}{cmyk}{0.4,1,0,0}

\definecolor{darkred}{rgb}{.6,0,0}
\definecolor{darkgreen}{rgb}{0,.4,0}
\definecolor{darkblue}{rgb}{0,0,.6}

\pgfplotscreateplotcyclelist{MyCyclelist}{%
	{UniBlau, line width = 2pt},
	{UniRot, line width = 2pt},
	{UniGruen, line width = 2pt},
	{UniOrange, line width = 2pt},
	{UniBlau!50!UniGruen, line width = 2pt},
	{UniGruen!50!UniOrange, line width = 2pt},
	{UniOrange!50!UniRot, line width = 2pt},
	{UniRot!50!UniBlau, line width = 2pt},
}

\pgfplotstableread[col sep=comma]{
	Name,   Ours,   ABY
	BW3,  	266,    23
	BW2,  	266,    15
	BW1,  	266,    8
}\linInftable

\pgfplotstableread[col sep=comma]{
	Name,   Ours,   ABY
	BW3,  	106,    3.5
	BW2,  	106,    2.3
	BW1,  	106,    1.2
}\logInftable

\pgfplotstableread[col sep=comma]{
	Name,   Ours,   ABY
	BW3,  	55.4,  0.21
	BW2,  	36.9,  0.14
	BW1,  	18.4,  0.07
}\nnInftable

\IEEEoverridecommandlockouts
\newif\ifsubmission
\submissionfalse
\submissiontrue    
\ifsubmission
	\newcommand{\ADDED}[1]{#1}  
	\newcommand{\EPRINT}[1]{#1}  
	\newcommand{\commentA}[1]{}
	\newcommand{\commentAJ}[1]{}
\else
	\newcommand{\ADDED}[1]{\textcolor{darkgreen}{#1}}
	\newcommand{\EPRINT}[1]{\textcolor{darkred}{#1}}
	\newcommand{\commentA}[1] {\textcolor{blue}  {{\sf (}{\sl{#1}} }}
	\newcommand{\commentAJ}[1]{\textcolor{violet}{{\sf (}{\sl{#1}} }}
\fi

\newcommand{\TOEDIT}[1]{\textcolor{violet}{{\sf (}{\sl{#1}} }}

\begin{document}
\date{}
\title{BLAZE: Blazing Fast Privacy-Preserving Machine Learning}
\author{
	\IEEEauthorblockN{Arpita Patra\IEEEauthorrefmark{1}, Ajith Suresh\IEEEauthorrefmark{1}}
	\IEEEauthorblockA{\IEEEauthorrefmark{1}Department of Computer Science and Automation, Indian Institute of Science, Bangalore India
		\\\{arpita, ajith\}@iisc.ac.in}
}
\maketitle
\begin{abstract}
Machine learning tools have illustrated their potential in many significant sectors such as healthcare and finance, to aide in deriving useful inferences. The sensitive and confidential nature of the data, in such sectors, raise natural concerns for the privacy of data. This motivated the area of Privacy-preserving Machine Learning (PPML) where privacy of the data is guaranteed. Typically, ML techniques require large computing power, which leads clients with limited infrastructure to rely on the method of Secure Outsourced Computation (SOC). In SOC setting, the computation is outsourced to a set of specialized and powerful cloud servers and the service is availed on a pay-per-use basis. In this work, we explore PPML techniques in the SOC setting for widely used ML algorithms-- Linear Regression, Logistic Regression, and Neural Networks.
 
\ADDED{We propose BLAZE,} a blazing fast PPML framework in the three server setting tolerating one malicious corruption over a ring ($\Z{\ell}$).  BLAZE achieves the stronger \ADDED{security} guarantee of fairness (all honest servers get the output whenever the corrupt server obtains the same).  Leveraging  an {\em input-independent} preprocessing phase, BLAZE has a fast  input-dependent online phase relying on efficient PPML primitives such as: (i) A dot product protocol for which the communication in the online phase is {\em independent} of the vector size, the first of its kind  in the three server setting; (ii) A method for truncation that shuns evaluating expensive circuit for Ripple Carry Adders (RCA) and achieves a constant round complexity. This improves over the truncation method of  ABY3 (Mohassel et al., CCS 2018) that uses RCA  and consumes a round complexity that is of the order of the depth of RCA (which is same as the underlying ring size).

An extensive benchmarking of BLAZE for the aforementioned ML algorithms over a 64-bit ring in both WAN and LAN settings shows massive improvements over ABY3. Concretely, we observe improvements up to $\mathbf{333\times}$ for Linear Regression, $\mathbf{53 \times}$ for Logistic Regression and $\mathbf{276\times}$ for Neural Networks over WAN. Similarly, we show improvements up to $\mathbf{2610\times}$ for Linear Regression, $\mathbf{54\times}$ for Logistic Regression and $\mathbf{278\times}$ for Neural Networks over LAN.
\end{abstract}

\section{Introduction}
\label{sec:Intro}
Machine learning (ML) is increasingly becoming one of the dominant research fields. Advancement in the domain has myriad real-life applications-- from smart keyboard predictions to more involved operations such as self-driving cars. It also finds useful applications in impactful fields such as healthcare and medicine, where ML tools are being used to assist healthcare specialists in better diagnosing abnormalities. This surge in interest in the field is bolstered by the availability of a large amount of data with the rise of companies such as Google and Amazon. This is also due to improved, more robust and accurate ML algorithms in use today. With better machinery and tools such as deep learning and reinforcement learning, ML techniques are starting to beat humans at some difficult tasks such as classifying echocardiograms~\cite{MadaniAMA17}.

In order to be deployed in practice, ML models face numerous challenges. The primary challenge is to provide a high level of accuracy and robustness, as it is imperative for the functioning of some mission-critical fields such as health care. Accuracy and robustness are contingent on a high amount of computing power and availability of data from more varied sources. Accumulating data from different and various sources is not practical for a single company/stake-holder to realize. Moreover, policies like the European Union General Data Protection Regulation (GDPR) or the EFF's call for information fiduciary rules for businesses have made it difficult and even illegal for companies to share datasets with each other without the prior consent of customers. In some cases, it might even be infeasible for companies to share their data with each other as it is proprietary information and sharing it may give rise to concerns such as competitive advantage. While in other cases, the data might be too sensitive, such as medical and financial records, that a breach of privacy cannot be tolerated. It is also possible that the companies providing ML services to clients risk leaking the model parameters rendering its services redundant, and the individual client's or company's data no longer private. In the light of huge interest in using ML and simultaneous requirement of security of data, the field of privacy-preserving machine learning (PPML) has emerged as a flourishing research area. These techniques can be used to ensure that no information about the query or dataset is leaked other than what is permissible by the algorithm, which in some cases might be only the prediction output. 

The primary challenge that inhibits widespread adoption of PPML is that the additional demand on privacy makes the already compute-intensive ML algorithms all the more demanding not just in terms of high compute power but also in terms of  other complexity measures such as communication complexity that the privacy-preserving techniques entail. Many everyday end-users are not equipped with computing infrastructure capable of efficiently executing these algorithms. It is economical and convenient for end-users to outsource an ML task  to more powerful and specialized systems. However, even while outsourcing to servers, the privacy of data must be ensured. Towards this, we use Secure Outsourced Computation (SOC) as a potential solution. SOC allows end-users to securely outsource computation to a set of specialized and powerful cloud servers and avail its services on a pay-per-use basis. SOC guarantees that individual data of the end-users remain private, tolerating reasonable collusion amongst the servers.

PPML, both for training and inference, can be realized in the SOC setting.  Firstly, an end-user  posing as a model-owner can host its trained machine learning model, in a secret-shared way, to a set of (untrusted) servers. An end-user as a customer can secret-share its query  amongst the same servers to allow the prediction to be computed in a shared fashion and to finally obtain the prediction result. Secondly, multiple data-owners can host their datasets in a shared way amongst a set of (untrusted) servers and can train a common model on their joint datasets while keeping their individual dataset private. Recently, many works \cite{MohasselZ17, MakriRSV19,RiaziWTS0K18, MR18, WaghGC19}, solve the aforementioned goals \ADDED{using} the techniques of Secure Multiparty Computation (MPC) where the untrusted servers are taken as the participants (or parties) of the MPC. 
\ADDED{
The corrupt server(s) can collude with an arbitrary number of data-owners in case of training and with either the model-owner or the customer in case of inference. Privacy of the end-users  is ensured leveraging the security guarantees of MPC.
}

MPC is arguably the standard-bearer problem in cryptography. It allows $n$ mutually distrusting parties to perform computations together on their private inputs, so that an adversary controlling at most $t$ parties, can not learn any information beyond what is already known and permissible by the computation. MPC for a small number of parties in the {\em honest majority} setting, specifically the setting of $3$ parties with one corruption,   has become popular over the last few years due to its spectacular performance~\cite{AFLNO16,FLNW17,ABFLLNOWW17,LN17,CGHIKLN18,MRZ15,IshaiKKP15,PatraR18,ByaliJPR18,NV18,BonehBCGI19},   leveraging the presence of single corruption.  Applications such as financial data analysis~\cite{BogdanovTW12}, email spam filtering~\cite{LaunchburyADM14}, distributed credential encryption\cite{MRZ15}, privacy-preserving statistical studies~\cite{BogdanovKKRST15} and popular MPC frameworks such as Sharemind~\cite{BogdanovLW08},  VIFF~\cite{Gei07} involve 3 parties.

In an effort to improve the practical efficiency, many recent works divide their protocol into two phases, namely-- i) {\em input-independent} preprocessing phase and ii) {\em input-dependent} online phase. This has become a prominent approach in both theoretical \cite{Bea91,Bea95,BH06,BH08,BFO12,CP17}  and practical \cite{DPSZ12,SPDZ2,SPDZ3,KOS16,BaumDTZ16,DamgardOS18,CramerDESX18,RiaziWTS0K18,KellerPR18}  domains. The preprocessing phase is used to perform a relatively expensive computation that is independent of the input. In the online phase, once  the  inputs  become available, the actual computation can be performed in a fast way making use of the pre-computed data. This paradigm suits scenarios analogous to our setting, where functions typically need to be evaluated a large number of times, and the function description is known beforehand.

There has been a recent paradigm shift of designing MPC over rings,  considering the fact that computer architectures use rings of size 32 or 64 bits. Designing and implementing  \ADDED{MPC protocols over rings can leverage CPU optimizations and have been proven to have a significant impact on  efficiency \cite{CramerFIK03,BogdanovLW08,DSZ15,DamgardOS18,Damgard0FKSV19}.}  Furthermore, operating over rings avoids the need to overload basic operations such as addition and multiplication during implementation, or rely on an external library as compared to working over prime order fields. 

Although MPC techniques can be used to realize SOC, the current best MPC techniques cannot be directly plugged into ML algorithms, largely due to the following reasons. Firstly,  in ML domain, most of the computation happens over decimal values, requiring us to embed the decimal values in 2's complement form  over a ring ($\Z{\ell}$), where the MSB represents the sign bit, followed by a designated number of bits representing the integer part and  fractional part. As a natural consequence of this embedding,  repeated multiplications cause an overflow in the ring, with the  fractional part  doubling up in size  after each multiplication and occupying double the number of its original bit assignment. The naive solution is to pick a ring large enough to avoid the overflow, but the number of sequential multiplications in a typical ML algorithm makes this solution impractical. The existing works \cite{MohasselZ17,MR18,WaghGC19} tackled this problem through a secure truncation, a very important primitive by now, which approximates the value by sacrificing the accuracy by an infinitesimal amount, performed after every multiplication.  Essentially,  the truncation applied in a privacy-preserving way gets the result of the multiplication back to the same format as that  of the inputs, by  right-shifting it and thereby slashing the expansion of the fractional part caused by the multiplication. Secondly, certain functions such as comparison or the widely used activation  such as ReLU or Sigmoid,  requiring extraction of MSB in a privacy-preserving manner,  needs involvement of the boolean world (over the ring $\Z{1}$), while functions such as addition, dot product  are more efficient when performed in the arithmetic domain (over the ring $\Z{\ell}$). The ML algorithms involve a mix of operations, constantly alternating between these two worlds. As shown in some of the recent works~\cite{DSZ15,MohasselZ17,MR18}, using mixed world computation is orders of magnitude more efficient as compared to most of the current best MPC techniques which operate only in either of the two worlds. Thirdly, while a typical MPC offers a way to tackle a multiplication gate,   ML algorithms invoke its variant dot product. A naive way of doing privacy-preserving dot product would invoke the method of multiplication $\ell$ times, with $\ell$ being the size of the input vectors. With ML algorithms dealing with humongous size data vectors, the naive approach may turn expensive and so customized way of performing dot product that attains independence from the vector size in its complexity is called for. In other words, PPML would need customized privacy-preserving building blocks, such as dot product, truncation, comparison, ReLU, Sigmoid etc.,  rather than the typical building blocks such as  addition and multiplication of MPC.

\ADDED{
\subsection{Related Work}
In the regime of PPML using MPC, earlier works considered the widely-used ML algorithms such as Decision Trees\cite{LindellP00}, Linear Regression\cite{DuA01,SanilKLR04}, k-means clustering\cite{JagannathanW05, BunnO07}, SVM Classification\cite{YuVJ06, VaidyaYJ08}, and Logistic Regression\cite{SlavkovicNT07}. However, these solutions are far from practical due to the high overheads that they incur.
SecureML~\cite{MohasselZ17} proposed a practically-efficient PPML framework in the two-server model using a mix of 2PC protocols that perform computation in {\bf A}rithmetic, {\bf B}oolean  and {\bf Y}ao style (aka. ABY framework~\cite{DSZ15}). One of their key contributions is a novel method for truncating decimal numbers.  They consider   training for linear regression, logistic regression, and neural network models. The work of Chameleon~\cite{RiaziWTS0K18} considered a 2PC setting where parties availed the help of a semi-trusted third party and consider SVMs and Neural Networks. Both SecureML and Chameleon considered semi-honest corruption only. 
The ABY framework was extended to the three-party setting by ABY3 \cite{MR18} and SecureNN~\cite{WaghGC19} (the latter consider neural networks only). These works consider malicious security and demonstrate that the honest-majority setting  can be leveraged to improve the performance by several orders of magnitude. Recently, ASTRA~\cite{CCPS19} furthered this line of work and improve upon ABY3. However, ASTRA presents a set of primitives to build protocols for Linear Regression and Logistic Regression inference. For the training of these ML algorithms and NN prediction, additional primitives like truncation, bit to arithmetic conversions are required, which are not considered in ASTRA. 
}
\subsection{Our Contribution}
We propose an efficient PPML framework  over the ring $\Z{\ell}$ in a SOC  setting,  with three servers amongst which at  most  one can be maliciously corrupt. The framework consists of a range of ML tools realized in a privacy-preserving way which is ensured via running computation in a secret-shared fashion.   We introduce a new secret-sharing semantics for three servers over a ring $\Z{\ell}$  tolerating up to one malicious corruption, which is the basis for all our constructions. We use the sharing  over both $\Z{\ell}$ and its special instantiation $\Z{1}$ and refer them  as {\em arithmetic} and  respectively {\em boolean} sharing.

\begin{figure}[htb!]
	\centering 
	\includegraphics[width=.7\textwidth]{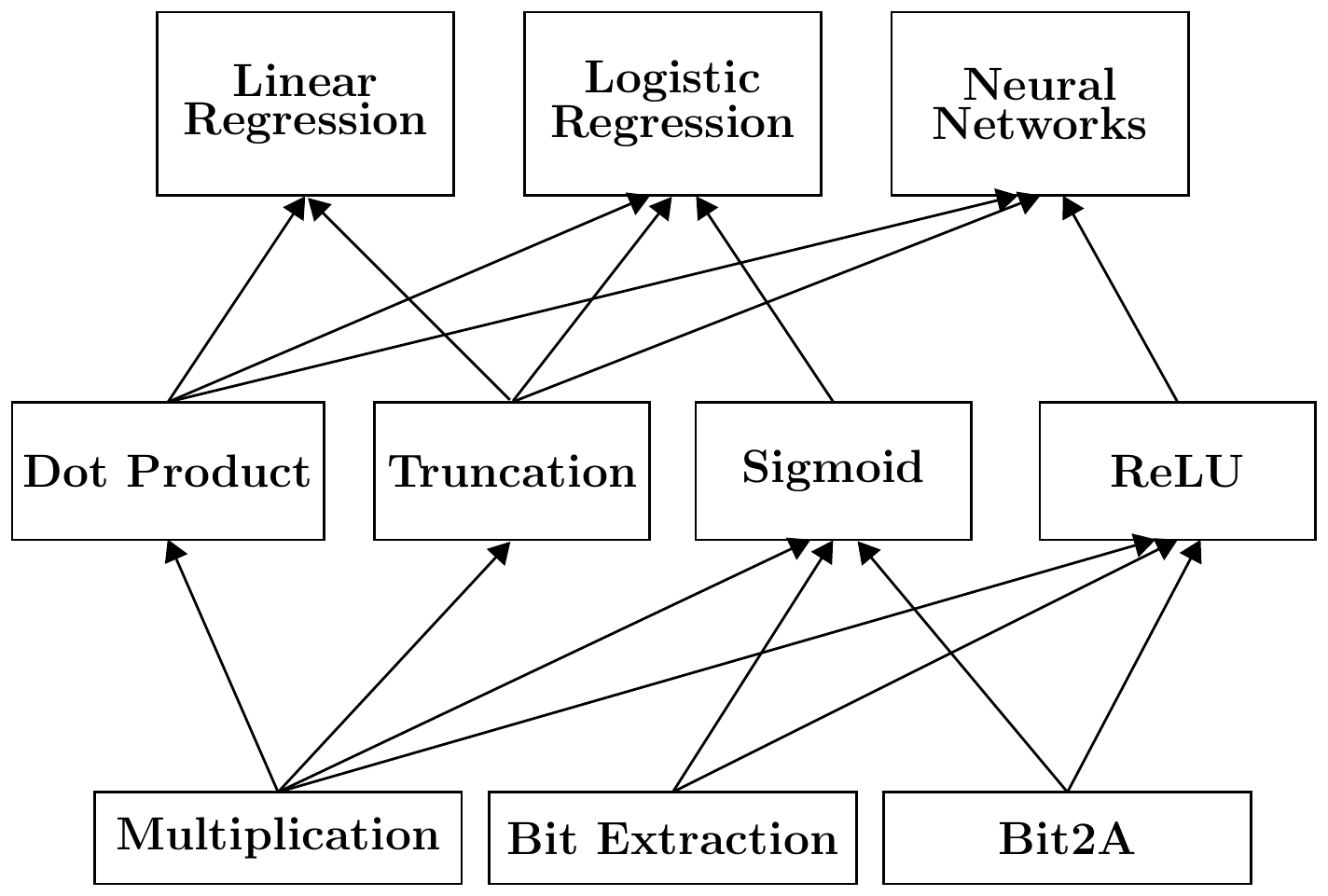}
	\caption{Hierarchy of primitives in  BLAZE Framework\label{fig:Hierarchy}}
\end{figure}

Our framework, as depicted in \figref{Hierarchy}, consists of three layers with the 3rd and final  layer consisting of the privacy-preserving realization of various ML algorithms and forming the end goal of our framework-- i) Linear Regression, ii) Logistic Regression, and iii) Neural Networks (NN).  The 3rd layer builds upon the privacy-preserving realisation of 2nd layer primitives-- (i) Dot Product: This is used to generate arithmetic sharing of $\vecX \band \vecY$, given the arithmetic sharing of each element of vectors $\vecX$ and $\vecY$, (ii) Truncation: Given the arithmetic sharing of a value $\val$, this generates the arithmetic sharing of truncated  version of the value for a publicly known truncation parameter,  and (iii) Non-linear Activation functions (Sigmoid and ReLU): Given the arithmetic sharing of a value, this is used to generate the arithmetic sharing of the resultant value obtained after applying the respective activation function on it.  The 2nd layer builds upon  the privacy-preserving realization of \ADDED{1st layer} primitives--  (i) Multiplication: This is used to generate arithmetic sharing of $\wx \cdot \wy$, given the arithmetic sharing of values $\wx$ and $\wy$, (ii) Bit Extraction: Given arithmetic sharing of a value $\val$,  this is used to generate boolean sharing of the most significant bit ($\MSB$) of the   value, and (iii) Bit to Arithmetic sharing Conversion (Bit2A):  This is used to convert the boolean sharing of a single bit value to its arithmetic sharing.   
The above tools, designed with a focus on practical efficiency, are cast in  {\em input-independent} preprocessing phase, and  {\em input-dependent} online phase. Our contributions, presented in top-down fashion starting with the end-goals (3rd layer), can be summed up as follows.
%
%
%
\begin{description}
	\item {\bf \noindent  Performance of  PPML Algorithms (Layer-III).}	
	\smallskip
	\item[--] Relying on our efficient Layer-II building blocks, our framework BLAZE results in blazing fast PPML protocols for Linear Regression, Logistic Regression, and NN. We consider both training and inference for Linear Regression and Logistic Regression and inference alone for NN. Our 2nd layer primitives are enough to provide support to build all the above. Extending our framework for NN training will require  Garbled circuit techniques,  and  seamless transitions across the three worlds arithmetic, boolean and Yao. We leave this as an interesting open direction. 
	
	We  illustrate the performance of BLAZE via thorough benchmarking and compare with its closest competitors ABY3~\cite{MR18} and ASTRA~\cite{CCPS19}.
	\ADDED{ 
	While the primitives built in ASTRA suffice for secure inference of Linear Regression and Logistic Regression, they do not suffice for secure training  of the aforementioned algorithms and secure NN inference. These require additional tools such as truncation, bit to arithmetic conversion. Also, in ASTRA, the inference phase of Logistic Regression produces a boolean sharing of the output (as an efficiency optimization), while an arithmetic sharing is needed to continue with further computation in case of training.
	For these reasons, we apply the same optimizations as proposed in ASTRA while comparing  the performance of  Linear Regression and Logistic Regression inferences with ASTRA. 
	}
	
	We provide benchmarking for both preprocessing and online phase separately over a 64-bit ring ($\Z{\ell}$) in both WAN and LAN settings. We use {\em throughput} as the benchmarking parameter, which denotes the number of operations (``iterations" for the case of training and ``queries" for the case of inference) that can be performed in unit time. For training, we benchmarked over batch sizes 128, 256 and 512 and feature size ranging from 100 to 900. For inference, in addition to the benchmarking over the aforementioned feature sizes, we benchmarked over real-world datasets as well. \tabref{Gain_Intro} \commentAJ{Edit with new values} summarizes the gain in throughput of our protocols over ABY3 for different ML algorithms. 
	\begin{table}[htb!]
		\centering
		\resizebox{.98\textwidth}{!}{
			\begin{tabular}{l | c | c | c | c }
				\toprule
				\multicolumn{5}{c}{Layer-III: PPML Algorithms}\\
				\toprule
				\multirow{2}[2]{*}{Algorithm} & \multicolumn{2}{c|}{Preprocessing} & \multicolumn{2}{c}{Online}\\ 
				\cmidrule{2-5}
				    & WAN & LAN & WAN & LAN\\
				\midrule
                \makecell[l]{Linear Regression\\(Training)}    & $\gainR{4.01}{4.08}$ & $\gainR{4.01}{4.08}$  & $\gainR{18.54}{333.72}$  & $\gainR{138.89}{2610.76}$ \\ \midrule
                \makecell[l]{Logistic Regression\\(Training)}  & $\gainR{1.97}{2.96}$ & $\gainR{1.92}{2.98}$  & $\gainR{6.13}{53.19}$   & $\gainR{6.11}{53.21}$ \\ \midrule
                \makecell[l]{Linear Regression\\(Inference)}   & $\gainR{4.02}{5.00}$ & $\gainR{4.02}{5.21}$  & $\gainR{2.81}{194.86}$   & $\gainR{52.00}{3600.00}$ \\ \midrule
                \makecell[l]{Logistic Regression\\(Inference)} & $\gainR{1.32}{2.36}$ & $\gainR{1.34}{2.41}$  & $\gainR{3.18}{27.52}$    & $\gainR{3.16}{27.04}$ \\ \midrule
                \makecell[l]{Neural Networks\\(Inference)}     & $\gainR{4.02}{4.07}$ & $\gainR{4.02}{4.07}$  & $\gainR{65.46}{276.31}$  & $\gainR{64.89}{276.84}$ \\
				\bottomrule
			\end{tabular}
		}
		\caption{\small Summary of BLAZE's Gain in Throughput over ABY3\label{tab:Gain_Intro}}
	\end{table}
	 
    In order to emphasise the improved communication of our protocols, we benchmarked over varied bandwidth from 25 to 75Mbps in WAN. 
    
    When compared with ASTRA, we observe improvements up to $194\times$ and $15\times$ over Linear Regression and Logistic Regression inference respectively over WAN. The respective improvements over LAN are $1800\times$ and $16\times$. Note that ASTRA has not considered the training of the above algorithms and NN inference.
\end{description}

\begin{description}
	\smallskip
	\item {\bf \noindent Primary Building Blocks for PPML (Layer-II).}
	\smallskip
	\item[--] {\em Dot Product}: Dot Product forms the fundamental building block for most of the ML algorithms and hence designing efficient constructions for the same are of utmost importance. We propose an efficient dot product protocol for which the communication in the online phase is independent of the size of the underlying vectors. Ours is the first solution in the three-party honest-majority and malicious setting, to achieve such a result. Concretely, our solution requires communication of $3 \nf$ and $3$ ring elements respectively in the preprocessing and online phases, where $\nf$ denotes the size of the underlying vectors.
	When compared with the dot product protocol of ABY3, which requires communication of $12 \nf$ and $9 \nf$ ring elements in the preprocessing phase and online phase, our protocol results in the corresponding improvement of $4\times$ and $3 \nf \times$. Similar comparison with ASTRA~\cite{CCPS19}, which requires communication of $21 \nf$ and $2 \nf + 2$ ring elements in the preprocessing phase and online phase, our protocol results in respective improvements of $7\times$ and $\approx 0.67 \nf\times$.
	\smallskip
	\item[--] {\em  Truncation}: For ML applications where the inputs are floating-point numbers, the protocol for truncation plays a crucial role in determining the overall efficiency of the proposed solution. Towards this, we propose an efficient truncation protocol for the three server setting. When incorporated into our dot product protocol, our truncation method adds a very minimal overhead of just two ring elements in the preprocessing phase of the dot product protocol and more importantly keeps its online complexity intact. In contrast, the state-of-the-art protocol of ABY3 requires expensive Ripple Carry Adder (RCA) circuits in the preprocessing phase which consumes rounds proportional to the underlying ring size. Moreover, their solution demands an additional round of communication with $3$ ring elements in the online phase.
	\smallskip		
	\item[--] {\em Non-linear Activation functions}: We provide efficient instantiation for Sigmoid and ReLU activation functions. The former is used in Logistic Regression, while the latter is used in Neural Networks. Our constructions require only constant round of communication ($\le 4$) in the online phase as opposed to ABY3. Moreover, we improve upon ABY3 in terms of online communication by a factor of $\approx 4.5\times$. 
	\smallskip
\end{description}
The performance comparison in terms of concrete cost for communication and round both for the preprocessing and online phase of these primitives appear in Table~\ref{tab:comparison_Intro}. 
\begin{table}[htb!]
	\centering
	\resizebox{.95\textwidth}{!}{
		\begin{tabular}{r | r | r | r | r | r}
			\multirow{2}[2]{*}{\makecell{Building\\Blocks}}\tnote{1} & \multicolumn{1}{c|}{\multirow{2}[2]{*}{Ref.}} 
			& \multicolumn{2}{c|}{Preprocessing} & \multicolumn{2}{c}{Online} \\ \cmidrule{3-6}
			& & R & C~($\ell$)\tnote{2} & R & C~($\ell$)\\ 
			\toprule 
			\multicolumn{6}{c}{Layer-II: Building Blocks for PPML}\\
			\midrule
			\multirow{3}{*}{Dot Product}     
			& ABY3        & 4 & $12 \nf$          & 1 & $9 \nf$     \\ 
			& ASTRA       & 6 & $21 \nf$          & $\mathbf{1}$ & $\approx 2 \nf$ \\ 
			& {\bf BLAZE} & 4 & $\mathbf{3 \nf}$  & 1 & $\mathbf{3}$ \\ \midrule
			\multirow{2}{*}{\makecell{Dot Product\\ with Truncation}}    
			& ABY3        & 2$\ell$ - 2  & $\approx 12 \nf + 84$  & 2 & $9 \nf + 3$         \\
			& {\bf BLAZE} & 4            & $\mathbf{3 \nf + 2}$   & 1 & $\mathbf{3}$ \\ \midrule
			\multirow{2}{*}{Sigmoid}     
			& ABY3        & 4 & $\approx 108$          & $\log \ell$ + 4 & $\approx 81$              \\
			& {\bf BLAZE} & 5 & $\approx 5\kappa+23$  & $\mathbf{5}$  & $\approx \kappa + 11$ \\ \midrule
			\multirow{2}{*}{ReLU}     
			& ABY3        & 4 & $60$           & $\log \ell$ + 3 & $45$              \\ 
			& {\bf BLAZE} & 5 & $\mathbf{\approx 5\kappa + 14}$  & $\mathbf{4}$               & $\approx \kappa + 7$ \\
			\midrule 
			\multicolumn{6}{c}{Layer-I: Privacy-preserving Primitives}\\
			\midrule
			\multirow{3}{*}{Multiplication}     
			& ABY3        & 4 & $12$          & 1 & $9$          \\ 
			& ASTRA       & 6 & $21$          & 1 & $4$          \\ 
			& {\bf BLAZE} & 4 & $\mathbf{3}$  & 1 & $\mathbf{3}$ \\ \midrule
			\multirow{3}{*}{\makecell[r]{Bit\\Extraction}}    
			& ABY3       & 4     & $24$          & $1 + \log \ell$ & $18$                 \\ 
			& ASTRA      & 7     & $46$          & 3           & $\approx 6$              \\ 
			& {\bf BLAZE} & 4    & $\mathbf{9}$  & $1 + \log \ell$           & $\mathbf{9}$ \\
			& {\bf BLAZE} & 5    & $\approx 5\kappa+2$  & $\mathbf{2}$           & $\approx \kappa$ \\ \midrule
			\multirow{2}{*}{Bit2A}     
			& ABY3        & 4        & $24$         & 2           & $18$             \\ 
			& {\bf BLAZE} & 5        & $\mathbf{9}$ & $\mathbf{1}$           & $\mathbf{4}$     \\ 
		

			\bottomrule
		\end{tabular}}
	    {\footnotesize
		\begin{tablenotes}
			\item[1] -- Notations: $\ell$ - size of ring in bits, $\kappa$ - computational security parameter, $\nf$ - size of vectors for dot product, `R' - number of rounds, `C' - total communication in units of $\ell$ bits.
			\item[2] -- ABY3, ASTRA and BLAZE requires an additional two rounds of interaction in the Online Phase for verification.
		\end{tablenotes}
	    }
		\caption{\small Comparison of ABY3\cite{MR18}, ASTRA\cite{CCPS19} and BLAZE in terms of Communication and Round Complexity\label{tab:comparison_Intro}}
\end{table}
\begin{description}
	\item {\bf \noindent Layer-I: Secondary Building Blocks for PPML}
	\smallskip
	\smallskip
	\item[--] {\em Multiplication}: We propose a new and efficient multiplication protocol for the 3 server setting that can tolerate at most one malicious corruption. Our construction invokes the multiplication protocol of \cite{BonehBCGI19} (which uses distributed Zero Knowledge) in the preprocessing phase to facilitate an efficient online phase. Concretely, our protocol requires an amortized communication of 3 ring elements in both the preprocessing and online phases. Apart from the improvement in communication, the asymmetric nature of our protocol enables one among the three servers to be idle majority of the time during the input-dependent phase. This construct serves as the primary building block for our dot product protocol.
	
	While the multiplication protocol of \cite{BonehBCGI19} performs better than ours with a communication complexity of 3 ring elements overall yet in an amortized sense, we choose our construct over it mainly due to the huge benefits it brings for the case of dot product protocol. The dot product for $\nf$-length vectors can be viewed as $\nf$ multiplications. Using \cite{BonehBCGI19} for the same will result in a communication of $3\nf$ (amortized) ring elements in the online phase. For the communication cost to get amortized, the protocol of \cite{BonehBCGI19}  requires a large number of multiplications to be performed together, which cannot be guaranteed for several instances such as inference phases of Linear Regression and Logistic Regression. Furthermore, their protocol makes use of expensive public-key cryptography, which is undesirable in settings similar to ours, where practical efficiency is of utmost importance in the online phase.
	
	On the other hand, our construct for multiplication when tweaked  to obtain a dot product protocol  requires communication of $3\nf + 3$ ring elements overall, where the preprocessing phase takes care of the expensive part involving invoking \cite{BonehBCGI19} and bearing heavy communication of  $3\nf$ elements.  This results in a blazing fast online phase for dot product which requires communication of just $3$ ring elements and symmetric key operations. Lastly, as our setting calls for the computation of many multiplication operations in the preprocessing phase, the protocol of \cite{BonehBCGI19} is used to perform them, and the communication cost gets amortized over many multiplication operations.
	
	\smallskip\smallskip
	\item[--] {\em Bit Extraction}: We provide two constructions based on the solutions proposed by ASTRA~\cite{CCPS19} and ABY3~\cite{MR18}.
	In the solution based on ASTRA, servers use a garbled circuit that computes a masked version of the most significant bit (MSB) of the input. This results in constant round complexity but the communication will be dependent on the security parameter $\kappa$. On the other hand, the solution based on ABY3 results in communication independent of $\kappa$ but with a round complexity of $1 + \log(\ell)$ where $\ell$ denotes the size of the underlying ring in bits.

	\smallskip	
	\item[--] {\em Bit to Arithmetic sharing Conversion (Bit2A)}:  The arithmetic equivalent of a bit $\bitb = \bitb_1 \xor \bitb_2$ can be written as $\arval{\bitb} = \arval{\bitb_1} + \arval{\bitb_2} - 2 \arval{\bitb_1} \arval{\bitb_2}$. Here $\arval{\bitb}$ denotes the value of bit $\bitb$ in ring $\Z{\ell}$. Thus the servers generate the arithmetic sharing of each of the shares of bit $\bitb$ and their product and  use the aforementioned relation to compute the final result. Our protocol, when compared to ABY3, gives $3\times$ and $4\times$ improvement with respect to the communication cost, in the preprocessing and online phase, respectively.
\end{description}
The performance comparison of these primitives appear in Table~\ref{tab:comparison_Intro}. 

\section{Preliminaries and Definitions}
\label{sec:Prelims}
%
We consider a set of three servers $\Partyset = \{ P_0, P_1, P_2 \}$ that are connected by pair-wise private and authentic channels in a synchronous network. 
\ADDED{
We consider a static and Byzantine adversary, who can corrupt at most one of the three servers. In the case of ML training, many data-owners who wish to jointly train the model, secret-shares (as per schemes discussed latter)  their data amongst the three servers. In the case of ML inference, a model-owner and a client secret-share the trained model and  query respectively among the three servers. Once all the inputs are available in shared fashion, servers perform the computation to generate the output in a shared format among them. For training, the output model is then reconstructed back to the data owners while for inference, the prediction result is reconstructed towards the client alone. We assume that an arbitrary number of data owners can collude with the corrupt server for training, while for inference, either the model-owner or the client can collude with the corrupt server. The same setting has been considered by
ASTRA~\cite{CCPS19}, ABY3~\cite{MR18}, and other related papers.
}

The ML algorithms to be evaluated (Layer-III), relevant to our setting, can be expressed as a circuit $\ckt$ with publicly known topology, consisting of the Layer-II gates-- Dot Product, Truncation, Sigmoid, and ReLU. The gates in the Layer-II are realized using the Layer-I primitives-- Multiplication, Bit Extraction, and Bit2A.

For a vector $\vecX$, $\vx_i$ denotes the $i^{th}$ element in the vector. For two vectors $\vecX$ and $\vecY$ of length $\nf$, the dot product is given by, $\vecX \band \vecY = \sum_{i = 1}^{\nf} \vx_i \vy_i$. 
Given two matrices $\Mat{X}, \Mat{Y}$, the operation $\Mat{X} \circ \Mat{Y}$ denotes the matrix multiplication.

\paragraph{Input-independent and  Input-dependent Phases}
The protocols of this work are cast into two phases: {\em input-independent} preprocessing phase and {\em input-dependent} online phase. 
\ADDED{
This approach is useful in  outsourced setting where the servers execute several instances of an agreed-upon function. The preprocessing for multiple instances can be executed in parallel. }
It is plausible for some of the protocols to have empty input-independent phase.

\paragraph{Shared Key Setup}
To facilitate non-interactive communication, parties use a one-time key setup that establishes pre-shared random keys for a pseudo-random function (PRF) among them. \ADDED{A similar setup for the three-party case was used in \cite{RiaziWTS0K18, FLNW17, ABFLLNOWW17, MR18, CCPS19}.} We model the above as functionality $\FSETUP$ (\boxref{fig:FSETUP}) and all our proofs are cast in $\FSETUP$-hybrid model.

\paragraph{Basic Primitives}
In our protocols, we make use of a {\em collision-resistant} hash function, denoted by $\Hash()$, to save communication. Also, we use a commitment scheme, denoted by $\commit()$, to boost the security of our constructions from abort to fairness.  We defer the formal details of key setup, hash function, and the commitment scheme to Appendix~\ref{app:Prelims}.

\ADDED{We use real-world / ideal-world simulation based approach to prove the security of our constructions and the details appear in Appendix~\ref{app:Security}.}

\section{Building Layer-I Primitives}
\label{sec:ThreePC}
In this section, we start with the sharing semantics that serve as the basis for all our primitives.  The computation in each primitive is executed in shared fashion to obtain the privacy-preserving property.
%
\subsection{Secret Sharing Semantics}
\label{sec:sematics}
We use three types of secret sharing, as detailed below. 
\paragraph{$\sqr{\cdot}$-sharing}
A value $\val \in \Z{\ell}$ is said to be $\sqr{\cdot}$-shared among servers $\ESet$, if the servers $P_1$ and $P_2$ respectively hold the values $\sqrA{\val} \in \Z{\ell}$ and $\sqrB{\val} \in \Z{\ell}$  such that $\val = \sqrA{\val} + \sqrB{\val} $. 
%
\paragraph{$\sgr{\cdot}$-sharing}
A value $\val \in \Z{\ell}$ is $\sgr{\cdot}$-shared among servers in $\Partyset$, if
\begin{myitemize}
	\item[--] there exist $\sqrA{\lv{\val}}, \sqrB{\lv{\val}} \in \Z{\ell}$ such that $\lv{\val} = \sqrA{\lv{\val}} + \sqrB{\lv{\val}}$.
	\item[--] $P_0$ holds $(\sqrA{\lv{\val}}, \sqrB{\lv{\val}})$, while $P_i$ for $i \in \EInSet$  holds $( \sqr{\lv{\val}}_i, \val+ \lv{\val})$
\end{myitemize}
%
\paragraph{$\shr{\cdot}$-sharing}
A value $\val \in \Z{\ell}$ is said to be $\shrd$-shared among servers in $\Partyset$, if 
\begin{myitemize}
	\item[--]  $\val$ is $\sgr{\cdot}$-shared i.e.   $P_0$ holds $(\sqrA{\av{\val}}, \sqrB{\av{\val}})$, while $P_i$ for $i \in \EInSet$  holds $( \sqr{\av{\val}}_i, \bv{\val})$ for $\av{\val}, \bv{\val} \in \Z{\ell}$ with $\bv{\val} = \val + \av{\val}$ and  $\av{\val} = \sqrA{\av{\val}} + \sqrB{\av{\val}}$
	
	\item[--] additionally, there exists $\gv{\val} \in \Z{\ell}$ such that $\ESet$ hold $\gv{\val}$, while $P_0$ holds $\bv{\val}+\gv{\val}$.
\end{myitemize}

The table below summarises the individual shares of the servers for the aforementioned secret sharings. $\sqr{\val}_{i}$, $\sgr{\val}_{i}$ and $\shr{\val}_{i}$ respectively denote the $i$th share held by $P_i$ for $\sqr{\val}$, $\sgr{\val}$ and $\shr{\val}$.
\begin{table}[htb!]
	\centering
	\resizebox{.9\textwidth}{!}{
		\begin{tabular}{c | c | c | c }
			\toprule
			
			& \makecell{$\sqr{\val}$}
			& \makecell{$\sgr{\val}$}
			& \makecell{$\shr{\val}$}\\
			\midrule
			$P_0$	 & $-$ 		& $(\sqrA{\lv{\val}}, \sqrB{\lv{\val}})$ & $(\sqrA{\av{\val}}, \sqrB{\av{\val}}, \bv{\val}+\gv{\val})$\\ 
			\midrule
			$P_1$	 & $\sqrA{\val}$ & $( \sqrA{\lv{\val}}, \val+ \lv{\val})$ & $(\sqrA{\av{\val}}, \bv{\val} = \val + \av{\val} , \gv{\val})$ \\ 
			\midrule
			$P_2$	 & $\sqrB{\val}$ & $( \sqrB{\lv{\val}}, \val+ \lv{\val})$ & $(\sqrB{\av{\val}}, \bv{\val} = \val + \av{\val}, \gv{\val})$\\ 
			\bottomrule
		\end{tabular}
	}
    \caption{\small Shares held by the parties under different  sharings\label{tab:shares}}
\end{table}

\paragraph{Arithmetic and Boolean Sharing} 
We use the sharing  over both $\Z{\ell}$ and $\Z{1}$ and refer them  as {\em arithmetic} and  respectively {\em boolean} sharing. The latter sharing is demarcated using a ${\mathbf B}$ in the superscript (e.g. $\shrB{\bitb}$).

\paragraph{Linearity of the secret sharing schemes}
Given the $\sqr{\cdot}$-sharing of $\wx, \wy$ and public constants $c_1, c_2$, servers can locally compute $\sqr{c_1 \wx + c_2 \wy}$ as $c_1 \sqr{\wx} + c_2 \sqr{\wy}$. Notice that linearity trivially extends to the case of $\sgr{\cdot}$-sharing and $\shrd$-sharing as well. 
Linearity allows the servers to perform the following operations {\em non-interactively}: i) addition of two shared values and ii) multiplication of the shared value with a public constant.

\subsection{Secret Sharing and Reconstruction protocols}
We dedicate this section to describe some of the secret sharing and reconstruction protocols that we need. We defer the communication complexity analysis and security proof of all the constructs to Appendix~\ref{appsec:secret} and Appendix~\ref{app:Security} respectively.
%

\paragraph{Sharing Protocol}
Protocol $\piSh$ (\boxref{fig:piSh}) enables server $P_i$ to generate $\shrd$-sharing of value $\val \in \Z{\ell}$. During the preprocessing phase, servers $P_0, P_1$ along with $P_i$ together sample random value $\sqrA{\av{\val}}$, while servers $P_0, P_2$ and $P_i$ sample $\sqrB{\av{\val}}$ using the shared randomness. This enables server $P_i$ to obtain the entire $\av{\val}$. Also, servers $\ESet$ together sample a random $\gv{\val} \in \Z{\ell}$. For the case when $P_i = P_0$, we optimize the protocol by making $P_0$ sample the $\gv{\val}$ value along with $\ESet$. This eliminates the need for servers $\ESet$ to send $\bv{\val} + \gv{\val}$ to $P_0$ during the online phase. Furthermore, the sharing does not need to hide the input from $P_0$ (who is the input contributor) by keeping $\gv{\val}$ private.

During the online phase, $P_i$ computes $\bv{\val}$ and sends it to $\ESet$ who then verify the sanity of the received value by exchanging its hash with the fellow recipient. To complete  the $\shrd$-sharing,  $P_1$ sends $\bv{\val}+\gv{\val}$ to $P_0$ while $P_2$ sends a hash of the same to $P_0$, who aborts if the received values mismatch.   
\begin{protocolsplitbox}{$\piSh(P_i, \val)$}{$\shrd$-sharing of a value $\val \in \Z{\ell}$ by server $P_i$}{fig:piSh}
	\justify
	\algoHead{Preprocessing:}  
	\begin{myitemize}
		\item[--] If $P_i = P_0$: Servers $P_0$, $P_j$ together sample random $\sqr{\av{\val}}_{j} \in \Z{\ell}$ for $j \in \EInSet$, while servers in $\Partyset$ sample random $\gv{\val} \in \Z{\ell}$.
		\item[--] If $P_i = P_1$: Servers $P_0$, $P_1$ together sample random $\sqr{\av{\val}}_{1} \in \Z{\ell}$, while servers in $\Partyset$ together sample random $\sqr{\av{\val}}_{2} \in \Z{\ell}$. Servers $\ESet$ together sample random $\gv{\val} \in \Z{\ell}$.
		\item[--] If $P_i = P_2$: Similar to the case of $P_i = P_1$.
	\end{myitemize}
	\justify
	\algoHead{Online:}
	\begin{myitemize}
		\item[--] $P_i$ computes $\bv{\val} = \val + \av{\val}$ and sends to both $P_1$ and $P_2$.
		\item[--] If $P_i = P_0$, servers $\ESet$ mutually exchange $\Hash(\bv{\val})$ and $\abort$ if there is a mismatch.
		\item[--] If $P_i \ne P_0$, $P_1$ computes and sends $\bv{\val}+\gv{\val}$ to $P_0$ while $P_2$ sends a hash of the same to $P_0$, who $\abort$ if the received values are inconsistent.
	\end{myitemize}        
\end{protocolsplitbox}

\EPRINT{
In the outsourced setting, input sharing is performed by the parties and not the servers. Concretely, for the case of ML training, data owners perform the input sharing while for the case of ML inference, input sharing is performed by the model owner and the client. For a party $P$ to perform the input sharing of value $\val$, server $P_j$ for $j \in \{1,2\}$ sends $\sqr{\av{\val}}_j$ to $P$ while $P_0$ sends a hash of the same to $P$. Party $P$ computes $\av{\val} = \sqrA{\av{\val}} + \sqrB{\av{\val}}$ if the received values are consistent and $\abort$ otherwise. $P$ then computes $\bv{\val} = \val + \av{\val}$ and sends to both $P_1$ and $P_2$. The rest of the protocol proceeds similar to $\piSh$ where servers $P_1, P_2$ mutually exchanges the hash of $\bv{\val}$ and verifies the consistency of $\bv{\val}$.
}

\paragraph{Joint Sharing Protocol}
Protocol $\piJSh(P_i, P_j, \val)$ (\boxref{fig:piJSh})  enables servers $P_i, P_j$ (an unordered pair)  to jointly generate $\shrd$-sharing of value $\val \in \Z{\ell}$, known to both of them. Towards this, server $P_i$ executes protocol $\piSh$ on the value $\val$ to generate its $\shr{\cdot}$-sharing. Server $P_j$ helps in verifying the correctness of the sharing performed by $P_i$. 

\begin{mypbox}{$\piJSh(P_i, P_j, \val)$}{$\shrd$-sharing of a value $\val \in \Z{\ell}$ by servers $P_i, P_j$}{fig:piJSh}
	\justify
	\begin{myitemize}
		\item[--] If $(P_i, P_j) = (P_1, P_0)$: Server $P_1$ executes protocol $\piSh(P_1, \val)$. $P_0$ computes $\bv{\val} = \val + \sqrA{\av{\val}} + \sqrB{\av{\val}}$. $P_0$ then sends $\Hash(\bv{\val})$ to $P_2$ who $\abort$s if the received value is inconsistent with the same received from $P_1$.
		\item[--] If $(P_i, P_j) = (P_2, P_0)$: Similar to the case above.
		\item[--] If $(P_i, P_j) = (P_1, P_2)$: During the preprocessing phase, $P_1, P_2$ together sample random $\gv{\val} \in \Z{\ell}$. Servers set $\sqrA{\av{\val}} = \sqrB{\av{\val}} = 0$ and $\bv{\val} = \val$. $P_1$ computes and sends $\val + \gv{\val}$ to $P_0$ while $P_2$ sends corresponding hash to $P_0$, who $\abort$s if the received values are inconsistent.
	\end{myitemize}
\end{mypbox}

Protocol $\piJSh$ can be made non-interactive for the case when the value $\val$ is available to both $P_i$ and $P_j$ in the preprocessing phase. Towards this, servers in $\Partyset$ sample random $\vr \in \Z{\ell}$ and locally set their shares as described in \tabref{SharesAssign}.  
Looking ahead, protocol $\piJSh$ offers tolerance against one active corruption, leveraging the fact that the secret to be shared is available amongst two servers, with one of them is guaranteed to be honest.   

\begin{table}[htb!]
	\centering
	\resizebox{.99\textwidth}{!}{
		\begin{tabular}{c | c | c | c }
			\toprule
			& $(P_1, P_2)$ & $(P_1, P_0)$   & $(P_2, P_0)$\\
			\midrule
			& $\begin{aligned} \sqrA{\av{\val}} = 0, &~\sqrB{\av{\val}} = 0\\ \bv{\val} = \val, &~\gv{\val} = \vr - \val \end{aligned}$ 
			& $\begin{aligned} \sqrA{\av{\val}} = -\val, &~\sqrB{\av{\val}} = 0\\ \bv{\val} = 0, &~\gv{\val} = \vr \end{aligned}$ 
			& $\begin{aligned} \sqrA{\av{\val}} = 0, &~\sqrB{\av{\val}} = -\val\\ \bv{\val} = 0, &~\gv{\val} = \vr  \end{aligned}$ \\
			\midrule
			
			$\begin{aligned} P_0 \\ P_1 \\ P_2 \end{aligned}$ 
			& $\begin{aligned} (0, ~0, ~\vr~~~~~) \\ (0, ~\val, ~\vr - \val) \\ (0, ~\val, ~\vr - \val) \end{aligned}$ 
			& $\begin{aligned} (-\val, ~0, ~\vr)  \\ (-\val, ~0, ~\vr) \\ (~~0, ~0, ~\vr)           \end{aligned}$ 
			& $\begin{aligned} (0, ~-\val, ~\vr)  \\ (0,~~~~0, ~\vr)  \\ (0, ~-\val, ~\vr)       \end{aligned}$  \\
			\bottomrule
		\end{tabular}
	}
	\caption{\small The columns consider the three distinct possibility of input contributing pairs. The first row shows the assignment to various components of the sharing. The last row (with three sub-rows) specifies the  shares held by the three servers.\label{tab:SharesAssign}}
\end{table}

\paragraph{Reconstruction Protocol}
Protocol $\piRec(\Partyset, \shr{\val})$ (\boxref{fig:piRec}) enables servers in $\Partyset$ to reconstruct the secret $\val$ from its $\shrd$-sharing. Towards this, each server receives her missing share from one of the other two servers and the hash of the same from the third one. If the received values are consistent, the server proceeds with the reconstruction and otherwise, it aborts. Reconstruction towards a single server $P_i$ can be viewed as a special case of this protocol and we use $\piRec(P_i, \val)$ to denote the same.
\begin{protocolbox}{$\piRec(\Partyset, \shr{\val})$}{Reconstruction of value $\val \in \Z{\ell}$ among servers in $\Partyset$}{fig:piRec}
	\justify
	\algoHead{Online:}
	\begin{mylist}
		\item[--] $P_1$ receives $\sqrB{\av{\val}}$ and $\Hash(\sqrB{\av{\val}})$ from $P_2$ and $P_0$ respectively. 
		\item[--] $P_2$ receives $\sqrA{\av{\val}}$ and $\Hash(\sqrA{\av{\val}})$ from $P_0$ and $P_1$ respectively. 
		\item[--] $P_0$ receives $\bv{\val}$  and $\Hash(\bv{\val})$  from $P_1$ and $P_2$ respectively.
	\end{mylist}
	Server $P_i$ for $i \in \PInSet$ $\abort$ if the received values are inconsistent. Else computes $\val = \bv{\val} - \sqrA{\av{\val}} - \sqrB{\av{\val}}$.       
\end{protocolbox}
\ADDED{In the outsourced setting where reconstruction happens towards the parties (data owners for ML training and client for ML inference), the servers will send their shares towards the parties directly. To reconstruct a value $\val$ towards party $P$, servers $P_0, P_1$ and $P_2$ sends $(\shareA{\av{\val}}, \Hash(\shareB{\av{\val}}))$, $(\bv{\val}, \Hash(\shareA{\av{\val}}))$ and $(\shareB{\av{\val}}, \Hash(\bv{\val}))$ respectively to $P$. Party $P$ will accept the shares if the corresponding hash match and $\abort$ otherwise.}

\paragraph{Fair Reconstruction Protocol}
\ADDED{
The security goal of fairness is well-motivated. Consider an outsourced setting where  a machine learning service  that is instantiated with a protocol {\em with abort} is offered against payment. 
Here, during the reconstruction of output, adversary can instruct the corrupt server to send inconsistent values (either shares or hash values) to honest parties and make them $\abort$. At the same time, adversary will learn the output from the honest shares received on behalf of the corrupt parties. This leads to a situation where some parties who have control over the corrupt server obtain the protocol output, while the other honest parties obtain nothing. This is a strong deterrent for the honest parties to participate in the protocol in the future. On the other hand, a system with fairness property guarantees that the honest parties will get the output whenever the corrupt parties gets the output.
In our 3PC setting, the presence of at least a single honest server ensures that all the participating honest parties will eventually get the output. This will attract more people to participate in the protocol and is crucial to applications like ML training where more data leads to a better-trained model.
}

\ADDED{
We use the techniques proposed by ASTRA~\cite{CCPS19} to achieve fairness and modify it for our sharing scheme. We defer formal details to the appendix (Section~\ref{app:fRec}).
}

\subsection{Layer-I Primitives}
\label{sec:build_blocks}
We are now ready to describe our Layer-I primitives-- Multiplication, Bit Extraction, and Bit2A. We defer the communication complexity analysis and security proof of all the constructs to Appendix~\ref{appsec:LayerI} and Appendix~\ref{app:Security} respectively.

\paragraph{Multiplication Protocol}
Protocol $\piMult(\Partyset, \shr{\wx}, \shr{\wy})$ enables the servers in $\Partyset$ to compute $\shrd$-sharing of $\wz = \wx  \wy$, given the $\shrd$-sharing of $\wx$ and $\wy$. We begin with a protocol for the semi-honest setting, which is a slightly modified variant of the protocol proposed by ASTRA. During the preprocessing phase, $P_0, P_j$ for $j \in \EInSet$ sample random $\sqr{\av{\wz}}_{j} \in \Z{\ell}$, while $\ESet$ sample random $\gv{\wz} \in \Z{\ell}$. In addition, $P_0$ locally computes $\Gammaxy = \av{\wx}\av{\wy}$ and generates $\sqr{\cdot}$-sharing of the same between $\ESet$. Since 

\resizebox{.94\linewidth}{!}{
	\begin{minipage}{\linewidth}
		\begin{align}
		\bv{\wz} &= \wz + \av{\wz} = \wx \wy + \av{\wz} = (\bv{\wx} - \av{\wx})\ADDED{(\bv{\wy} - \av{\wy})} + \av{\wz} \nonumber\\
		           &= \bv{\wx}\bv{\wy} - \bv{x}\av{\wy} - \bv{y}\av{\wx} + \Gammaxy + \av{\wz} \label{eq:betaz}
		\end{align}
  	\end{minipage}}

holds,  servers $\ESet$ locally compute $\sqr{\bv{\wz}}_j = (j-1)\bv{\wx}\bv{\wy} - \bv{x}\sqr{\av{\wy}}_{j} - \bv{y}\sqr{\av{\wx}}_{j} + \GammaxyV{j} + \sqr{\av{\wz}}_{j}$ during the online phase and  mutually exchange their shares to reconstruct $\bv{\wz}$. Server $P_1$ then computes and sends $\bv{\wz} + \gv{\wz}$ to $P_0$, completing the semi-honest protocol. The  correctness that asserts $\wz = \wx \wy$ or in other words $\bv{\wz} - \av{\wz}= \wx \wy$ holds due to Equation~\ref{eq:betaz}. 

In the malicious setting, we observe that the aforementioned protocol suffers from three issues:
\begin{myenumerate}
	\item When $P_0$ is corrupt, the $\sqr{\cdot}$-sharing of $\Gammaxy$ performed by $P_0$ during the preprocessing phase might not be correct, i.e.\ $\Gammaxy \neq \av{\wx}\av{\wy}$.
	\item When $P_1$ (or $P_2$) is corrupt, the $\sqr{\cdot}$-share of $\bv{\wz}$ handed over to the fellow honest evaluator during the online phase might not be correct, causing reconstruction of an incorrect $\bv{\wz}$. 
	\item When $P_1$ is corrupt, the value $\bv{\wz} + \gv{\wz}$ that is sent to $P_0$ during the online phase may not be correct.
\end{myenumerate}

While the first two issues in the above list are inherited from the protocol of ASTRA, the third one is due to our new sharing semantics (compared to ASTRA \ADDED{where $\gv{\val}$ and $\bv{\val} + \gv{\val}$ were not part of the shares}) that imposes an additional component of $\bv{\wz} + \gv{\wz}$ held by $P_0$. We begin with solving the last issue first. In order to verify the correctness of $\bv{\wz} + \gv{\wz}$ sent by $P_1$, server $P_2$ computes a hash of the same and send it to $P_0$, who $\abort$s if the received values are inconsistent. 

For the remaining two issues, though they are quite distinct in nature, we make use of the asymmetric roles played by the servers $\{P_0\}$ and $\{\ESet\}$ to introduce a single check that solves both the issues at the same time. Though the check is inspired from the protocol of ASTRA, our technical innovation lies in the way in which the check is performed. In ASTRA, servers first execute the semi-honest protocol and the correctness of the computation is verified with the help of $\sgr{\cdot}$-sharing of a multiplication triple generated in the preprocessing phase.  Unlike ASTRA, we perform a single multiplication (and nothing additional) in the preprocessing phase to generate the correct preprocessing data required for a multiplication gate in the online phase. This brings down the communication in the preprocessing phase drastically from $21$ ring elements to $3$ ring elements. The details of our method are provided next.

To solve the second issue, where a corrupt $P_1$ (or $P_2$) sends an incorrect $\sqr{\cdot}$-share of $\bv{\wz}$, we make use of server $P_0$ as follows: Using the values $\starbeta{\wx} = \bv{\wx} + \gv{\wx}$ and $\starbeta{\wy} = \bv{\wy} + \gv{\wy}$, $P_0$ computes $\starbeta{\wz} = - \starbeta{\wx}\av{\wy} - \starbeta{\wy}\av{\wx} + 2\Gammaxy + \av{\wz}$. Now $\starbeta{\wz} $ can be written as below:~

\resizebox{.94\linewidth}{!}{
	\begin{minipage}{\linewidth}
		\begin{align*}
		\starbeta{\wz} &= - \starbeta{\wx}\av{\wy} - \starbeta{\wy}\av{\wx} + 2\Gammaxy + \av{\wz}\\
		               &= -(\bv{\wx} + \gv{\wx})\av{\wy} - (\bv{\wy} + \gv{\wy})\av{\wx} + 2\Gammaxy + \av{\wz}\\
		               &= (-\bv{\wx}\av{\wy} - \bv{\wy}\av{\wx} + \Gammaxy + \av{\wz}) - (\gv{\wx}\av{\wy} + \gv{\wy}\av{\wx} - \Gammaxy)\\
		               &= (\bv{\wz} - \bv{\wx}\bv{\wy}) - (\gv{\wx}\av{\wy} + \gv{\wy}\av{\wx} - \Gammaxy + \psi) + \psi \hspace*{11pt}[\text{Equation~\ref{eq:betaz}}]\\
		               &= (\bv{\wz} - \bv{\wx}\bv{\wy} + \psi) - \Chi   \hspace*{13pt}[\text{where } \Chi = \gv{\wx}\av{\wy} + \gv{\wy}\av{\wx} - \Gammaxy + \psi]
		\end{align*}
	\end{minipage}}

Assuming that (a) $\psi \in \Z{\ell}$ is a random value sampled together by $P_1$ and $P_2$ (and unknown to $P_0$) and (b) $P_0$ knows the value $\Chi$, $P_0$ can send $\starbeta{\wz} + \Chi$ to $P_1$ and $P_2$ who using  the knowledge of $\bv{\wx}, \bv{\wy}$ and $\psi$ can verify the correctness of $\bv{\wz}$ by computing $\bv{\wz} - \bv{\wx}\bv{\wy} + \psi$ and checking against the value $\starbeta{\wz} + \Chi$ received from $P_0$.
Now we describe how to enable $P_0$ \ADDED{to} obtain the value $\Chi$. Note that server $P_j$ for $j \in \EInSet$ can locally compute $\sqr{\Chi}_j = \gv{\wx}\sqr{\av{\wy}}_j + \gv{\wy}\sqr{\av{\wx}}_j - \sqr{\Gammaxy}_j + \sqr{\psi}_j$ where $\sqr{\psi}_j$ can be generated non-interactively by $\ESet$ using shared randomness. $\ESet$ can then send their $\sqr{\cdot}$-shares of $\Chi$ to $P_0$ to enable him obtain the value $\Chi$. To verify if $P_0$ computed $\Chi$ correctly, we leverage the following relation.    The values $\md = \gv{\wx}-\av{\wx}, \me = \gv{\wy}-\av{\wy}$ and $\mf = (\gv{\wx}\gv{\wy} + \psi)- \Chi$ should satisfy $\mf = \md \me$ if and only if $\Chi$ is correctly computed, because:~

\resizebox{.94\linewidth}{!}{
	\begin{minipage}{\linewidth}
		\begin{align*}
		\md \me &= (\gv{\wx}-\av{\wx})(\gv{\wy}-\av{\wy}) = \gv{\wx}\gv{\wy} - \gv{\wx}\av{\wy} - \gv{\wy}\av{\wx} + \Gammaxy\\
		        &= (\gv{\wx}\gv{\wy} + \psi) - (\gv{\wx}\av{\wy} + \gv{\wy}\av{\wx} - \Gammaxy + \psi)\\
		        &= (\gv{\wx}\gv{\wy} + \psi)- \Chi = \mf 
		\end{align*}
	\end{minipage}}

Therefore,  the correctness of $\Chi$ reduces  to verifying if the triple $(\md, \me, \mf)$ is a multiplication triple or not. Interestingly, the same check suffices to resolve the first issue of corrupt $P_0$ generating incorrect $\sqr{\Gammaxy}$-sharing. This is because, if $P_0$ would have shared $\Gammaxy + \Delta$ where $\Delta$ denotes the error introduced, then $\md \me = \mf + \Delta \ne \mf$.

\ADDED{
Equipped with the aforementioned observations \ADDED{(\tabref{shareZK})}, our final trick, that distinguishes BLAZE's multiplication from that of ASTRA's,  is to compute a $\sgr{\cdot}$-sharing of $\mf$ starting with $\sgr{\cdot}$-sharing of \ADDED{$\md,\me$} using the efficient   maliciously secure multiplication protocol of \cite{BonehBCGI19} referred to as $\piZKPC$ henceforth and described in Appendix~\ref{app:ZKCrypto}  for completeness,  and extract out the values for $\Gammaxy, \psi$ and $\Chi$ from $\mf$ which are bound to be correct.}  This  can be executed entirely in the preprocessing phase. 
Protocol $\piZKPC$ works over $\sgr{\cdot}$-sharing (\secref{sematics}), which is why this part our computation is done over this type of sharing,    and requires a per party communication of $1$ ring element, when amortized over large circuits (ref. Theorem $1.4$ of \cite{BonehBCGI19}\footnote{ \url{https://eprint.iacr.org/2019/188}}). Concretely, given the $\shrd$-sharing of the inputs $\wx$ and $\wy$ of the multiplication protocol, servers locally compute $\sgr{\cdot}$-sharing of values $\md$ and $\me$ as follows. (The sharing semantics for $\sqr{\val}$ for any $\val$ is recalled below.)
%
\begin{table}[htb!]
	\centering
	\resizebox{.9\textwidth}{!}{
	\begin{tabular}{c | c | c | c}
		& $P_0$ & $P_1$ & $P_2$ \\ 
		\scriptsize{$\sgr{\val}$} & \scriptsize{$(\sqrA{\lv{\val}}, \sqrA{\lv{\val}})$} 
		& \scriptsize{$(\sqrA{\lv{\val}}, \val+ \lv{\val})$} & \scriptsize{$(\sqrB{\lv{\val}}, \val+ \lv{\val})$} \\ \toprule
		$\sgr{\md}$	 & $(\sqr{\av{\wx}}_{1},\sqr{\av{\wx}}_{2})$ &$(\sqr{\av{\wx}}_{1},\gv{\wx})  $& $(\sqr{\av{\wx}}_{2},\gv{\wx})  $\\ \midrule
		$\sgr{\me}$	 & $(\sqr{\av{\wy}}_{1},\sqr{\av{\wy}}_{1})$ &$(\sqr{\av{\wy}}_{1},\gv{\wy})  $& $(\sqr{\av{\wy}}_{2},\gv{\wy})  $\\ \bottomrule
	\end{tabular}
    }
    \caption{\small The $\sgr{\cdot}$-sharing of values $\md$ and $\me$ \label{tab:shareZK}}
\end{table}
%

Upon executing protocol $\piZKPC(\Partyset, \md, \me)$, servers obtain $\sgr{\mf} = (\sqr{\lv{\mf}}, \mf + \lv{\mf})$. To be precise, $P_0$ obtains $(\sqrA{\lv{\mf}}, \sqrB{\lv{\mf}})$ while $P_j$ for $j \in \EInSet$ obtains $(\sqr{\lv{\mf}}_j, \mf + \lv{\mf})$. Servers then map the values $\sqr{\Chi}$ and $\gv{\wx}\gv{\wy} + \psi$ to $\sqr{\lv{\mf}}$ and $\mf + \lv{\mf}$ respectively followed by extracting the required values as:

\resizebox{.94\linewidth}{!}{
	\begin{minipage}{\linewidth}
		\begin{align*}
		\sqrA{\Chi} = \sqrA{\lv{\mf}} \text{and} \sqrB{\Chi} &= \sqrB{\lv{\mf}} \hspace*{19pt} \rightarrow \Chi ~= \sqrA{\lv{\mf}} + \sqrB{\lv{\mf}}\\
		\gv{\wx}\gv{\wy} + \psi &= \mf + \lv{\mf} \hspace*{15pt} \rightarrow \psi = \mf + \lv{\mf} - \gv{\wx}\gv{\wy}\\[3pt]
		\sqr{\Gammaxy}_j &= \gv{\wx}\sqr{\av{\wy}}_j + \gv{\wy}\sqr{\av{\wx}}_j + \sqr{\psi}_j - \sqr{\Chi}_j \hspace*{5pt}[j \in \EInSet]
		\end{align*}
	\end{minipage}}

where $\sqr{\psi}$ is generated non-interactively by servers $\ESet$ by sampling a random value $\vr \in \Z{\ell}$ together and setting $\sqrA{\psi} = \vr$ and $\sqrB{\psi} = \psi - \vr$. We claim that after extracting the values as mentioned above, servers $\ESet$ hold $\sqr{\Gammaxy} = \sqr{\av{\wx} \av{\wy}}$. To see this, note that

\resizebox{.94\linewidth}{!}{
	\begin{minipage}{\linewidth}
		\begin{align*}
		\Gammaxy &= \gv{\wx}\av{\wy} + \gv{\wy}\av{\wx} + \psi - \Chi\\
		         &= (\md + \lv{\md})\lv{\me} + (\me + \lv{\me})\lv{\md} + (\mf + \lv{\mf} - \gv{\wx}\gv{\wy}) - \lv{\mf}\\
		         &= (\md + \lv{\md})(\me + \lv{\me}) - \md \me + \lv{\md} \lv{\me} + (\mf - \gv{\wx}\gv{\wy})\\
		         &= \gv{\wx}\gv{\wy} - \mf + \lv{\md} \lv{\me} + (\mf - \gv{\wx}\gv{\wy}) = \lv{\md} \lv{\me} = \av{\wx} \av{\wy}
		\end{align*}
	\end{minipage}}

This concludes the informal discussion. Our protocol $\piMult$ appears in \boxref{fig:piMult}.
\begin{protocolbox}{$\piMult(\Partyset, \shr{\wx}, \shr{\wy})$}{Multiplication Protocol}{fig:piMult}
	\justify
	\algoHead{Preprocessing:}
	\begin{myitemize}
		\item[--] Servers $P_0, P_j$ for $j \in \EInSet$ together sample a random $\sqr{\av{\wz}}_{j} \in \Z{\ell}$, while $\ESet$ sample a random $\gv{\wz} \in \Z{\ell}$. 
		\item[--] Servers in $\Partyset$ locally compute $\sgr{\cdot}$-sharing of $\md = \gv{\wx}-\av{\wx}$ and $\me = \gv{\wy}-\av{\wy}$ by setting the shares as (as per \ADDED{\tabref{shareZK}}):
		\begin{align*}
			\sqrA{\lv{\md}} = \sqr{\av{\wx}}_{1},~~ \sqrB{\lv{\md}} = \sqr{\av{\wx}}_{2},~~ (\md + \lv{\md}) = \gv{\wx}\\
			\sqrA{\lv{\me}} = \sqr{\av{\wy}}_{1},~~ \sqrB{\lv{\me}} = \sqr{\av{\wy}}_{2},~~ (\me + \lv{\me}) = \gv{\wy}
		\end{align*}
		\item[--] Servers in $\Partyset$ execute $\piZKPC(\Partyset, \md, \me)$ to generate $\sgr{\mf} = \sgr{\md \me}$.
		\item[--] $P_0, P_j$ for $j \in \EInSet$ locally set $\sqr{\Chi}_j = \sqr{\lv{\mf}}_j$, while $P_1, P_2$ set $\psi = \mf + \lv{\mf} - \gv{\wx}\gv{\wy}$. $P_0$ then computes $\Chi = \sqr{\Chi}_1 + \sqr{\Chi}_2$.
		\item[--] $P_1, P_2$ sample random $\vr \in \Z{\ell}$ and set $\sqrA{\psi} = \vr, \sqrB{\psi} = \psi - \vr$. 
		\item[--] $P_j$ for $j \in \EInSet$ set $\sqr{\Gammaxy}_j = \gv{\wx}\sqr{\av{\wy}}_j + \gv{\wy}\sqr{\av{\wx}}_j + \sqr{\psi}_j - \sqr{\Chi}_j$
	\end{myitemize} 
	\algoHead{Online:}
	\begin{myitemize}
		\item[--] $P_j$ for $j \in \EInSet$ computes and exchanges $\sqr{\bv{\wz}}_{j} = (j-1)\bv{\wx}\bv{\wy} - \bv{\wx}\sqrV{\av{\wy}}{j} - \bv{\wy}\sqrV{\av{\wx}}{j} + \GammaxyV{j} + \sqr{\av{\wz}}_{j}$ to reconstruct $\bv{\wz} = \sqr{\bv{\wz}}_{1} + \sqr{\bv{\wz}}_{2}$.
		\item[--] $P_0$ computes $\starbeta{\wz} = -(\bv{\wx}+\gv{\wx})\av{\wy} - (\bv{\wy}+\gv{\wy})\av{\wx} + \av{\wz} +2\Gammaxy + \Chi$ and sends $\Hash(\starbeta{\wz})$ to both $P_1$ and $P_2$. 
		\item[--] $P_j$ for $j \in \EInSet$ $\abort$s if $\Hash(\bv{\wz} - \bv{\wx}\bv{\wy} + \psi) \neq \Hash(\starbeta{\wz})$. 
		\item[--] $P_1$ sends $\bv{\wz}+\gv{\wz}$ and $P_2$ sends the $\Hash(\bv{\wz}+\gv{\wz})$ to $P_0$. $P_0$ will $\abort$ if it receives inconsistent values. 
	\end{myitemize}
\end{protocolbox}
%
Looking ahead, our multiplication protocol lends its technical strength to all our layer-II primitives, especially the dot product. The preprocessing phase of dot product protocol invokes its preprocessing phase in a black-box way many times that lets its optimal complexity (of 3 elements per multiplication) kick in. However, the online phase of dot product is not plain  invocation of online phase of multiplication protocol in a black-box way. In fact, the tweaks here are crucial for achieving a complexity that is independent  of the feature length. The other layer-II primitives use multiplication protocol in a block-box way.  
%
\paragraph{Bit Extraction Protocol}
Protocol $\piBitExt(\Partyset, \shr{\val})$ (Fig.~\ref{fig:piBitExt}) enables servers in $\Partyset$ to compute the boolean sharing ($\shrB{\cdot}$) of most significant bit ($\MSB$) of value $\val \in \Z{\ell}$, given its arithmetic sharing $\shr{\val}$. The first approach is to use an optimized Parallel Prefix Adder (PPA) proposed by ABY3~\cite{MR18}. The PPA circuit consists of $2\ell$ AND gates and has a multiplicative depth of $\log(\ell)$. We refer readers to ABY3 for more details. The next approach is to use a garbled circuit that results in a constant round solution. We provide details for the latter approach below. 

Let $GC = (\vu_1, \vu_2, \vu_3, \vu_4, \vu_5)$ denote a garbled circuit with inputs $\vu_1, \vu_2, \vu_3 \in \Z{\ell}$ and $\vu_4, \vu_5 \in \{0,1\}$ and output $\vy = \MSB(\vu_1 - \vu_2 - \vu_3) \xor \vu_4 \xor \vu_5$. Note that the MSB calculation portion of the circuit can be instantiated using the optimized PPA of ABY3. Let $\vu_1 = \bv{\val}, \vu_2 = \sqr{\av{\val}}_1$ and $\vu_3 =  \sqr{\av{\val}}_2$ so that $\vu_1 - \vu_2 - \vu_3 = \val$. Let $\vu_4 = \vr_1, \vu_5 = \vr_2$ where $\vr_1$ and $\vr_2$ denote random bits sampled by $P_0, P_1$ and $P_0, P_2$ respectively.

\begin{protocolbox}{$\piBitExt(\Partyset, \shr{\val})$}{Extraction of MSB bit of value $\val \in \Z{\ell}$}{fig:piBitExt}
	\justify
	Let $GC = (\vu_1, \vu_2, \vu_3, \vu_4, \vu_5)$ denote a garbled circuit with inputs $\vu_1, \vu_2, \vu_3 \in \Z{\ell}$ and $\vu_4, \vu_5 \in \{0,1\}$ and output $\vy = \MSB(\vu_1 - \vu_2 - \vu_3) \xor \vu_4 \xor \vu_5$. Let $\vu_1 = \bv{\val}, \vu_2 = \sqr{\av{\val}}_1$ and $\vu_3 =  \sqr{\av{\val}}_2$ so that $\vu_1 - \vu_2 - \vu_3 = \val$.
	\justify
	\algoHead{Preprocessing:}
	\begin{myitemize}
		\item[--] $P_0, P_j$ for $j \in \EInSet$ sample random $\vr_j \in \{0,1\}$ and execute $\piJSh$ on $\vr_j$ to generate $\shrB{\vr_j}$. Let $\vu_4 = \vr_1$ and $\vu_5 = \vr_2$.
		\item[--] $P_0, P_1$ garbles the circuit $GC$ and sends $GC$ to $P_2$ along with the decoding information. Note that the values $\vu_2$ and $\vu_4$ are embedded in the $GC$ itself since they are known to $P_0, P_1$.
		\item[--] Corresponding to each bit of $\vu_3$, $P_0, P_1$ compute commitments for both the keys (zero key and one key) using common randomness and send these commitments to $P_2$. In addition, $P_0$ sends the decommitment of actual key corresponding to the bits of $\vu_3$ to $P_2$ who $\abort$ if the values are inconsistent. Similar steps are executed for the bit $\vu_5 = \vr_2$.
	\end{myitemize}  
	\justify
	\algoHead{Online Phase:}
	\begin{myitemize}
		\item[--] $P_0, P_1$ compute commitments for both the keys corresponding to the bits of $\vu_1$ similar to the case of $\vu_3$. $P_1$ opens the right commitment towards $P_2$.
		\item[--] $P_2$ evaluates $GC$ and obtains $\vv = \MSB(\val) \xor \vr_1 \xor \vr_2$ in clear. $P_2$ sends $\vv$ to $P_1$ along with a hash of the key corresponding to $\vv$. $P_1$ $\abort$ if the received values are inconsistent.
		\item[--] $P_1, P_2$ execute $\piJSh$ on $\vv$ to generate $\shrB{\vv}$. Servers locally compute $\shrB{\MSB(\val)} = \shrB{\vv} \xor \shrB{\vr_1} \xor \shrB{\vr_2}$.
	\end{myitemize}        
\end{protocolbox}
%
On a high level, protocol proceeds as follows: $P_0, P_1$ garbles the circuit $GC$ and send $GC$ to $P_2$ along with the keys corresponding to the inputs and the decoding information. $P_2$ upon evaluating $GC$ obtains $\vv = \MSB(\val) \xor \vr_1 \xor \vr_2$ in clear and sends $\vv$ along with a hash of the actual key corresponding to $\vv$ to $P_1$. $P_1, P_2$ then jointly generate $\shrB{\vv}$. Servers then XOR $\shrB{\vv}$ to $\shrB{\vr_1}$ and $\shrB{\vr_2}$ that are generated in the preprocessing phase to obtain the final result.
%
\paragraph{Bit2A} 
Protocol $\PiBitA(\Partyset, \shrB{\bitb})$ (Fig.~\ref{fig:piBit2A}) enables servers in $\Partyset$ to compute the arithmetic sharing of a single bit $\bitb$, given its  $\shrB{\cdot}$-sharing. We denote   the value of bit $\bitb$ in the ring $\Z{\ell}$ as $\arval{\bitb}$. Now observing that $\arval{\bitb} = \arval{\bv{\bitb} \xor \av{\bitb}} = \arval{\bv{\bitb}} + \arval{\av{\bitb}} - 2 \arval{\bv{\bitb}} \arval{\av{\bitb}}$, we compute an arithmetic sharing of $\arval{\bv{\bitb}}$,  $\arval{\av{\bitb}}$ and their product $\arval{\bv{\bitb}} \arval{\av{\bitb}}$  to obtain arithmetic sharing of $\arval{\bitb}$. To compute an arithmetic sharing of  $\arval{\av{b}}$, we use  $\arval{\av{b}} = \arval{\sqr{\av{b}}_{1} \xor \sqr{\av{b}}_{2}} = \arval{\sqr{\av{b}}_{1}}+ \arval{\sqr{\av{b}}_{2}} - 2 \arval{\sqr{\av{b}}_{1}} \arval{\sqr{\av{b}}_{2}}$ and compute an arithmetic sharing of $\arval{\sqr{\av{b}}_{1}}$,  \arval{\sqr{\av{b}}_{2}} and their product $ \arval{\sqr{\av{b}}_{1}} \arval{\sqr{\av{b}}_{2}}$ as follows.  $P_0, P_j$ for $j \in \EInSet$ execute $\piJSh$ on $\arval{\sqr{\av{b}}_{j}}$ to generate $\shr{\arval{\sqr{\av{b}}_{j}}}$. Servers then execute $\piMult$ on $\shr{\arval{\sqr{\av{b}}_{1}}}$ and $\shr{\arval{\sqr{\av{b}}_{2}}}$ to generate $\shr{\arval{\sqr{\av{b}}_{1}} \arval{\sqr{\av{b}}_{2}}}$, followed by locally computing the result. The computation of $\shr{\arval{\bitb}}$ follows similarly.
\begin{protocolsplitbox}{$\PiBitA(\Partyset, \shrB{\bitb})$}{Bit2A Protocol}{fig:piBit2A}
	\justify
	\algoHead{Preprocessing:}
	\begin{myitemize}
		\item[--] $P_0, P_j$ for $j \in \EInSet$ execute $\piJSh$ on $\arval{\sqr{\av{\bitb}}_{j}}$ to generate $\shr{\arval{\sqr{\av{\bitb}}_{j}}}$.
		\item[--] Servers in $\Partyset$ execute $\piMult(\Partyset, \arval{\sqr{\av{\bitb}}_{1}}, \arval{\sqr{\av{\bitb}}_{2}})$ to generate $\shr{\vu}$ where $\vu =\arval{ \sqr{\av{\bitb}}_{1}}\arval{ \sqr{\av{\bitb}}_{2}}$, followed by locally computing $\shr{\arval{\av{\bitb}}} = \shr{\arval{\sqr{\av{\bitb}}_{1}}} + \shr{\arval{\sqr{\av{\bitb}}_{2}}} - 2 \shr{\vu}$.
		\item[--] Servers in $\Partyset$ execute the preprocessing phase of $\piMult(\Partyset, \arval{\bv{\bitb}}, \arval{\av{\bitb}})$ where $\vv = \arval{\bv{\bitb}} \arval{\av{\bitb}}$.
	\end{myitemize}  
	\justify
	\algoHead{Online:}
	\begin{myitemize}
		\item[--] $P_1, P_2$ execute  $\piJSh$ on $\arval{\bv{\bitb}}$ to generate $\shr{\arval{\bv{\bitb}}}$.
		\item[--] Servers in $\Partyset$ execute online phase of $\piMult(\Partyset, \arval{\bv{\bitb}}, \arval{\av{\bitb}})$ to generate $\shr{\vv}$ where $\vv = \arval{ \bv{\bitb}} \arval{ \av{\bitb}}$, followed by locally computing $\shr{\arval{\bitb}} = \shr{\arval{\bv{\bitb}}} + \shr{\arval{\av{\bitb}}} - 2 \shr{\vv}$.
	\end{myitemize}        
\end{protocolsplitbox}
%
\vspace{-2mm}

\section{Building Layer-II Primitives}
\label{sec:privML}
Since ML algorithms involve operating over decimals, we use signed two's \ADDED{complement} form~\cite{MohasselZ17, MR18, CCPS19} over the ring $\Z{\ell}$ to represent the decimal numbers. Here, the most significant bit ($\MSB$) denotes the sign and the last $d$ bits are reserved for the fractional part. We choose $\ell = 64$ and $d = 13$, which leaves $50$ bits for the integer part. The $\ell$-bit strings are treated as elements of $\Z{\ell}$. A product of two numbers from this domain requires $d$ to be $26$ bits if we do not want to compromise on the accuracy. However, for training tasks which require many sequential multiplications, this might lead to an overflow. Hence, a method for truncation is required in order to cast the product result back in the aforementioned format. Also, typically ML algorithms perform multiplication in the form of dot product. We present below protocols for-- \ADDED{(a) dot product, (b) truncation, (c) dot product with truncation, (d) secure comparison, and (e) non-linear activation functions}. We defer the communication complexity analysis and security proof of all our constructs to Appendix~\ref{appsec:LayerII} and Appendix~\ref{app:Security} respectively.
\paragraph{Dot Product}
Protocol $\piDotP$  (Fig.~\ref{fig:piDotP}) enables servers in $\Partyset$ to generate $\shrd$-sharing of $\vecX \band \vecY$, given the $\shrd$-sharing of vectors $\vecX$ and $\vecY$. By $\shrd$-sharing of a vector $\vecX$ of size $\nf$, we mean each element $\vx_i \in \Z{\ell}$ of it, for $i \in [\nf]$, is $\shrd$-shared. A naive solution is to view the problem as $\nf$ instances of $\piMult$, where the $i^{th}$ instance computes $\wz_i = \wx_i \cdot \wy_i$. The final result can then be obtained by locally adding the shares of $\wz_i$ corresponding to all the instances. But this would require a communication that is linearly dependent on the size of the vectors (i.e. $\nf$). We make the communication of $\piDotP$ in the online phase independent of $\nf$ as follows: Instead of reconstructing each $\bv{\wz_{i}}$ separately to compute $\bv{\wz}$ with $\wz = \vecX \band \vecY$, $P_1, P_2$ locally compute $\sqr{\bv{\wz}} = \sqr{\bv{{\wz}_1}} + \ldots + \sqr{\bv{{\wz}_\nf}}$ and reconstruct $\bv{\wz}$. Moreover, instead of sending $\starbeta{{\wz}_i}$ for each $\wz_i = \wx_i \cdot \wy_i$, $P_0$ can ``combine" all the $\starbeta{{\wz}_i}$ values and send a single $\starbeta{{\wz}}$ to $P_1, P_2$ for verification. In detail, $P_0$ computes $\starbeta{{\wz}} = \sum_{i=1}^{\nf} \starbeta{{\wz}_i}$ and sends a hash of the same to both $P_1$ and $P_2$, who then can cross check with a hash of $\bv{\wz} - \sum_{i=1}^{\nf} (\bv{{\wx}_i} \cdot \bv{{\wy}_i} - \psi_i)$.
%
\begin{protocolbox}{$\piDotP(\Partyset, \{\shr{\wx_i},\shr{\wy_i}\}_{i \in [\nf]})$}{Dot Product Protocol}{fig:piDotP}
	\justify
	\algoHead{Preprocessing:}
	\begin{myitemize}
		\item[--] Servers in $\Partyset$ execute preprocessing phase of $\piMult(\Partyset, \wx_i, \wy_i)$ for each pair $(\wx_i, \wy_i)$ where $i \in [\nf]$ and $\wz_i = \wx_i \wy_i$. $P_0$ obtains $\Chi_i$, while $P_j$, for $j \in \EInSet$, obtains $\GammaxyiV{j}$ and $\psi_i$.
		\item[--] $P_0$ computes $\Chi = \sum_{i=1}^{\nf} \Chi_i , \Gammaxy = \sum_{i=1}^{\nf} \GammaV{\wx_{i}\wy_{i}}$, while $P_j$ for $j \in \EInSet$ computes  $\GammaxyV{j} = \sum_{i=1}^{\nf} \GammaxyiV{j}, \psi = \sum_{i} \psi_i$. 
		\item[--] $P_0, P_j$ for $j \in \EInSet$ compute $\sqr{\av{\wz}}_{j} = \sum_{i=1}^{\nf} \sqr{\av{\wz_{i}}}_{j}$. 
	\end{myitemize}  
    \vspace{-4mm}
	\justify
	\algoHead{Online Phase:}
	\begin{myitemize}
		\item[--] $P_j$ for $j \in \EInSet$ computes $\sqr{\bv{\wz}}_{j} = \sum_{i=1}^{\nf} ((j-1)\bv{\wx_{i}}\bv{\wy_{i}}-  \bv{\wx_{i}}\sqr{\av{\wy_{i}}}_{j}- \bv{\wy_{i}}\sqr{\av{\wx_{i}}}_{j}) + \GammaxyV{j} + \sqr{\av{\wz}}_{j}$ and mutually exchanges $\sqr{\bv{\wz}}_{j}$ to reconstruct $\bv{\wz}$.  
		\item[--] $P_0$ computes $\starbeta{\wz} = -\sum_{i=1}^{\nf} (\bv{\wx_i}+\gv{\wx_i})\av{\wy_i} - \sum_{i=1}^{\nf} (\bv{\wy_i}+\gv{\wy_i})\av{\wx_i} + \av{\wz} + 2\Gammaxy + \Chi$ and sends $\Hash(\starbeta{\wz})$ to $\ESet$.
		\item[--] $P_j$ for $j \in \EInSet$ $\abort$ if $\Hash(\starbeta{\wz}) \ne \Hash(\bv{\wz}-\sum_{i=1}^{\nf}\bv{\wx_i}\bv{\wy_i} + \psi)$.
		\item[--] $P_1$ sends $\bv{\wz}+\gv{\wz}$ and $P_2$ sends the $\Hash(\bv{\wz}+\gv{\wz})$ to $P_0$. $P_0$ will $\abort$ if it receives inconsistent values. 
	\end{myitemize}        
\end{protocolbox}
\vspace{-3mm}
%
\paragraph{Truncation}
A truncation protocol enables the servers to compute $\shr{\val^d}$ from $\shr{\val}$, where $\val^d$ denotes the truncated   value of $\val$ (right-shifted value of $\val$ by $d$ bit positions, where $d$ is the number of bits allocated for the fractional part).  
SecureML\cite{MohasselZ17} proposed an efficient truncation method for 2 parties  where the parties locally truncate their shares after every multiplication. ABY3\cite{MR18} showed that this method fails when extended to 3-party, and proposed an alternative way using a shared truncated pair $(\vr, \vrt)$, for a random $\vr$, to achieve truncation. Their method of truncating the shares  of the product after evaluating a multiplication gate preserves the underlying truncated value with very high probability. We follow the technique of ABY3 and primarily differ in the way in which $(\vr, \vrt)$ is generated. With the random truncation pair $(\vr, \vrt)$ and a value $\val$ to be truncated, both available in $\shrd$-shared form,  the truncated $\val$ in $\shrd$-shared format can be obtained by opening $(\val - \vr)$, truncating it and then adding it to $\shr{\vrt}$. Below we present a protocol that prepares the random truncation pair.

%

\begin{mypbox}{$\piTrunc(\Partyset)$}{Generating Random Truncated Pair $(\vr, \vrt)$}{fig:piTr}
	\justify	
	\begin{myitemize}
		\item[--] $P_0, P_j$ for $j \in \EInSet$ sample random $R_j \in \Z{\ell}$. $P_0$ sets $\vr = R_1 + R_2$ while $P_j$ sets $\sqr{\vr}_{j} = R_j$. $P_j$ sets $\sqr{\vrd}_{j}$ as the ring element that has last $d$ bits of $\vr_j$ in the last $d$ positions and $0$ elsewhere.	
		\item[--] $P_0$ locally truncates $\vr$ to obtain $\vrt$ and executes $\piSh(P_0,\vrt)$ to generate $\shr{\vrt}$. $P_1$ locally sets $\sqrA{\vrt} = \bv{\vrt}-\sqr{\av{\vrt}}_{1}$, while $P_2$ sets $\sqrB{\vrt} =-\sqr{\av{\vrt}}_{2}$.
		\item[--] $P_1$ computes $\vu = \sqr{\vr}_{1} - 2^d \sqr{\vrt}_{1} - \sqr{\vrd}_{1}$ and sends $\Hash(\vu)$ to $P_2$.
		\item[--] $P_2$ locally computes $\vv = 2^d \sqr{\vrt}_{2} + \sqr{\vrd}_{2} - \sqr{\vr}_{2}$ and $\abort$ if $\Hash(\vu) \ne \Hash(\vv)$.
	\end{myitemize}  
\end{mypbox}
\vspace{-3mm}
Protocol $\piTrunc(\Partyset)$ (Fig.~\ref{fig:piTr}) generates a pair $(\sqr{\vr}, \shr{\vrt})$ for a random $\vr$.  Servers $P_0, P_j$ for $j \in \EInSet$ sample random value $R_j \in \Z{\ell}$ followed by $P_0$ locally truncating $\vr = R_1 + R_2$ to obtain $\vrt$. Note that $\vr = 2^d \vrt + \vrd$ where $\vrd$ denotes the ring element that has last $d$ bits of $\vr$ in the last $d$ positions and $0$ elsewhere. $P_0$ then generates $\shr{\vrt}$ by executing the sharing protocol $\piSh$. To verify the correctness of sharing performed by $P_0$, servers $P_1, P_2$ compute a $\sqr{\cdot}$-sharing of $\va = (\vr - 2^d \vrt + \vrd)$, given $(\sqr{\vr}, \shr{\vrt})$ and checks if $\va = 0$. To optimize communication, $P_1$ sends a hash of his share $\Hash(\sqr{\va}_1)$ to $P_2$, who $\abort$s if the received hash value mismatches with $\Hash(-\sqr{\va}_1)$. 

To see the correctness, it suffices to show that $\vu = \vv$ where $\vu = \sqr{\vr}_{1} - 2^d \sqr{\vrt}_{1} - \sqr{\vrd}_{1}$ and $\vv = 2^d \sqr{\vrt}_{2} + \sqr{\vrd}_{2} - \sqr{\vr}_{2}$. We start from the observation that $\vr = 2^d \vrt + \vrd$.

\resizebox{.94\linewidth}{!}{
	\begin{minipage}{\linewidth}
		\begin{align*}
		\vr                                &= 2^d \vrt + \vrd\\
		\sqr{\vr}_{1} + \sqr{\vr}_{2}      &= 2^d (\sqr{\vrt}_{P_1} + \sqr{\vrt}_{P_2}) + (\sqr{\vrd}_{P_1} + \sqr{\vrd}_{P_2})\\
		\sqr{\vr}_{1} - 2^d \sqr{\vrt}_{1} - \sqr{\vrd}_{1} &= 2^d \sqr{\vrt}_{2} + \sqr{\vrd}_{2} - \sqr{\vr}_{2}\\
		\vu &= \vv
		\end{align*}
	\end{minipage}}
$\piTrunc(\Partyset)$ can entirely be run in the preprocessing phase. Our dot product with truncation, presented below, will invoke it in the preprocessing phase. 
\paragraph{Dot Product with Truncation}
%
Protocol $\piDotPTr(\Partyset,\newline \{\shr{\wx_i},\shr{\wy_i}\}_{i \in [\nf]})$ (\boxref{fig:piDotPTr}) enables servers in $\Partyset$ to generate $\shrd$-sharing of truncated value of $\wz = \vecX \band \vecY$  denoted as $\trunc{\wz}$, given the $\shrd$-sharing of vectors $\vecX$ and $\vecY$. To achieve the goal,  we modify our dot product protocol $\piDotP$ in a way  that does not inflate the online cost. This is unlike ABY3, which requires an additional reconstruction in the online phase.

In the preprocessing phase, along with the steps of $\piDotP$, the servers execute $\piTrunc$ to generate a truncation pair $(\vr, \vrt)$. In the online phase, the servers $\ESet$ locally compute $\sqd$-sharing of $(\wz - \vr)$ (instead of $\sqr{\bv{\wz}}$) where $\wz = \vecX \band \vecY$. This is followed by $\ESet$ locally truncating $(\wz - \vr)$ to obtain $\trunc{(\wz - \vr)}$ and generating $\shrd$-sharing of the same by executing $\piJSh$ protocol. Finally, the servers locally compute $\shrd$-sharing of $\wz$ by adding the shares of $\trunc{(\wz - \vr)}$ and $\shr{\vrt}$. To ensure the correctness of the computation, the steps of $P_0$ are modified such that $P_0$ will be computing ${(\wz - \vr)}^\star$ instead of $\starbeta{\wz}$.

\begin{protocolbox}{$\piDotPTr(\Partyset, \{\shr{\wx_i},\shr{\wy_i}\}_{i \in [\nf]})$}{Dot Product Protocol with Truncation}{fig:piDotPTr}
	\justify
	\algoHead{Preprocessing:}
    \begin{myitemize}
    	\item[--] Servers in $\Partyset$ execute preprocessing phase of $\piDotP(\Partyset, \{\shr{\wx_i},\shr{\wy_i}\}_{i \in [\nf]})$.                 
    	\item[--] In parallel, servers execute $\piTrunc(\Partyset)$ to generate the truncation pair $(\sqr{\vr}, \shr{\vrt})$. Moreover $P_0$ obtains the value $\vr$ in clear.             
    \end{myitemize}
	\justify
	\algoHead{Online:}
	\begin{myitemize}
		\item[--] $P_j$ for $j \in \EInSet$ computes $\sqr{(\wz - \vr)}_j = \sqr{\wz}_j - \sqr{\vr}_j$ where $\sqr{\wz}_j = \sqr{\bv{\wz}}_{j} - \sqr{\av{\wz}}_{j}= \sum_{i=1}^{\nf} ((j-1)\bv{\wx_{i}}\bv{\wy_{i}}-  \bv{\wx_{i}}\sqr{\av{\wy_{i}}}_{j}- \bv{\wy_{i}}\sqr{\av{\wx_{i}}}_{j}) + \GammaxyV{j}$.
		\item[--] $P_j$ for $j \in \EInSet$ mutually exchange $\sqr{(\wz - \vr)}_j$ to reconstruct $(\wz - \vr)$, followed by locally truncating it to obtain $\trunc{(\wz - \vr)}$.
		\item[--] $\ESet$ execute $\piJSh(P_1, P_2, \trunc{(\wz - \vr)})$ to generate $\shr{\trunc{(\wz - \vr)}}$.
		\item[--] Servers in $\Partyset$ locally compute $ \shr{\wz} = \shr{\trunc{(\wz - \vr)}} + \shr{\vrt}$ 
		\item[--] $P_0$ computes $\Psi = -\sum_{i=1}^{\nf} (\bv{\wx_i}+\gv{\wx_i})\av{\wy_i} - \sum_{i=1}^{\nf} (\bv{\wy_i}+\gv{\wy_i})\av{\wx_i} + 2\Gammaxy - \vr$, sets ${(\wz - \vr)}^\star = \Psi + \Chi$ and sends $\Hash({(\wz - \vr)}^\star)$ to both $P_1$ and $P_2$. 
		\item[--] $P_j$ for $j \in \EInSet$ $\abort$s if $\Hash((\wz - \vr) - \sum_{i=1}^{\nf}\bv{\wx_i}\bv{\wy_i} + \psi) \neq \Hash({(\wz - \vr)}^\star)$. 
	\end{myitemize}
\end{protocolbox}
%

%
\paragraph{Secure Comparison}
Given two values $\wx, \wy \in \Z{\ell}$ in $\shrd$-shared format, secure comparison allows parties to check whether $\wx < \wy$ or not. In fixed-point arithmetic representation, this can be accomplished by checking the sign of $\val = \wx - \wy$, which is stored in its $\MSB$ position. Towards this, servers locally compute $\shr{\val} = \shr{\wx} - \shr{\wy}$ followed by extracting the $\MSB$ using protocol $\piBitExt$ on $\shr{\val}$. For the cases that demand the result in arithmetic sharing format, servers can apply the Bit2A  protocol $\PiBitA$ on the outcome of $\piBitExt$.

\paragraph{Activation Functions}
We consider two widely used activation functions-- i) Rectified Linear Unit (ReLU) and ii) Sigmoid (Sig).

{\em -- ReLU:}
The ReLU function, defined as $\ReLU(\val) = \maxv(0, \val)$ can be viewed as $\ReLU(\val) = \overline{\bitb} \cdot \val$ where the bit $\bitb = 1$ if $\val < 0$ and $0$ otherwise. Here $\overline{\bitb}$ denotes the complement of bit $\bitb$. Protocol $\piReLU(\Partyset, \shr{\val})$ enables servers in $\Partyset$ to compute $\shrd$-sharing of $\ReLU(\val)$ given the $\shrd$-sharing of $\val \in \Z{\ell}$.

For this, servers first execute the $\MSB$ extraction protocol $\piBitExt$ on $\val$ to obtain $\shrB{\bitb}$. Given $\shrB{\bitb}$, servers locally compute $\shrB{\overline{\bitb}}$ by  setting $\bv{\overline{\bitb}} = 1 \xor \bv{\bitb}$. Servers then execute Bit2A protocol $\PiBitA$ on $\shrB{\overline{\bitb}}$ to generate $\shr{\overline{\bitb}}$. Lastly, servers execute multiplication protocol $\piMult$ on $\overline{\bitb}$ and $\val$ to generate $\shrd$-sharing of the result.

{\em -- Sig:}
We use the MPC-friendly version of the Sigmoid function~\cite{MohasselZ17, MR18, CCPS19}, which is defined as:

\begin{align*}
\Sig(\val) = \left\{
               \begin{array}{lll}
                   0                  & \quad \val < -\frac{1}{2} \\
                   \val + \frac{1}{2} & \quad - \frac{1}{2} \leq \val \leq \frac{1}{2} \\
                   1                  & \quad \val > \frac{1}{2}
               \end{array}
             \right.
\end{align*}

Note that $\Sig(\val) = \overline{\bitb_1} \bitb_2 (\val + 1/2) + \overline{\bitb_2}$, where $\bitb_1 = 1$ if $\val + 1/2 < 0$ and $\bitb_2 = 1$ if $\val - 1/2 < 0$. Protocol $\piSig(\Partyset, \shr{\val})$ is similar to that of $\piReLU$ and therefore we omit the details.

\section{Building PPML and Benchmarking}
\label{sec:Implementation}
We  consider three widely used ML algorithms for our benchmarking and compare with their closest competitors-- i) Linear Regression (training and inference), ii) Logistic Regression (training and inference) and iii) Neural Networks (inference). Training for NN requires conversions to and from Garbled Circuits (for tackling some functions) which are not considered in this work. To obtain fairness in our protocols, the final outcome is reconstructed via fair reconstruction protocol $\pifRec(\Partyset, \shr{\val})$~(\boxref{fig:pifRec}). In addition to the above, we also benchmark the dot product protocol separately as it is a major building block for PPML. We start with the experimental setup.

\paragraph{Benchmarking Environment}
We use a 64-bit ring ($\Z{64}$). The benchmarking is performed over a LAN of $1$Gbps bandwidth and a WAN of $75$Mbps bandwidth. Over the LAN, we use machines equipped with 3.6 GHz Intel Core i7-7700 CPU processor and 32 GB of RAM Memory. The WAN was instantiated using n1-standard-8 instances of Google Cloud\footnote{https://cloud.google.com/} with machines located in East Australia ($P_0$), South Asia ($P_1$) and South East Asia ($P_2$). Over the WAN, machines are equipped with 2.3 GHz Intel Xeon E5 v3 (Haswell) processors supporting hyper-threading, with 8 vCPUs, and 30 GB of RAM Memory. The average round-trip time ($\rtt$) was taken as the time for communicating 1 KB of data between a pair of parties. Over the LAN, the $\rtt$ turned out to be $0.296 ms$. In the WAN, the $\rtt$ values for the pairs $P_0$-$P_1$, $P_0$-$P_2$ and $P_1$-$P_2$ are $152.3 ms$, $60.19 ms$ and $92.63 ms$ respectively. 

\paragraph{Software Details}
We implement our protocols
using the publicly available ENCRYPTO library~\cite{ENCRYPTO} in C++17. We implemented the code of ABY3~\cite{MR18} and ASTRA~\cite{CCPS19} in our environment since they were not publicly available. The collision-resistant hash function was instantiated using SHA-256. We have used multi-threading and our machines were capable of handling a total of 32 threads. Each experiment is run for 20 times and the average values are reported.

\paragraph{Benchmarking Parameter}
\ADDED{We use {\em throughput} ($\TP$) as the benchmarking parameter following ABY3~\cite{MR18} and ASTRA~\cite{CCPS19} as it would help to analyse the effect of improved communication and round complexity in a single shot. Here $\TP$ denotes the number of operations (``iterations" for the case of training and ``queries" for the case of inference) that can be performed in unit time.} 
%
%
We consider minute as the unit time since  most of our protocols over WAN requires more than a second to complete. To analyse the performance of our protocols under various bandwidth settings, we report the performance under the following bandwidths: 25 Mbps, 50Mbps and, 75Mbps. 

We provide the benchmarking for the WAN setting below and defer the same for the LAN setting to Appendix~\ref{app:MicroBenchLAN}. 
\subsection{Dot Product} 
\label{sec:Bench_DOTP}
Here the throughput is computed as the number of dot products performed per minute (\#dotp/min) and the same is computed for both preprocessing and online phases separately.  

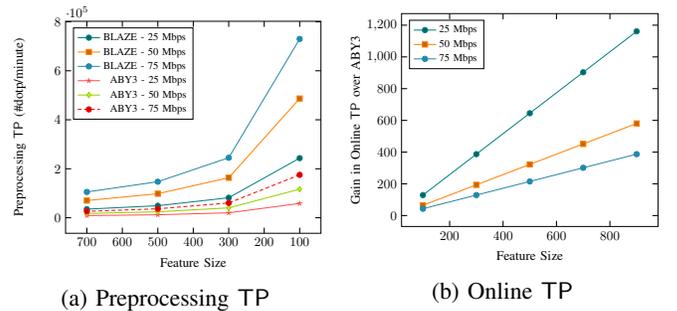
\begin{figure}[htb!]
	\centering
	\begin{subfigure}{.49\textwidth}
		\centering
		\resizebox{.98\textwidth}{!}{
			\begin{tikzpicture}
			\begin{axis}[legend pos=north west, xlabel={Feature Size}, ylabel={Preprocessing $\TP$ (\#dotp/minute)}, x dir=reverse, ,cycle list name=exotic]
			\addplot plot coordinates { (100, 243326.73) (300, 81647.84) (500, 49053.89) (700, 35058.49)};
			\addlegendentry{{\footnotesize BLAZE - 25 Mbps}}
			\addplot plot coordinates { (100, 486653.47) (300, 163295.68) (500, 98107.78) (700, 70116.98)};
			\addlegendentry{{\footnotesize BLAZE - 50 Mbps}}
			\addplot plot coordinates { (100, 729980.20) (300, 244943.52) (500, 147161.68) (700, 105175.46)};
			\addlegendentry{{\footnotesize BLAZE - 75 Mbps}}
			\addplot plot coordinates { (100, 58375.3) (300, 20127.76) (500, 12160.32) (700, 8711.80)};
			\addlegendentry{{\footnotesize ~~ABY3 - 25 Mbps}}
			\addplot plot coordinates { (100, 116750.59) (300, 40255.53) (500, 24320.63) (700, 17423.61)};
			\addlegendentry{{\footnotesize ~~ABY3 - 50 Mbps}}
			\addplot plot coordinates { (100, 175125.89) (300, 60383.29) (500, 36480.95) (700, 26135.41)};
			\addlegendentry{{\footnotesize ~~ABY3 - 75 Mbps}}
			\end{axis}
			\node[align=center,font=\bfseries, xshift=2.5em, yshift=-2em] (title) at (current bounding box.north) {};
			\end{tikzpicture}
		}
		\caption{\small Preprocessing $\TP$}\label{fig:DOTPPreWAN}
	\end{subfigure}
	\begin{subfigure}{.49\textwidth}
		\centering
		\resizebox{.98\textwidth}{!}{
			\begin{tikzpicture}
			\begin{axis}[legend pos=north west, xlabel={Feature Size}, ylabel={Gain in Online $\TP$ over ABY3}, cycle list name=exotic]
			\addplot plot coordinates { (100, 129.01) (300, 387.03) (500, 645.06) (700, 903.08) (900, 1161.10)};
			\addlegendentry{{\footnotesize 25 Mbps}}
			\addplot plot coordinates { (100, 64.51) (300, 193.52) (500, 322.53) (700, 451.54) (900, 580.55)};
			\addlegendentry{{\footnotesize 50 Mbps}}
			\addplot plot coordinates { (100, 43) (300, 129.01) (500, 215.02) (700, 301.03) (900, 387.03)};
			\addlegendentry{{\footnotesize 75 Mbps}}
			\end{axis}
			\node[align=center,font=\bfseries, xshift=2.5em, yshift=-2em] (title) at (current bounding box.north) {};
			\end{tikzpicture}
		}
		\caption{\small Online $\TP$}\label{fig:DOTPWANOn}
	\end{subfigure}
	\vspace{-3mm}
	\caption{\small Throughput ($\TP$) Comparison of ABY3 and BLAZE over varying Bandwidths}\label{fig:DOTPWAN}
\end{figure}

For the preprocessing phase, we plot the throughput of the dot product protocol of BLAZE (\figref{DOTPPreWAN}) and ABY3  over vectors of length ranging from 100 to 1000. We note at least a gain of $4\times$, which is a consequence of  $4\times$ improvement in communication, over ABY3. An interesting observation to be made here is that our protocol, over the bandwidth of 25Mbps, gives better throughput when compared to ABY3 even over a higher bandwidth of 75Mbps. 

For the online phase (\figref{DOTPWANOn}), we plot the gain over ABY3 in terms of throughput. We observe an appreciable gain in throughput which is a direct corollary of the communication cost our protocol being independent of the vector size. Concretely, for a bandwidth of 50 Mbps, our gain ranges from $64\times$ to $580\times$. Note that, with an increase in bandwidth there is a drop in the gain. This is because even at a bandwidth of 25 Mbps the maximum attainable throughput cannot be handled by our processors. For a bandwidth of 8 Mbps, the maximum attainable throughput is within our processing capacity, where we observe throughput gain ranging from $400\times$ to $3600\times$. This showcases the practicality of our constructions over low-end networks. 

In the preprocessing phase, over all the bandwidths under consideration, the maximum attainable throughput lies well within the processing capacity of our machines. Consequentially, we do not observe a drop in the throughput gain with increasing bandwidth, as is seen in the online phase. This is the reason why we choose to plot the actual throughput values instead of the gain in the case of the preprocessing phase. On increasing the processing capacity we expect a consistent gain in online throughput with increasing bandwidth.    

\subsection{ML Training} 
\label{sec:Bench_MLTrain}
In this section, we explore the training phase of Linear Regression and Logistic Regression algorithms. The training phase can be divided into two stages-- (i) a {\em forward propagation} phase, where the model computes the output given the input; (ii) a {\em backward propagation} phase, where the model parameters are adjusted according to the difference in the computed output and the actual output. For our benchmarking, we define one {\em iteration} in the training phase as one forward propagation followed by a backward propagation. Our performance improvement over ABY3 is reported in terms of the number of iterations over feature size varying from 100 to 1000, and a batch size of $B \in\{128, 256,512\}$. Batching~\cite{MohasselZ17,MR18} is a common optimization where $n$ samples are divided into batches of size B and a combined update function is applied to the weight vectors. 
\ADDED{
In order to analyse the performance over a wide range of features and batch sizes, we choose to benchmark over synthetic datasets following ABY3~\cite{MR18}.
}

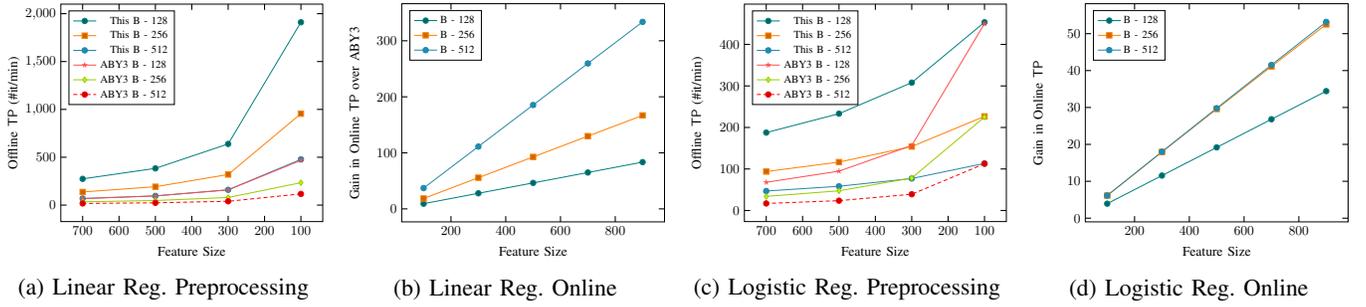
\begin{figure*}
	\centering
	\begin{subfigure}{.24\textwidth}
		\centering
		\resizebox{.99\textwidth}{!}{
			\begin{tikzpicture}
			\begin{axis}[legend pos=north west, xlabel={Feature Size}, ylabel={Offline $\TP$ (\#it/min)}, x dir=reverse, ,cycle list name=exotic]
			\addplot plot coordinates { (100, 1910.45) (300, 638.94) (500, 383.62) (700, 274.04)};
			\addlegendentry{{\footnotesize ~~~This B - 128}}
			\addplot plot coordinates { (100, 955.22) (300, 319.47) (500, 191.81) (700, 137.04)};
			\addlegendentry{{\footnotesize ~~~This B - 256}}
			\addplot plot coordinates { (100, 477.61) (300, 159.73) (500, 95.90) (700, 68.52)};
			\addlegendentry{{\footnotesize ~~~This B - 512}}
			\addplot plot coordinates { (100, 467.72) (300, 158.61) (500, 95.50) (700, 68.32)};
			\addlegendentry{{\footnotesize ABY3 B - 128}}
			\addplot plot coordinates { (100, 233.86) (300, 79.31) (500, 47.75) (700, 34.16)};
			\addlegendentry{{\footnotesize ABY3 B - 256}}
			\addplot plot coordinates { (100, 116.93) (300, 39.65) (500, 23.87) (700, 17.08)};
			\addlegendentry{{\footnotesize ABY3 B - 512}}
			\end{axis}
			\node[align=center,font=\bfseries, xshift=2.5em, yshift=-2em] (title) at (current bounding box.north) {};
			\end{tikzpicture}
		}
		\caption{\small Linear Reg. Preprocessing}\label{fig:RegWANa}
	\end{subfigure}
	\begin{subfigure}{.24\textwidth}
		\centering
		\resizebox{.99\textwidth}{!}{
			\begin{tikzpicture}
			\begin{axis}[legend pos=north west, xlabel={Feature Size}, ylabel={Gain in Online $\TP$ over ABY3}, cycle list name=exotic]
			\addplot plot coordinates { (100, 9.27) (300, 27.81) (500, 46.35) (700, 64.89) (900, 83.43)};
			\addlegendentry{{\footnotesize B - 128}}
			\addplot plot coordinates { (100, 18.54) (300, 55.62) (500, 92.70) (700, 129.78) (900, 166.86)};
			\addlegendentry{{\footnotesize B - 256}}
			\addplot plot coordinates { (100, 37.08) (300, 111.24) (500, 185.40) (700, 259.56) (900, 333.72)};
			\addlegendentry{{\footnotesize B - 512}}
			\end{axis}
			\node[align=center,font=\bfseries, xshift=2.5em, yshift=-2em] (title) at (current bounding box.north) {};
			\end{tikzpicture}
		}
		\caption{\small Linear Reg. Online}\label{fig:RegWANb}
	\end{subfigure}
	\begin{subfigure}{.24\textwidth}
		\centering
		\resizebox{.99\textwidth}{!}{
			\begin{tikzpicture}
			\begin{axis}[legend pos=north west, xlabel={Feature Size}, ylabel={Offline $\TP$ (\#it/min)}, x dir=reverse, ,cycle list name=exotic]
			\addplot plot coordinates { (100, 453.36) (300, 307.94) (500, 233.15) (700, 187.59)};
			\addlegendentry{{\footnotesize ~~~This B - 128}}
			\addplot plot coordinates { (100, 226.68) (300, 153.97) (500, 116.57) (700, 93.80)};
			\addlegendentry{{\footnotesize ~~~This B - 256}}
			\addplot plot coordinates { (100, 113.34) (300, 76.98) (500, 58.29) (700, 46.90)};
			\addlegendentry{{\footnotesize ~~~This B - 512}}
			\addplot plot coordinates { (100, 450.14) (300, 156.54) (500, 94.74) (700, 67.93)};
			\addlegendentry{{\footnotesize ABY3 B - 128}}
			\addplot plot coordinates { (100, 225.07) (300, 78.27) (500, 47.37) (700, 33.96)};
			\addlegendentry{{\footnotesize ABY3 B - 256}}
			\addplot plot coordinates { (100, 112.54) (300, 39.13) (500, 23.69) (700, 16.98)};
			\addlegendentry{{\footnotesize ABY3 B - 512}}
			\end{axis}
			\node[align=center,font=\bfseries, xshift=2.5em, yshift=-2em] (title) at (current bounding box.north) {};
			\end{tikzpicture}
		}
		\caption{\small Logistic Reg. Preprocessing}\label{fig:RegWANc}
	\end{subfigure}
	\begin{subfigure}{.24\textwidth}
		\centering
		\resizebox{.99\textwidth}{!}{
			\begin{tikzpicture}
			\begin{axis}[legend pos=north west, xlabel={Feature Size}, ylabel={Gain in Online $\TP$}, cycle list name=exotic]
			\addplot plot coordinates { (100, 3.93) (300, 11.55) (500, 19.18) (700, 26.81) (900, 34.43)};
			\addlegendentry{{\footnotesize B - 128}}
			\addplot plot coordinates { (100, 6.13) (300, 17.93) (500, 29.59) (700, 41.12) (900, 52.51)};
			\addlegendentry{{\footnotesize B - 256}}
			\addplot plot coordinates { (100, 6.14) (300, 18.01) (500, 29.80) (700, 41.53) (900, 53.19)};
			\addlegendentry{{\footnotesize B - 512}}
			\end{axis}
			\node[align=center,font=\bfseries, xshift=2.5em, yshift=-2em] (title) at (current bounding box.north) {};
			\end{tikzpicture}
		}
		\caption{\small Logistic Reg. Online}\label{fig:RegWANd}
	\end{subfigure}
	\vspace{-4mm}
	\caption{\small Throughput ($\TP$) Comparison of ABY3 and BLAZE for ML Training}\label{fig:RegWAN}
\end{figure*}

\paragraph{\bf{Linear Regression}}
In Linear Regression, one iteration comprises of \ADDED{updating} the weight vector $\vecW$ using the gradient descent algorithm (GD). It is updated according to the following function:
\begin{align*}
	\vecW = \vecW - \frac{\alpha}{B} \Mat{X}_i^{T}  \circ ((\Mat{X}_i \circ \vecW) - \Mat{Y}_i)
\end{align*}
where $\alpha$ denotes the learning rate and $\Mat{X}_i$ denotes a subset of batch size B, randomly selected from the entire dataset in the $i$th iteration. The forward propagation involves computing $\Mat{X}_i \circ \vecW$, while the backward propagation consists of updating the weight vector. The update function requires computation of a series of matrix multiplications, which can be achieved using dot product protocols. The update function, as mentioned earlier, can be computed entirely using $\shrd$ shares as:
\begin{align*}
	\shr{\vecW} = \shr{\vecW} - \frac{\alpha}{B} \shr{\Mat{X}_j^{T}}  \circ ((\shr{\Mat{X}_j} \circ \shr{\vecW}) - \shr{\Mat{Y}_j})
\end{align*}
The operations of subtraction as well as multiplication by a public constant can be performed locally.

We compare the throughput for Linear Regression in \figref{RegWAN}. \figref{RegWANa} depicts throughput in the preprocessing phase, and \figref{RegWANb} illustrates the online throughput gain over ABY3. Since Linear Regression primarily involves computing multiple dot products, the underlying efficient dot product protocol improves the performance drastically. As a result, in the preprocessing phase, we observe a gain of $4\times$ and, in the online phase, performance gain for a batch size of 128 ranges from $9.2 \times$ to $83.4\times$. The performance gain in the online phase increases significantly for larger batch sizes and goes all the way up to $333\times$ for a batch size of 512.

\paragraph{\bf{Logistic Regression}}

The training in Logistic Regression, is similar to the case of Linear Regression, with an additional application of sigmoid activation function over  $\Mat{X}_i \circ \vecW$ in the forward propagation. Precisely, the update function for $\vecW$ is as follows:

\begin{align*}
	\vecW = \vecW - \frac{\alpha}{B} \Mat{X}_i^{T}  \circ \left(\Sig(\Mat{X}_i \circ \vecW)-\Mat{Y}_i\right)
\end{align*}

The performance of the training phase in Logistic Regression is analysed in \figref{RegWAN}. The throughput in the preprocessing phase is depicted in \figref{RegWANc}, while \figref{RegWANd} showcases the online throughput gains. The improvements seen in Linear Regression are carried over to this case as well. An overall drop in the throughput is observed both in the preprocessing as well as in the online phase because of the overhead caused by the sigmoid activation function. In the preprocessing phase, our protocol for the largest batch size under consideration outperforms that of ABY3 over the smallest batch size. In the online phase, improvements range from $4.17\times$ to $36.60\times$ for the batch size of 128, compared to ABY3. The primary reason for this gain is our efficient method for $\MSB$ extraction, which requires $\mathbf{2}$ rounds in the online phase, as opposed to $1 + \log\ell$ rounds of communication in ABY3. Similar to Linear Regression, an increase in the online throughput gain can be observed for larger batch sizes.

\tabref{MLTrainTP} provides concrete details for the training phase of Linear Regression and Logistic Regression algorithms over a batch size of 128 and a feature size of 784. More details are provided in Appendix~\ref{app:MicroBenchWAN}.
\begin{table}[htb!]
	\centering
	\resizebox{.9\textwidth}{!}{
		\begin{tabular}{c | c | r | c | r | c }
			\toprule
			\multirow{2}[2]{*}{Algorithm} & \multirow{2}[2]{*}{Ref.} & \multicolumn{2}{c|}{Preprocessing} & \multicolumn{2}{c}{Online}\\ 
			\cmidrule{3-6}
			& & $\TP$ & Gain & $\TP$ & Gain\\
			\midrule
			\multirow{2}{*}{\makecell{Linear\\Regression}}
			& ABY3       & 61.02 & \multirow{2}{*}{$\mathbf{4.01\times}$} & 30.61  & \multirow{2}{*}{$\mathbf{145.35\times}$}\\ 
			& {\bf BLAZE} & $\mathbf{244.74}$ &  & $\mathbf{4449.55}$ & \\ 
			\midrule
			\multirow{2}{*}{\makecell{Logistic\\Regression}}
			& ABY3       & 60.71 & \multirow{2}{*}{$\mathbf{4.02\times}$} 
			& 60.99  & \multirow{2}{*}{$\mathbf{31.89\times}$}\\ 
			& {\bf BLAZE} & $\mathbf{243.81}$ &  & $\mathbf{1945.24}$ & \\ 
			\bottomrule
		\end{tabular}
	}
	\caption{\small Throughput ($\TP$) for ML Training for a batch size B-128 and feature size $\nf$-784\label{tab:MLTrainTP}}
\end{table}

\begin{figure}[htb!]
	\centering
	\begin{subfigure}{.49\textwidth}
		\centering
		\resizebox{.98\textwidth}{!}{
			\begin{tikzpicture}
			\begin{axis}[legend pos=north west, xlabel={Feature Size}, ylabel={Gain in Online $\TP$ over ABY3}, cycle list name=exotic]
			\addplot plot coordinates { (100, 18.54) (300, 55.62) (500, 92.70) (700, 129.78) (900, 166.86)};
			\addlegendentry{{\footnotesize 25 Mbps}}
			\addplot plot coordinates { (100, 9.27) (300, 27.81) (500, 46.35) (700, 64.89) (900, 83.43)};
			\addlegendentry{{\footnotesize 50 Mbps}}
			\addplot plot coordinates { (100, 6.18) (300, 18.54) (500, 30.90) (700, 43.26) (900, 55.62)};
			\addlegendentry{{\footnotesize 75 Mbps}}
			\end{axis}
			\node[align=center,font=\bfseries, xshift=2.5em, yshift=-2em] (title) at (current bounding box.north) {};
			\end{tikzpicture}
		}
		\caption{\small Linear Regression}\label{fig:RegWANBWa}
	\end{subfigure}
	\begin{subfigure}{.49\textwidth}
		\centering
		\resizebox{.98\textwidth}{!}{
			\begin{tikzpicture}
			\begin{axis}[legend pos=north west, xlabel={Feature Size}, ylabel={Gain in Online $\TP$ over ABY3}, cycle list name=exotic]
			\addplot plot coordinates { (100, 6.11) (300, 17.77) (500, 29.17) (700, 40.31) (900, 51.21)};
			\addlegendentry{{\footnotesize 25 Mbps}}
			\addplot plot coordinates { (100, 4.17) (300, 12.28) (500, 20.38) (700, 28.49) (900, 36.60)};
			\addlegendentry{{\footnotesize 50 Mbps}}
			\addplot plot coordinates { (100, 2.78) (300, 7.70) (500, 12.79) (700, 17.87) (900, 22.95)};
			\addlegendentry{{\footnotesize 75 Mbps}}
			\end{axis}
			\node[align=center,font=\bfseries, xshift=2.5em, yshift=-2em] (title) at (current bounding box.north) {};
			\end{tikzpicture}
		}
		\caption{\small Logistic Regression}\label{fig:RegWANBWb}
	\end{subfigure}
	\vspace{-3mm}
	\caption{\small Online Throughput ($\TP$) Comparison of ABY3 and BLAZE over varying Bandwidths}\label{fig:RegWANBW}
\end{figure}
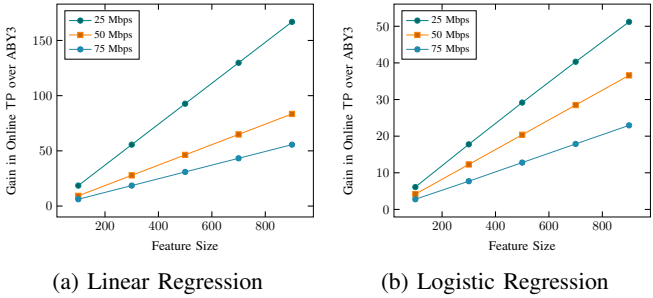

\paragraph{Comparison over varying Bandwidths}
Here, we analyse the performance of the training algorithms in the online phase over varying bandwidths. In \figref{RegWANBWa}, we plot the gain in online throughput of Our protocol over ABY3 for the Linear Regression algorithm for the bandwidths-- 25Mbps, 50Mbps, and 75Mbps. We observe that the improvement in communication cost is even more conspicuous for lower bandwidth. The gain over the bandwidth of 75Mbps ranges from $6.18\times$ to $ 55.62\times$ over various feature sizes, while over a bandwidth of 25Mbps it ranges from $18.54\times$ to $ 166.86\times$. This shows the practicality of our protocol over low bandwidths.  

A similar trend is observed for the case of Logistic Regression and the plot is presented in \figref{RegWANBWb}. For a bandwidth of 75Mbps, the gain in online throughput ranges from $2.78\times$ to $22.95\times$, while for 25Mbps, the range is from $6.11$ to $51.21$.

\subsection{ML Inference} 
\label{sec:Bench_MLInf}
In this section, we benchmark the inference phase of Linear Regression, Logistic Regression, and NN. For inference, the benchmarking parameter is the number of queries processed per minute (\#queries/min). While the details for the online phase are presented here, we defer the details for the preprocessing phase to Appendix~\ref{app:Bench_MLInf_WAN}. 

Like ABY3 and SecureML~\cite{MohasselZ17}, our method for truncation introduces a bit-error at the least significant bit position. The accuracy of the prediction itself, however, ranges from $93.2\%$ for linear regression to $97.8\%$ for NN.

\paragraph{Linear Regression and Logistic Regression}
Inference in the case of Linear Regression and Logistic Regression can be viewed as a single pass of the forward propagation phase. Below we provide the benchmarking for the same over the protocols of ABY3. 

\begin{figure}[htb!]
	\centering
	\begin{subfigure}{.49\textwidth}
		\centering
		\resizebox{.98\textwidth}{!}{
			\begin{tikzpicture}
			\begin{axis}[legend pos=north west, xlabel={Feature Size}, ylabel={Gain in Online $\TP$ over ABY3}, cycle list name=exotic]
			\addplot plot coordinates { (100, 43.30) (300, 129.91) (500, 216.51) (700, 303.12) (900, 389.73)};
			\addlegendentry{{\footnotesize 25 Mbps}}
			\addplot plot coordinates { (100, 21.65) (300, 64.95) (500, 108.26) (700, 151.56) (900, 194.86)};
			\addlegendentry{{\footnotesize 50 Mbps}}
			\addplot plot coordinates { (100, 14.43) (300, 43.30) (500, 72.17) (700, 101.04) (900, 129.91)};
			\addlegendentry{{\footnotesize 75 Mbps}}
			\end{axis}
			\node[align=center,font=\bfseries, xshift=2.5em, yshift=-2em] (title) at (current bounding box.north) {};
			\end{tikzpicture}
		}
		\caption{\small Linear Regression}\label{fig:InfWANva}
	\end{subfigure}
	\begin{subfigure}{.49\textwidth}
		\centering
		\resizebox{.98\textwidth}{!}{
			\begin{tikzpicture}
			\begin{axis}[legend pos=north west, xlabel={Feature Size}, ylabel={Gain in Online $\TP$ over ABY3}, cycle list name=exotic]
			\addplot plot coordinates { (100, 3.16) (300, 9.13) (500, 15.10) (700, 21.07) (900, 27.04)};
			\addlegendentry{{\footnotesize 25 / 50 / 75 Mbps}}
			\end{axis}
			\node[align=center,font=\bfseries, xshift=2.5em, yshift=-2em] (title) at (current bounding box.north) {};
			\end{tikzpicture}
		}
		\caption{\small Logistic Regression}\label{fig:InfWANvb}
	\end{subfigure}
	\vspace{-3mm}
	\caption{\small Online Throughput ($\TP$) Comparison of ABY3 and BLAZE for Linear Regression and Logistic Regression Inference}\label{fig:InfWAN}
\end{figure}
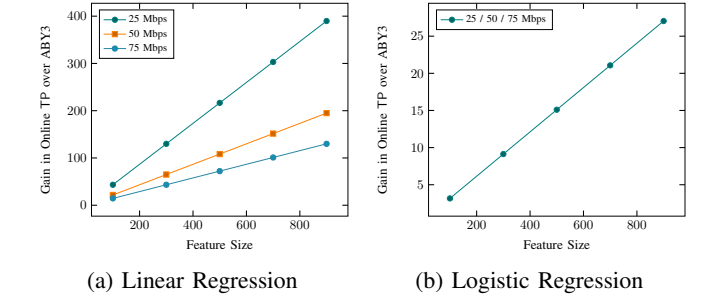

For Linear Regression, we observe that the gain in online throughput over ABY3 ranging from $14\times$ to $216\times$ across different bandwidths. The respective gain for Logistic Regression ranges from $3\times$ to $27\times$.

\ADDED{
In ASTRA, the inference phase of Linear and Logistic Regression are optimized further. For instance, the output of Logistic Regression is the boolean sharing of a single bit. Hence, we benchmarked our protocol with the optimizations of ASTRA and the benchmarking appears in Section~\ref{sec:ASTRAInfWAN}. 
}

\begin{figure}[htb!]
	\centering
	\resizebox{.6\textwidth}{!}{
		\begin{tikzpicture}
		\begin{axis}[legend pos=north west, xlabel={Feature Size}, ylabel={Gain in Online $\TP$ over ABY3}, cycle list name=exotic]
		\addplot plot coordinates { (100, 65.46) (300, 118.18) (500, 170.89) (700, 223.60) (900, 276.31)};
		\addlegendentry{{\footnotesize 25 / 50 / 75 Mbps}}
		\end{axis}
		\node[align=center,font=\bfseries, xshift=2.5em, yshift=-2em] (title) at (current bounding box.north) {};
		\end{tikzpicture}
	}
	\caption{\small Comparison of Online Throughput (TP) of BLAZE and ABY3 for Neural Network Inference}\label{fig:MLInf_Online_NN_WAN}
\end{figure}
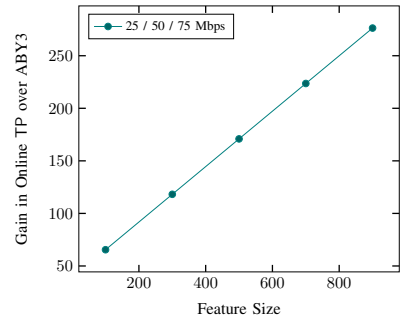

\paragraph{Neural Networks}
Here we consider a NN with two hidden layers, each having 128 nodes followed by an output layer of 10 nodes. The activation function ReLU ($\ReLU$) is applied after the evaluation of each layer.  
For a bandwidth of $25$Mbps, our protocol could process the online phase of $16,866$ queries in a minute and the throughput goes all the way up to $50,602$ queries/min for a bandwidth of $75$Mbps. ABY3, on the other hand, can process $70$ and $210$ queries/min for $25$Mbps and $75$Mbps, respectively. 

\figref{MLInf_Online_NN_WAN} plots the gain in online throughput of BLAZE over ABY3 for varying feature sizes. Unlike Linear Regression and Logistic Regression, the gain is not dropped with the increase in bandwidth. This is because of the huge communication incurred for NN which makes the maximum attainable throughput within our processing capacity.

\tabref{InfTP} provides concrete details for the inference phase of the aforementioned algorithms for feature size of 784.
\begin{table}[htb!]
	\centering
	\resizebox{.9\textwidth}{!}{
		\begin{tabular}{c | c | r | c | r | c }
			\toprule
			\multirow{2}[2]{*}{Algorithm} & \multirow{2}[2]{*}{Ref.} & \multicolumn{2}{c|}{Preprocessing} & \multicolumn{2}{c}{Online}\\ 
			\cmidrule{3-6}
			& & $\TP$ ($\times$$10^3$) & Gain & $\TP$ ($\times$$10^3$) & Gain\\
			\midrule
			\multirow{2}{*}{\makecell{Linear\\Regression}}
			& ABY3       & 15.57 & \multirow{2}{*}{$\mathbf{4.02\times}$} & 15.67  & \multirow{2}{*}{$\mathbf{169.75\times}$}\\ 
			& {\bf BLAZE} & $\mathbf{62.61}$ &  & $\mathbf{2660.53}$ & \\ 
			\midrule
			\multirow{2}{*}{\makecell{Logistic\\Regression}}
			& ABY3       & 15.41 & \multirow{2}{*}{$\mathbf{4.03\times}$} 
			& 15.55  & \multirow{2}{*}{$\mathbf{23.57\times}$}\\ 
			& {\bf BLAZE} & $\mathbf{62.13}$ &  & $\mathbf{366.68}$ & \\ 
			\midrule
			\multirow{2}{*}{\makecell{Neural\\Networks}}
			& ABY3       & 0.10 & \multirow{2}{*}{$\mathbf{4.01\times}$} & 0.14  & \multirow{2}{*}{$\mathbf{245.74\times}$}\\ 
			& {\bf BLAZE} & $\mathbf{0.41}$ &  & $\mathbf{33.74}$ & \\ 
			\bottomrule
		\end{tabular}
	}
	\caption{\small Throughput ($\TP$) for ML Inference for a feature size of $\nf$-784\label{tab:InfTP}}
\end{table}

\subsubsection{ML Inference on Real World Datasets} 
\label{sec:Bench_MLInfR}
Here we benchmark the online phase of ML inference of all the three algorithms over real-world datasets (\tabref{InfWANa}). The datasets are obtained from UCI Machine Learning Repository~\cite{UCIDataset} and the details are provided in \tabref{realD}.

\begin{table}[htb!] 
	\centering
	\resizebox{0.9\textwidth}{!}
	{
		\begin{tabular}{l l r r}
			\toprule
			\multicolumn{1}{c}{Algorithm} & \multicolumn{1}{c}{Dataset} & $\#$features & $\#$samples\\
			\midrule
			\makecell[l]{Linear\\Regression}   & \makecell[l]{{\bf Superconductivity} Critical\\Temperature Data Set~\cite{SUPER18}} 	& 81   & 21263\\ \midrule
			\makecell[l]{Logistic\\Regression} & \makecell[l]{{\bf FMA} Music Analysis\\Dataset~\cite{BenziDVB16}}					    & 518  & 106574\\ \midrule
			\makecell[l]{Neural\\Networks}     & \makecell[l]{{\bf Parkinson} Disease\\Classification Dataset~\cite{Sakar2019ACA}} 	    & 754  & 754\\
			\bottomrule 
		\end{tabular}
	}
    \caption{\small Real World Datasets used for ML Inference\label{tab:realD}}
\end{table}

\begin{table}[ht]
	\centering
	\resizebox{.985\textwidth}{!}{
		\begin{tabular}{c|r|r|r|r|r|r}
			\toprule
			\multirow{2}[4]{*}{Bandwidth} 
			& \multicolumn{2}{c|}{Linear Regression}  & \multicolumn{2}{c|}{Logistic Regression} 
			& \multicolumn{2}{c}{Neural Networks} \\
			& \multicolumn{2}{c|}{\bf (Superconductivity)}
			& \multicolumn{2}{c|}{\bf (FMA)}
			& \multicolumn{2}{c}{\bf (Parkinson)}\\
			 \cmidrule{2-7}
			& ABY3 & BLAZE & ABY3 & BLAZE & ABY3 & BLAZE\\ \midrule
			25 Mbps  & $75852$  & $\mathbf{2660532}$ & $11725$ & $\mathbf{183339}$ & $70$ & $\mathbf{16867}$ \\
			50 Mbps  & $151704$  & $\mathbf{2660532}$ & $23450$ & $\mathbf{366678}$ & $140$ & $\mathbf{33735}$ \\
			75 Mbps  & $227556$  & $\mathbf{2660532}$ & $35175$ & $\mathbf{550017}$ & $210$ & $\mathbf{50603}$ \\
			\bottomrule
		\end{tabular}
	}
	\caption{\small Comparison of Online $\TP$ of ABY3 and BLAZE for Inference over Real World Datasets (Datasets are given in Brackets). Values are given in \#queries/min.} \label{tab:InfWANa}
\end{table}

In \tabref{InfWANa},  we observe that the online throughput of our protocols for the case of Linear Regression is not increasing with the increase in bandwidth. This can be justified as the processing capacity becomes the bottleneck and prevents our protocols from reaching the maximum attainable throughput even for a bandwidth of 25Mbps. This can be prevented by introducing more computing power to the environment.

\subsubsection{Comparison with ASTRA}\label{sec:ASTRAInfWAN}
Here we compare Linear Regression and Logistic Regression inference of BLAZE and ASTRA. For a fair comparison, we apply the optimizations proposed by ASTRA in our protocols. Since Linear Regression inference essentially reduces to a dot product, the benchmarking for the former can be used to analyse the performance of dot product of BLAZE and ASTRA. Hence we omit a separate benchmarking for dot product. 

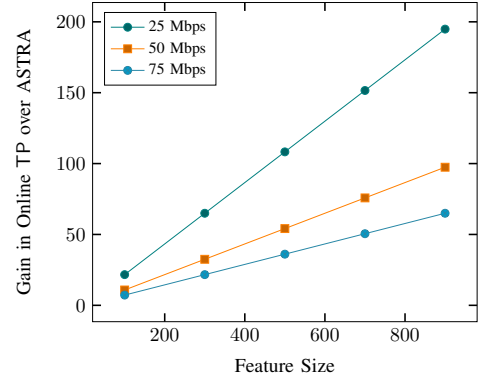
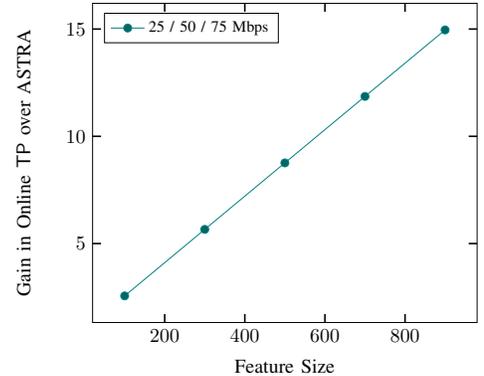
\begin{figure}[htb!]
	\centering
	\begin{subfigure}{.8\textwidth}
		\centering
		\resizebox{.9\textwidth}{!}{
			\begin{tikzpicture}
			\begin{axis}[legend pos=north west, xlabel={Feature Size}, ylabel={Gain in Online $\TP$ over ASTRA}, cycle list name=exotic]
			\addplot plot coordinates { (100, 21.65) (300, 64.95) (500, 108.26) (700, 151.56) (900, 194.86)};
			\addlegendentry{{\footnotesize 25 Mbps}}
			\addplot plot coordinates { (100, 10.83) (300, 32.48) (500, 54.13) (700, 75.78) (900, 97.43)};
			\addlegendentry{{\footnotesize 50 Mbps}}
			\addplot plot coordinates { (100, 7.22) (300, 21.65) (500, 36.09) (700, 50.52) (900, 64.95)};
			\addlegendentry{{\footnotesize 75 Mbps}}
			\end{axis}
			\node[align=center,font=\bfseries, xshift=2.5em, yshift=-2em] (title) at (current bounding box.north) {};
			\end{tikzpicture}
		}
		\caption{\small Linear Regression Inference}\label{fig:ASTRARegWANBWa}
	\end{subfigure}
	\begin{subfigure}{.8\textwidth}
		\centering
		\resizebox{.9\textwidth}{!}{
			\begin{tikzpicture}
			\begin{axis}[legend pos=north west, xlabel={Feature Size}, ylabel={Gain in Online $\TP$ over ASTRA}, cycle list name=exotic]
			\addplot plot coordinates { (100, 2.56) (300, 5.66) (500, 8.76) (700, 11.86) (900, 14.96)};
			\addlegendentry{{\footnotesize 25 / 50 / 75 Mbps}}
			\end{axis}
			\node[align=center,font=\bfseries, xshift=2.5em, yshift=-2em] (title) at (current bounding box.north) {};
			\end{tikzpicture}
		}
		\caption{\small Logistic Regression Inference}\label{fig:ASTRARegWANBWb}
	\end{subfigure}
	\caption{\small Online Throughput ($\TP$) Comparison of ASTRA and BLAZE for Linear Regression and Logistic Regression Inference}\label{fig:ASTRARegWANBW}
\end{figure}

In \figref{ASTRARegWANBW}, we plot the online throughput of BLAZE and ASTRA for Linear Regression (\figref{ASTRARegWANBWa}) and Logistic Regression (\figref{ASTRARegWANBWb}) inference. Concretely, we plot the gain in online throughput over ASTRA over different batch sizes and bandwidths. For the preprocessing phase, our protocols clearly outperforms that of ASTRA and hence we omit the plot for the same.

For both Linear Regression and Logistic Regression inference, we observe that the gain in online throughput over ASTRA drops with an increase in bandwidth. This is because of our limited processing capacity which prevents our protocols from attaining the maximum attainable throughput even for a bandwidth of $25$ Mbps. On the other hand, the throughput of ASTRA increases with the increase in bandwidth. To see this, we limited the bandwidth further to $3$Mbps. At $3$Mbps, the gain in online throughput over ASTRA ranges from $180\times$ to $1623\times$ for Linear Regression inference. 

\section{Conclusion}
\label{sec:Conclusions}
In this work, we presented a blazing fast  framework, BLAZE, for PPML. Our framework, designed for three servers tolerating at most one malicious corruption, works seamlessly over a ring $\Z{\ell}$. Cast in the preprocessing model, our constructs outperform the state-of-the-art solutions by several orders of magnitude both in the round and communication complexity. We showcased the application of our framework in Linear Regression, Logistic Regression, and Neural Networks.
We leave open the problem of extending our framework to support 
training of Neural Networks. 
\ADDED{
Another interesting line of work is to explore the potential of Trusted Execution Environments (TEE)~\cite{RiaziWTS0K18} in improving the overall efficiency of the framework.
}


\section*{Acknowledgement}
We thank our shepherd Matt Fredrikson, and anonymous reviewers for their valuable feedback.

\bibliographystyle{IEEEtran}
\bibliography{main_ndss}

\appendices

\section{Preliminaries}
\label{app:Prelims}
\paragraph{Shared Key Setup}
Let $F : {0, 1}^{\csec} \times {0, 1}^{\csec} \rightarrow X$ be a secure  pseudo-random function PRF, with co-domain $X$ being $\Z{\ell}$. The set of keys established among the servers are:
\begin{myitemize}
	\item[--] One key shared between every pair-- $k_{01}, k_{02},\allowbreak k_{12}$ for the parties $(P_0, P_1), (P_0, P_2), (P_1, P_2)$ respectively.
	\item[--] One shared key amongst all-- $k_{\Partyset}$. 
\end{myitemize}
If the servers $P_0, P_1$ wish to sample a random value $r \in \Z{\ell}$ non-interactively, they invoke $F_{k_{01}}(id_{01})$ to obtain $r$, where $id_{01}$ is a counter which the servers update locally after every PRF invocation. The key used to sample a value will be clear from the context (from the identities of the pair that samples or from the fact that it is sampled by all) and will  be omitted. We model the key setup via a functionality $\FSETUP$ (\boxref{fig:FSETUP}) that  can be realised using any secure  MPC protocol. 
\vspace{-3mm}
\begin{systembox}{$\FSETUP$}{Functionality for Shared Key Setup}{fig:FSETUP}
	\justify
	$\FSETUP$ interacts with the servers in $\Partyset$ and the adversary $\Sim$. $\FSETUP$ picks random keys $\Key{ij}$ for $i,j \in \{0,1,2\}$ and $\Key{\Partyset}$. Let $\wy_i$ denote the keys corresponding to server $P_i$. Then
	\begin{myitemize}
		\item[--] $\wy_i = (k_{01}, k_{02}$ and $\Key{\Partyset})$ when $P_i = P_0$.
		\item[--] $\wy_i = (k_{01}, k_{12}$ and $\Key{\Partyset})$ when $P_i = P_1$.
		\item[--] $\wy_i = (k_{02}, k_{12}$ and $\Key{\Partyset})$ when $P_i = P_2$.
	\end{myitemize}
	\begin{description}
		\item {\bf Output to adversary: }  If $\Sim$ sends $\abort$, then send $(\OUTPUT, \bot)$ to all the servers. Otherwise,  send $(\OUTPUT, \wy_i)$ to the adversary $\Sim$, where $\wy_i$ denotes the keys corresponding to the corrupt server.
		
		\item {\bf Output to selected honest servers: } Receive $(\SELECT, \{I\})$ from adversary $\Sim$, where $\{I\}$ denotes a subset of the honest servers. If an honest server $P_i$ belongs to $I$, send $(\OUTPUT, \bot)$, else send $(\OUTPUT, \wy_i)$.
	\end{description}
\end{systembox}
\vspace{-3mm}
\paragraph{Collision Resistant Hash Function}
Consider a hash function family $\Hash = \mathcal{K}\times \mathcal{L} \rightarrow \mathcal{Y}$. The hash function $\Hash$ is said to be collision resistant if for all probabilistic polynomial-time adversaries $\Adv$, given the description of $\Hash_k$ where {$k \in_R \mathcal{K}$}, there exists a negligible function $\negl()$ such that $\Pr[ (x_1,x_2) \leftarrow \Adv(k):(x_1 \ne x_2) \wedge \Hash_k(x_1)=\Hash_k(x_2)] \leq \negl(\csec)$, where $m = \mathsf{poly}(\csec)$ and $x_1,x_2 \in_R \{0,1\}^m$.

\paragraph{Commitment Scheme}
Let $\commit(x)$ denote the commitment of a value $x$. The commitment scheme $\commit(x)$ possesses two properties; {\em hiding} and {\em binding}. The former ensures that given just the commitment,  privacy of value $x$ is guaranteed. The latter prevents a corrupt party from opening the commitment to a different value $x' \ne x$. The commitment scheme can be instantiated using a hash function $\mathcal{H}()$, whose  security can be proved in the random-oracle model (ROM). For instance, $(c, o) =  (\mathcal{H}(x||r), \allowbreak x||r) = \commit (x; r)$.

\section{Multiplication Protocol of \cite{BonehBCGI19}}
\label{app:ZKCrypto}
In this section, we provide details of the multiplication protocol proposed by \cite{BonehBCGI19} on $\sgr{\cdot}$-shared values. For a value $\val$, the $\sgr{\cdot}$-sharing is defined as:

\resizebox{.94\linewidth}{!}{
	\begin{minipage}{\linewidth}
		\begin{align*}
		\sgr{\val}_0 = (\sqrA{\lv{\val}}, \sqrB{\lv{\val}}),
		\sgr{\val}_1 = (\sqrA{\lv{\val}}, \val + \lv{\val}),
		\sgr{\val}_2 = (\sqrB{\lv{\val}}, \val + \lv{\val})
		\end{align*}
\end{minipage}}

Given the $\sgr{\cdot}$-shares of $\md$ and $\me$, $\piZKPC(\Partyset, \sgr{\md}, \sgr{\me})$ (\boxref{fig:piZKPC}) computes $\sgr{\cdot}$-share of $\mf = \md \me$.

\begin{protocolbox}{$\piZKPC(\Partyset, \sgr{\md}, \sgr{\me})$}{$\sgrd$-shared Multiplication Protocol of \cite{BonehBCGI19}}{fig:piZKPC}
	\justify
	\algoHead{Computation:}  
	\begin{myitemize}
		\item[--] Servers $P_0, P_j$ for $j \in \{1,2\}$ locally sample a random $\LValV{\mf}{j} \in \Z{\ell}$. Also, $P_0, P_1$ samples a random $\LValA{\md,\me} \in \Z{\ell}$
		\item[--] $P_0$ computes $\LVal{\md,\me} = \LVal{\md} \cdot \LVal{\me}$ and sets $\LValB{\md,\me} = \LVal{\md,\me} - \LValA{\md,\me}$. $P_0$ sends $\LValB{\md,\me}$ to $P_2$. 
		\item[--] Server $P_j$ for $j \in \{1,2\}$ computes and mutually exchanges $\sqr{\mf + \LVal{\mf}}_j = (j-1)(\md + \LVal{\md})(\me + \LVal{\me}) - \LValV{\md}{j}(\me + \LVal{\me}) - \LValV{\me}{j}(\md + \LVal{\md}) + \LValV{\md,\me}{j} + \LValV{\mf}{j}$ to reconstruct $(\mf + \LVal{\mf})$.  
	\end{myitemize}
	\justify
	\algoHead{Verification:} 
	Using distributed zero-knowledge, each server proves the correctness of the following statement to the other two:
	\begin{myitemize}
		\item[--] Server $P_0$: $\LVal{\md,\me} = \LVal{\md} \cdot \LVal{\me}$.
		\item[--] Server $P_j$: $\sqr{\mf + \LVal{\mf}}_j = (j-1)(\md + \LVal{\md})(\me + \LVal{\me}) - \LValV{\md}{j}(\me + \LVal{\me}) - \LValV{\me}{j}(\md + \LVal{\md}) + \LValV{\md,\me}{j} + \LValV{\mf}{j}$. Here $j \in \EInSet$.
	\end{myitemize} 
\end{protocolbox}
%

Now we explain the verification method of \cite{BonehBCGI19} in detail. The technique enables prover $\Prover$ to prove to the verifiers $\Verifier_1, \Verifier_2$ in zero knowledge that it knows $w$ such that $(x,w) \in \mathcal{R}$. Let $\ckt$ denotes the circuit corresponding to the statement being verified such that $\ckt(x,w) = 0$ iff $(x,w) \in \mathcal{R}$. The statement $x$ is shared among the verifiers; $x_1$ with $\Verifier_1$ and $x_2$ with $\Verifier_2$ such that $x = x_1 || x_2$, where $||$ denotes concatenation, $|x_1| = n_1$, $|x_2| = n_2$ and $n = n_1 + n_2$. Let $M$ be the number of multiplication gates in $\ckt$.

\begin{figure}[htb!]
	\centering 
	\includegraphics[width=.8\textwidth]{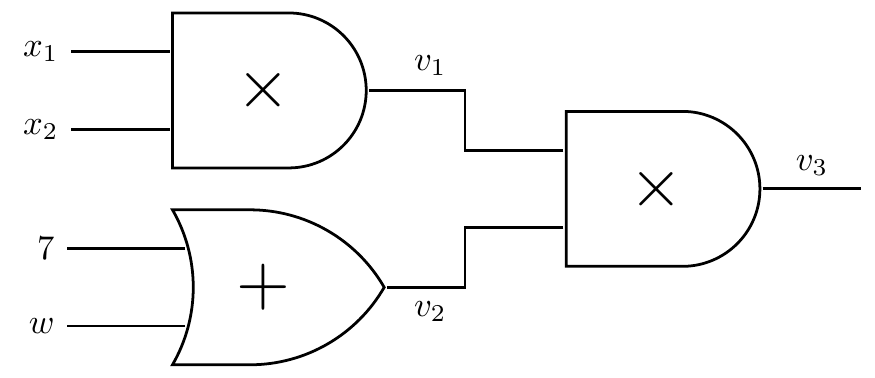}
	\caption{Example circuit\label{fig:ZKPC}}
\end{figure}

Without loss of generality and for easy explanation, we consider the circuit given in Figure~\ref{fig:ZKPC} and the prover and verifiers being $P_0$ and $(P_1, P_2)$ respectively. 

The prover $P_0$ first constructs three polynomials $f()$, $g()$ and $h()$ on the values on left, right and output wires, respectively, of the multiplication gates. The constant terms of $f()$ and $g()$ are set as random ring elements $z_1$ and $z_2$ respectively, while the constant term for $h()$ is set as $z_1 z_2$. More precisely, 

\resizebox{.94\linewidth}{!}{
	\begin{minipage}{\linewidth}
		\begin{align*}
		f(0) &= z_1,  & & g(0) = z_2, & & h(0) = z_1 z_2\\
		f(1) &= x_2,  & & g(1) = x_1, & & h(1) = x_1 x_2\\
		f(2) &= v_2 = 7 + w,  & & g(2) = v_1 = x_1 x_2, & & h(2) = v_1 v_2\\
		\end{align*}
\end{minipage}}

With the above values set, $P_0$ interpolates the polynomials $f()$, $g()$ and $h()$. While $f(), g()$ are polynomials of degree at most $M$, $h()$ is a polynomial of degree at most $d = 2M$. The proof $\pi$ is then defined as $(w,z_1,z_2,c_h)\in \Z{\ell}^{\sigma}$ where $c_h$ denotes the coefficients of $h()$. The size of proof is denoted by $\sigma = s + d + 3$ where $s$ is the size of the witness. $P_0$ then provides additive shares of $\pi$, denoted by $\pi_i$, to the verifiers; $\pi_i$ to $P_i$ for $i \in \EInSet$. Note that if $P_0$ is honest then $\forall r \in \Z{\ell},~ h(r) = f(r)g(r)$ and $h(M) = 0$. 

Verifiers $P_1, P_2$ together sample a random value $r \in \Z{\ell} \setminus \{z_1,z_2\}$ and generate corresponding query vectors $q_f$, $q_g$ and $q_h \in \Z{\ell}^{n+\sigma}$. Each verifier $P_i$ for $i \in \EInSet$ then construct three query vectors from $q_f$, $q_g$ and $q_h$. More precisely, corresponding to polynomial $f()$, verifier $P_i$ constructs vector $Q^i_f \in \mathbb{F}^{n_i+\sigma}$ from $q_f$ such that the first $n_i$ positions are reserved for entries corresponding to $x_i$ followed by $q_f$. $P_i$ for $i \in \{1,2\}$ then locally computes the dot product ($\band$) of the vectors $(x_i||\pi_i)$ and $Q^i_f$ as $f_i(r) = (x_i||\pi_i) \band Q^i_f$ and sends it to verifier $P_1$. 
$P_1$ after receiving the shares of $f_i(r)$ computes the value $f(r) = f_1(r) + f_2(r)$. This comes from the fact that each query vector $q$ defines a linear combination of the input $x$ and proof $\pi$. Hence, the verifiers can form additive shares of the answers to the queries, which is the polynomials evaluated at $r$, using their parts of input $x_i$ and additive share $\pi_i$. This comes from the fact that each query vector $q$ defines a linear combination of the input $x$ and proof $\pi$. Hence, the verifiers can form additive shares of the answers to the queries, which is the polynomials evaluated at $r$, using their parts of input $x_i$ and additive share $\pi_i$. Similar steps are done for polynomials $g()$ and $h()$ which enables $P_1$ to obtain $g(r)$ and $h(r)$. $P_1$ $\abort$s if $h(r) \ne f(r) g(r)$. A cheating prover $P_0$ will pass this check with probability at most $\frac{2M-1}{2^\ell-2}$, which for a large enough $\ell$ is negligible.       

Now for the second check, ie.\ $h(M)=0$, verifiers generate query vector $q$ in a similar fashion. More precisely, $P_i$ for $i \in \{1,2\}$ forms $Q^i$, computes $h_i(M)$ and sends his share of $h(M)$ to $P_1$. Verifier $P_1$ $\abort$ if $h(M) \ne h_1(M) + h_2(M) = 0$.

\cite{BonehBCGI19} propose two variants of the above technique. The first variant gives a 2 round fully linear interactive oracle proofs with query complexity $\BigO{\sqrt{n}}$, where $n$ is the size of the input. The second variant gives $\BigO{\log(M)}$ rounds fully linear interactive oracle proofs with query complexity of $\BigO{\log(M)}$, where $M$ is the number of multiplication gates in $\ckt$. We use the former result in our work. We refer the readers to \cite{BonehBCGI19} for a more detailed description of the verification and its optimizations.

\begin{lemma}[Communication]
	\label{app:piZKPC}
	Protocol $\piZKPC$ requires $4$ rounds and an amortized communication of $3 \ell$ bits.
\end{lemma}

The ideal-world functionality realising $\piZKPC$ protocol is presented in \boxref{fig:FZKPC}.

\begin{systembox}{$\FZKPC$}{Functionality for $\piZKPC$}{fig:FZKPC}
	\justify
	$\FZKPC$ interacts with the servers in $\Partyset$ and the adversary $\Sim$. $\FZKPC$ receives the $\sgr{\cdot}$-shares of values $\md$ and $\me$ from the servers where,
	
	\resizebox{.98\textwidth}{!}{
		\begin{tabular}{ccc}
			$\sgr{\md}_0 = (\sqrA{\lv{\md}}, \sqrB{\lv{\md}})$, &$\sgr{\md}_1 = (\sqrA{\lv{\md}}, \md + \lv{\md})$, &$\sgr{\md}_2 = (\sqrB{\lv{\md}}, \md + \lv{\md})$\\
			$\sgr{\me}_0 = (\sqrA{\lv{\me}}, \sqrB{\lv{\me}})$, &$\sgr{\me}_1 = (\sqrA{\lv{\me}}, \me + \lv{\me})$, &$\sgr{\me}_2 = (\sqrB{\lv{\me}}, \me + \lv{\me})$
		\end{tabular}
	}
	
	If the functionality receives $\bot$ from $\Sim$, then send $\bot$ to every server, else do the following: 
	
	\begin{description}
		\item {\bf Computation of output: } Compute $\md = (\md + \lv{\md}) - \sqrA{\lv{\md}} - \sqrB{\lv{\md}}$ and $\me = (\me + \lv{\me}) - \sqrA{\lv{\me}} - \sqrB{\lv{\me}}$ followed by computing $\mf = \md \me$. Randomly select $\sqrA{\lv{\mf}}, \sqrB{\lv{\mf}}$ from $\Z{\ell}$ and set $\lv{\mf} = \sqrA{\lv{\mf}} + \sqrB{\lv{\mf}}$. The output shares are set as
		
		\resizebox{.98\textwidth}{!}{
			\begin{tabular}{ccc}
				$\sgr{\mf}_0 = (\sqrA{\lv{\mf}}, \sqrB{\lv{\mf}})$, &$\sgr{\mf}_1 = (\sqrA{\lv{\mf}}, \mf + \lv{\mf})$, &$\sgr{\mf}_2 = (\sqrB{\lv{\mf}}, \mf + \lv{\mf})$
			\end{tabular}
		}
		
		\item {\bf Output to adversary: }  If $\Sim$ sends $\abort$, then send $(\OUTPUT, \bot)$ to all the servers. Otherwise,  send $(\OUTPUT, \sgr{\mf}_{\Sim})$ to the adversary $\Sim$, where $\sgr{\mf}_{\Sim}$ denotes the share of $\mf$ corresponding to the corrupt server.
		
		\item {\bf Output to selected honest servers: } Receive $(\SELECT, \{I\})$ from adversary $\Sim$, where $\{I\}$ denotes a subset of the honest servers. If an honest server $P_i$ belongs to $I$, send $(\OUTPUT, \bot)$, else send $(\OUTPUT, \sgr{\mf}_{i})$,  where $\sgr{\mf}_{i}$ denotes the share of $\mf$ corresponding to the honest server $P_i$.
	\end{description}
\end{systembox}

\section{Building Blocks}
\label{app:Blocks}
\subsection{Secret Sharing and Reconstruction Protocols}\label{appsec:secret}

\subsubsection{Sharing Protocol}
\begin{lemma}[Communication]
	\label{app:piSh}
	Protocol $\piSh$ (\boxref{fig:piSh}) is non-interactive in the preprocessing phase and requires $1$ round and an amortized communication of $2 \ell$ bits in the online phase.
\end{lemma}
\begin{IEEEproof}
	During the preprocessing phase, servers sample the shares of $\AVal{}$ and $\GVal{}$ values non-interactively using the shared key setup. During the online phase, when $P_i = P_0$, he/she computes and sends $\BVal{}$ to both $P_1$ and $P_2$ resulting in $1$ round and a communication of $2 \ell$ ring elements. This is followed by $P_1, P_2$ mutually exchanging hash of $\BVal{}$ value received from $P_0$. Servers can combine $\BVal{}$-values for several instances into a single hash and hence the cost gets amortized over multiple instances. For the case when $P_i = P_1$, she sends $\BVal{}$ and $\BVal{} + \GVal{}$ to $P_2$ and $P_0$ respectively, resulting in $1$ round and a communication of $2 \ell$ ring elements. This is followed by $P_2$ sending a hash of $\BVal{} + \GVal{}$ to $P_0$. As mentioned earlier, combining values for multiple instances into a single hash amortizes the cost. The case for $P_i = P_2$ follows similarly.
\end{IEEEproof}

\subsubsection{Joint Sharing Protocol}
\begin{lemma}[Communication]
	\label{app:piJSh}
	Protocol $\piJSh$ (\boxref{fig:piJSh})  is non-interactive in the preprocessing phase and requires $1$ round and an amortized communication of at most $2 \ell$ bits in the online phase.
\end{lemma}
\begin{IEEEproof}
	In this protocol, one of the servers executes $\piSh$ protocol. In parallel, another server sends a hash of $\BVal{}$ to both $P_1$ and $P_2$, whose cost gets amortized over multiple instances. Hence the overall cost follows from that of an instance of the sharing protocol (Lemma~\ref{app:piSh}).
\end{IEEEproof}

\subsubsection{Reconstruction Protocol}

\begin{lemma}[Communication]
	\label{app:piRec}
	Protocol $\piRec$ (\boxref{fig:piRec}) requires $1$ round and an amortized communication of $3 \ell$ bits in the online phase.
\end{lemma}
\begin{IEEEproof}
	During the protocol, each server receives her missing share from another server, resulting in $1$ round and a communication of $3 \ell$ bits. Also, each server receives a hash of the missing share from another server for verification. The hash for multiple instances can be combined to a single hash and thus this cost gets amortized over multiple instances.
\end{IEEEproof}

\subsubsection{Fair Reconstruction Protocol}\label{app:fRec}
Protocol $\pifRec(\Partyset, \shr{\val})$ (\boxref{fig:pifRec}) ensures fair reconstruction of the secret $\val$ for servers in $\Partyset$. This implies that the honest servers are guaranteed to obtain the secret $\val$ whenever the corrupt server obtains the same. The techniques for fair reconstruction introduced in ASTRA to achieve fairness are adapted for our sharing scheme. 

\begin{protocolsplitbox}{$\pifRec(\Partyset, \shr{\val})$}{Fair Reconstruction of value $\val \in \Z{\ell}$ among $\Partyset$}{fig:pifRec}
	\justify
	\algoHead{Preprocessing:}  
	\begin{myitemize}
		\item[--] Servers $P_0, P_j$ for $j \in \{1,2\}$ locally sample a random $r_j \in \Z{\ell}$, prepare commitments of $\AValV{\val}{j}$ and $r_j$. $P_0, P_j$ then send ($\commit(\AValV{\val}{j})$ , $\commit(r_j)$) to $P_{2-j}$. 
		\item[--] $P_j$ for $j \in \{1,2\}$ $\abort$ if the received commitments mismatch.
	\end{myitemize}
	\vspace{-3mm}
	\justify
	\algoHead{Online:} 
	\begin{myitemize}
		\item[--] $P_1, P_2$ compute a commitment of $\BVal{\val}$ and send it to $P_0$.
		\item[--] If the commitments do not match, $P_0$ sends $(\abort, o_j)$ to $P_{2-j}$ for $j \in \{1,2\}$ and aborts, where $o_j$ denotes opening information for the commitment of $r_j$. Else $P_0$ sends $\continue$  to both $P_1$ and $P_2$.
		\item[--] $P_1, P_2$ exchange the messages received from $P_0$. 
		\item[--] $P_1$ aborts if he receives either (i) $(\abort, o_2)$ from $P_0$ and $o_2$ opens the commitment of $r_2$ or (ii) $(\abort, o_1)$ from $P_2$ and $o_1$ is the correct opening information of $r_1$. The case for $P_2$ is similar to that of $P_1$
		\item[--] If no abort happens, servers obtain their missing share of $\val$ as follows:
		\begin{myitemize}
			\item[--] $P_0, P_1$ open $\AValV{\val}{1}$ towards $P_2$.
			\item[--] $P_0, P_2$ open $\AValV{\val}{2}$ towards $P_1$.
			\item[--] $P_1, P_2$ open $\BVal{\val}$ towards $P_0$.
		\end{myitemize}
		\item[--] Servers reconstruct the value $\val$ using missing share that matches with the agreed upon commitment.
	\end{myitemize}
\end{protocolsplitbox}
%

The protocol proceeds as follows: In order to fairly reconstruct $\val$, servers together commit to their common shares. Concretely, in the preprocessing phase, the servers $P_0,P_1$ commit $\AValV{\val}{1}$ to $P_2$ and $P_0,P_2$ commit $\AValV{\val}{2}$ to $P_1$. In the online phase, the servers $P_1,P_2$ commit $\BVal{\val}$ to $P_0$. The recipient in each case can abort if the received commitments do not match. In the case of no abort, $P_0$  signals $P_1$ and $P_2$ to start opening  the commitments which provides each server with the missing share so that they can reconstruct $\val$. It is fair because at least one honest party would have provided the missing share that would allow reconstruction. Lastly, if the protocol aborts before, then none receive the output.
Note that a corrupt $P_0$ can send distinct signals to $P_1$ and $P_2$ (abort to one and continue to the other), breaching unanimity. To resolve this without relying on a broadcast channel, $P_0,P_1$ together commit a value $r_1$ to $P_2$ and $P_0,P_2$ together commit a common value $r_2$ to $P_1$ in the preprocessing phase. In the online phase, if $P_0$ aborts, it sends opening of $r_2$ to $P_1$ and $r_1$ to $P_2$, along with the abort signal. Now a server, say $P_1$ on receiving the abort can convince $P_2$ that it has indeed received abort from $P_0$, using $r_2$ as the {\em proof of origin} for the abort message. This is because $P_1$ can secure $r_2$ only via $P_0$. A single pair of $(r_1,r_2)$ can be used as a proof of origin for multiple instances of reconstruction running in parallel.

\ADDED{In the outsourced setting, the fair reconstruction of a value $\val$ proceeds as follows: Servers execute all the steps of fair reconstruction protocol (\boxref{fig:pifRec}) except the opening of the commitments in the online phase. If no $\abort$ happens, then each of the three servers sends the commitment of $\shareA{\av{\val}}, \shareB{\av{\val}}$, and $\bv{\val})$ to the party $P$ towards which the output needs to be reconstructed. Since we are in the honest majority setting, there will be a majority value among each of the commitment which the party $P$ accepts. In the next round, servers open the shares towards party $P$ as follows: $P_0, P_1$ open $\AValV{\val}{1}$; $P_0, P_2$ open $\AValV{\val}{2}$; $P_1, P_2$ open $\BVal{\val}$. For each of share, party $P$ will accept the opening that matches with the commitment that it accepted.}

\begin{lemma}[Communication]
	\label{app:pifRec}
	Protocol $\pifRec$ (\boxref{fig:pifRec}) requires $4$ rounds and an amortized communication of $6 \ell$ bits in the online phase.
\end{lemma}
\begin{IEEEproof}
	During the preprocessing phase, servers $P_0, P_1$ prepare and send commitment of $\AVal{}$ shares corresponding to $P_2$. The commitment can be instantiated using a hash function (Appendix~\ref{app:Prelims}) and the commitment for multiple instances can be clubbed together amortizing its cost.
	
	The online phase proceeds as follows: In round 1, servers $P_1, P_2$ prepare a commitment of $\BVal{}$ value and send it to $P_1$. In round 2, based on the consistency of the values received, $P_0$ sends back either an abort or continue signal. This is followed by round 3, where $P_1, P_2$ mutually exchange the value received from $P_0$. If no mismatch is found, servers exchange their missing share for the output in round 4, resulting in the communication of $6$ ring elements. Note that the communication for the first three rounds can be clubbed together and thus the communication cost gets amortized over multiple instances.
\end{IEEEproof}

\subsection{Layer-I Primitives}\label{appsec:LayerI}

\subsubsection{Multiplication Protocol}\label{appsec:piMult}
\begin{lemma}[Communication]
	\label{app:piMult}
	Protocol $\piMult$ (\boxref{fig:piMult}) requires $4$ rounds and an amortized communication of $3 \ell$ bits in the preprocessing phase and requires $1$ round and an amortized communication of $3 \ell$ bits in the online phase.
\end{lemma}
\begin{IEEEproof}
	During the preprocessing phase, servers non-interactively generate $\GVal{}$ value and shares for $\AVal{}$ value corresponding to the output using the shared key setup. This is followed by servers executing one instance of $\piZKPC$ protocol, which requires $4$ rounds and an amortized communication of $3$ ring elements (Lemma~\ref{app:piZKPC}). The rest of the steps in the preprocessing phase are non-interactive.
	
	During the online phase, servers $P_1, P_2$ mutually exchange their shares of $\BVal{}$ corresponding to the output, resulting in $1$ round and communication of $2$ ring elements. In parallel, $P_0$ computes and sends hash of $\starbeta{}$ corresponding to the output to both $P_1$ and $P_2$. $P_1$ then sends $\BVal{} + \GVal{}$ corresponding to the output to $P_0$ resulting in an additional communication of $1$ ring element. In parallel, $P_2$ sends a hash of the same to $P_0$. The hash for multiple instances can be combined and hence the cost gets amortized over multiple instances. Note that the communication to and from $P_0$ can be deferred to the end but before the output reconstruction.
\end{IEEEproof}

\subsubsection{Bit Extraction protocol}\label{appsec:piBitExt}

\begin{lemma}[Communication]
	\label{app:piBitExt}
	Protocol $\piBitExt$ (\boxref{fig:piBitExt}) requires $5$ rounds and an amortized communication of $5 \ell \kappa + \kappa$ bits in the preprocessing phase and requires $2$ round and an amortized communication of $\ell \kappa + 2$ bits in the online phase. Here $\kappa$ denotes the computational security parameter.
\end{lemma}
\begin{IEEEproof}
	During the preprocessing phase, servers execute two instances of $\piJSh$ protocol on boolean values. The protocol can be made non-interactive as mentioned in Section~\ref{sec:privML}. The garbled circuit $GC$ consists of $2\ell$ AND gates (cf. optimized PPA of ABY3~\cite{MR18}) and requires a communication of $4\ell \kappa$ bits for the communication from garbler to the evaluator. Moreover, communicating the keys corresponding to inputs $\vu_3 =  \sqr{\av{\val}}_2$ and $\vu_5 = \vr_{2}$ require a communication of $\ell\kappa$ and $\kappa$ bits respectively. So, in total, preprocessing phase requires an overall communication of $5 \ell \kappa + \kappa$ bits. This is equivalent to $(5 \kappa + 2)\ell$ bits in our setting where $\kappa = 128 = 2 \ell$ for $\ell = 64$. As pointed out in ABY3~\cite{MR18}, the communication cost for the commitments can be amortised if the number of bits for which the garbled share needs to be generated is more than the statistical security parameter.

	During the online phase, garbler communicate the key corresponding to $\vu_1 = \bv{\val}$ to evaluator which results in one round and a communication of $\ell \kappa$ bits. After evaluation, $P_2$ sends the resultant bit to $P_1$ which results in another round and a communication of $1$ bit.
	This is followed by servers $P_1, P_2$ executing one instance of $\piJSh$ on a boolean value resulting in an additional communication of $1$ bit. Note that the communication to $P_0$ can be deferred to the end of the protocol but before output reconstruction.  
\end{IEEEproof}
\subsubsection{Bit2A Conversion protocol}\label{appsec:piBitA}

\begin{lemma}[Communication]
	\label{app:piBitA}
	Protocol $\PiBitA$ (\boxref{fig:piBit2A}) requires $5$ rounds and an amortized communication of $9 \ell$ bits in the preprocessing phase and requires $1$ round and an amortized communication of $4 \ell$ bits in the online phase.
\end{lemma}
\begin{IEEEproof}
	During the preprocessing phase, servers execute two instances of $\piJSh$ protocol both of which can be made non-interactive as mentioned in Section~\ref{sec:privML}. This is followed by entire multiplication of two ring values which costs $5$ rounds and communication of $6$ ring elements (Lemma~\ref{app:piMult}). In parallel, servers perform the preprocessing corresponding to a multiplication resulting in an additional communication of $3$ ring elements (Lemma~\ref{app:piMult}).
	
	During the online phase, servers $P_1, P_2$ executing one instance of $\piJSh$ protocol resulting in the communication of $1$ ring element. Note that the communication to $P_0$ in this step can be deferred to the end of the protocol but before output reconstruction. This is followed by servers executing the online phase corresponding to a multiplication resulting in $1$ round and communication of $3$ ring elements. 
\end{IEEEproof}

\subsection{Layer-II Primitives}\label{appsec:LayerII}

\subsubsection{Dot Product Protocol}\label{appsec:piDotP}
\begin{lemma}[Communication]
	\label{app:piDotP}
	Protocol $\piDotP$ (\boxref{fig:piDotP}) requires $4$ rounds and an amortized communication of $3 \nf \ell$ bits in the preprocessing phase and requires $1$ round and an amortized communication of $3 \ell$ bits in the online phase. Here $\nf$ denotes the size of the underlying vectors.
\end{lemma}
\begin{IEEEproof}
	During the preprocessing phase, servers execute the preprocessing phase of $\piMult$ corresponding to each of the $\nf$ multiplications in parallel resulting in $4$ rounds and an amortized communication of $3 \nf$ ring elements (Lemma~\ref{app:piMult}).
	
	The online phase is similar to that of $\piMult$ protocol apart from servers combine their shares corresponding to all the $\nf$ multiplications into one and then exchange. This results in $1$ round and an amortized communication of $3$ ring elements.
\end{IEEEproof}

\subsubsection{Truncation}\label{appsec:piTr}
\begin{lemma}[Communication]
	\label{app:piTrunc}
	Protocol $\piTrunc$ (\boxref{fig:piTr}) requires $2$ rounds and an amortized communication of $2 \nf \ell$ bits.
\end{lemma}
\begin{IEEEproof}
	In this protocol, servers first non-interactively sample additive shares of $\vr$ using the shared randomness. Server $P_0$ then executes $\piSh$ protocol on truncated value of $\vr$ resulting in $1$ round and a communication of $2$ ring elements (Lemma~\ref{app:piSh}). This is followed by $P_1$ sending a hash of $\vu$ to $P_2$. Note that the cost for hash gets amortized over multiple instances.
\end{IEEEproof}

\subsubsection{Dot Product with Truncation}\label{appsec:piDotPTr}
\begin{lemma}[Communication]
	\label{app:piDotPTr}
	Protocol $\piDotPTr$ (\boxref{fig:piDotPTr}) requires $4$ rounds and an amortized communication of $3 \nf \ell + 2 \ell$ bits in the preprocessing phase and requires $1$ round and an amortized communication of $3 \ell$ bits in the online phase.
\end{lemma}
\begin{IEEEproof}
	During the preprocessing phase, servers execute the preprocessing phase corresponding to $\nf$ instances of $\piMult$ protocol, resulting in a communication of $3 \nf$ ring elements (Lemma~\ref{app:piDotP}). In parallel, servers execute one instance of $\piTrunc$ protocol resulting in an additional communication of $2$ ring elements (Lemma~\ref{app:piTrunc}).
	
	The online phase is similar to that of $\piDotP$ protocol apart from servers $\ESet$ computing additive shares of $\wz - \vr$, where $\wz = \vecX \band \vecY$, which results in a communication of $2$ ring elements. This is followed by servers $P_1, P_2$ executing one instance of $\piJSh$ protocol on the truncated value of $\wz$ to generate its arithmetic sharing. This incurs a communication of $1$ ring element. This is followed by servers locally adding their shares. Hence, the online phase requires $1$ round and an amortized communication of $3$ ring elements.
\end{IEEEproof}

\subsubsection{Activation Functions}\label{appsec:actFunc}

\begin{lemma}[Communication]
	\label{app:piReLU}
	Protocol $\ReLU$ requires $5$ rounds and an amortized communication of $12 \ell + 9p$ bits in the preprocessing phase and requires $3$ rounds and an amortized communication of $7 \ell + 3p + 1$ bits in the online phase. Here $p$ denotes the size of the larger field which is $p = \ell + 25$ in this work.
\end{lemma}
\begin{IEEEproof}
	One instance of $\ReLU$ protocol involves the execution of one instance of $\piBitExt$, $\PiBitA$, and $\piMult$. Hence the cost follows from Lemma~\ref{app:piBitExt}, Lemma~\ref{app:piBitA} and Lemma~\ref{app:piMult}. 
\end{IEEEproof}

\begin{lemma}[Communication]
	\label{app:piSig}
	Protocol $\Sig$ requires $5$ rounds and an amortized communication of $21 \ell + 18p + 3$ bits in the preprocessing phase and requires $4$ rounds and an amortized communication of $11 \ell + 6p + 5$ bits in the online phase.
\end{lemma}
\begin{IEEEproof}
	One instance of $\Sig$ protocol involves the execution of the following protocols in order-- i) two instances of $\piBitExt$ protocol, ii) once instance of $\piMult$ protocol over boolean value, iii) two instances of $\PiBitA$ protocol, and iv) one instance of $\piMult$ protocol over ring elements. The cost follows from Lemma~\ref{app:piBitExt}, Lemma~\ref{app:piBitA} and Lemma~\ref{app:piMult}. 
\end{IEEEproof}

\section{Micro benchmarking over WAN}
\label{app:MicroBenchWAN}
In this section, we provide detailed benchmarking of ML algorithms over the WAN setting.

\subsection{ML Training} \label{app:Bench_MLTrain_WAN}

In \tabref{MicroBench_TrainPre}, we tabulate the performance in the preprocessing phase of the protocol of BLAZE and ABY3 for Linear Regression and Logistic Regression Training. The data for batch size B $\in \{128, 256, 512\}$ and feature sizes $\{100, 500, 900\}$ are provided. The values in the table shows the number of iterations in the preprocessing phase that can be completed in a minute. A higher value in the table corresponds a protocol with lower latency.
\begin{table}[htb!]
	\centering
	\resizebox{\textwidth}{!}{
		\begin{tabular}{c|c|c|r|r|r}
			\toprule
			\multirow{2}[2]{*}{Algorithm} 
			& \multirow{2}[2]{*}{\makecell{Batch\\Size}} & \multirow{2}[2]{*}{Ref.} 
			& \multicolumn{3}{c}{Feature Size} \\ \cmidrule{4-6}
			& & & $\nf = 100$ & $\nf = 500$ & $\nf = 900$\\
			\midrule
			\multirow{6}[4]{*}{\makecell{Linear\\Regression\\(\#iterations/min)}} 
			& \multirow{2}{*}{128}   & ABY3         
			& $73.50$ 	 & $68.42$ 	 & $64.72$ 	 \\
			&                      & {\bf BLAZE}   
			& $97.08$ 	 & $97.47$ 	 & $91.72$ 	 \\
			\cmidrule{2-6}
			& \multirow{2}{*}{256} & ABY3         
			& $72.38$ 	 & $64.36$ 	 & $55.96$ 	 \\
			&                      & {\bf BLAZE}   
			& $96.60$ 	 & $91.31$ 	 & $84.38$ 	 \\
			\cmidrule{2-6}
			& \multirow{2}{*}{512} & ABY3         
			& $70.23$ 	 & $55.04$ 	 & $43.30$ 	 \\
			&                      & {\bf BLAZE}   
			& $95.65$ 	 & $83.67$ 	 & $70.55$ 	 \\
			\midrule
			\multirow{6}[4]{*}{\makecell{Logistic\\Regression\\(\#iterations/min)}} 
			& \multirow{2}{*}{128}   & ABY3         
			& $19.77$ 	 & $19.38$ 	 & $19.07$ 	 \\
			&                        & {\bf BLAZE}   
			& $31.91$ 	 & $31.51$ 	 & $31.31$ 	 \\
			\cmidrule{2-6}
			& \multirow{2}{*}{256}   & ABY3         
			& $19.68$ 	 & $19.04$ 	 & $18.23$ 	 \\
			&                        & {\bf BLAZE}   
			& $31.86$ 	 & $31.26$ 	 & $30.40$ 	 \\
			\cmidrule{2-6}
			& \multirow{2}{*}{512} 	 & ABY3         
			& $19.52$ 	 & $18.13$ 	 & $16.64$ 	 \\
			&                        & {\bf BLAZE}   
			& $31.75$ 	 & $30.31$ 	 & $28.40$ 	 \\
			\bottomrule
		\end{tabular}
	}
	\caption{Preprocessing Phase: Comparison of ABY3 and BLAZE for ML Training (higher = better)}\label{tab:MicroBench_TrainPre}
\end{table}

Similarly, in \tabref{MicroBench_TrainOn}, we tabulate the performance in the online phase of the protocol of BLAZE and ABY3 for Linear Regression and Logistic Regression Training. The values in the table shows the number of online iterations that can be completed in a minute.
\begin{table}[htb!]
	\centering
	\resizebox{\textwidth}{!}{
		\begin{tabular}{c|c|c|r|r|r}
			\toprule
			\multirow{2}[2]{*}{Algorithm} 
			& \multirow{2}[2]{*}{\makecell{Batch\\Size}} & \multirow{2}[2]{*}{Ref.} 
			& \multicolumn{3}{c}{Feature Size} \\ \cmidrule{4-6}
			& & & $\nf = 100$ & $\nf = 500$ & $\nf = 900$\\
			\midrule
			\multirow{6}[4]{*}{\makecell{Linear\\Regression\\(\#iterations/min)}} 
			& \multirow{2}{*}{128}   & ABY3         
			& $97.57$  	 & $95.07$ 	 & $89.94$  \\
			&                      & {\bf BLAZE}   
			& $139.05$  	 & $139.05$ 	 & $139.05$  \\
			\cmidrule{2-6}
			& \multirow{2}{*}{256} & ABY3         
			& $97.11$  	 & $89.94$ 	 & $80.09$  \\
			&                      & {\bf BLAZE}   
			& $139.05$  	 & $139.05$ 	 & $139.05$  \\
			\cmidrule{2-6}
			& \multirow{2}{*}{512} & ABY3         
			& $95.08$  	 & $80.10$ 	 & $68.94$  \\
			&                      & {\bf BLAZE}   
			& $139.05$  	 & $139.05$ 	 & $139.05$  \\
			\midrule
			\multirow{6}[4]{*}{\makecell{Logistic\\Regression\\(\#iterations/min)}} 
			& \multirow{2}{*}{128}   & ABY3         
			& $20.54$  	 & $20.43$ 	 & $20.18$  \\
			&                        & {\bf BLAZE}   
			& $60.79$  	 & $60.79$ 	 & $60.79$  \\
			\cmidrule{2-6}
			& \multirow{2}{*}{256}   & ABY3         
			& $20.52$  	 & $20.18$ 	 & $19.63$  \\
			&                        & {\bf BLAZE}   
			& $60.79$  	 & $60.79$ 	 & $60.79$  \\
			\cmidrule{2-6}
			& \multirow{2}{*}{512} 	 & ABY3         
			& $20.40$  	 & $19.61$ 	 & $18.87$  \\
			&                        & {\bf BLAZE}   
			& $60.79$  	 & $60.79$ 	 & $60.79$  \\
			\bottomrule
		\end{tabular}
	}
	\caption{Online Phase: Comparison of ABY3 and BLAZE for ML Training (higher = better)}\label{tab:MicroBench_TrainOn}
\end{table}

\subsection{ML Inference} \label{app:Bench_MLInf_WAN}

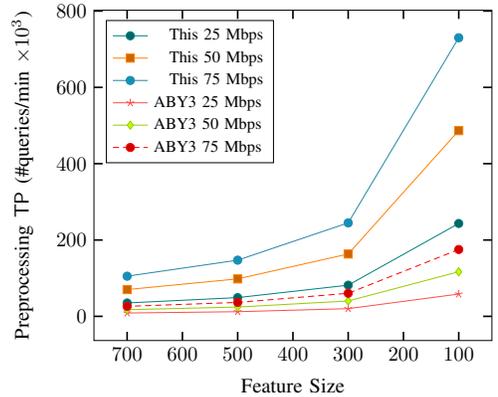
\begin{figure}[htb!]
	\centering
	\resizebox{.75\textwidth}{!}{
		\begin{tikzpicture}
		\begin{axis}[legend pos=north west, xlabel={Feature Size}, ylabel={Preprocessing $\TP$ (\#queries/min $\times10^3$)}, x dir=reverse, ,cycle list name=exotic]
		\addplot plot coordinates { (100, 243.33) (300, 81.65) (500, 49.05) (700, 35.06)};
		\addlegendentry{{\footnotesize ~~~This 25 Mbps}}
		\addplot plot coordinates { (100, 486.65) (300, 163.30) (500, 98.11) (700, 70.12)};
		\addlegendentry{{\footnotesize ~~~This 50 Mbps}}
		\addplot plot coordinates { (100, 729.98) (300, 244.94) (500, 147.16) (700, 105.18)};
		\addlegendentry{{\footnotesize ~~~This 75 Mbps}}
		\addplot plot coordinates { (100, 58.38) (300, 20.13) (500, 12.16) (700, 8.71)};
		\addlegendentry{{\footnotesize ABY3 25 Mbps}}
		\addplot plot coordinates { (100, 116.75) (300, 40.26) (500, 24.32) (700, 17.42)};
		\addlegendentry{{\footnotesize ABY3 50 Mbps}}
		\addplot plot coordinates { (100, 175.13) (300, 60.38) (500, 36.48) (700, 26.14)};
		\addlegendentry{{\footnotesize ABY3 75 Mbps}}
		\end{axis}
		\node[align=center,font=\bfseries, xshift=2.5em, yshift=-2em] (title) at (current bounding box.north) {};
		\end{tikzpicture}
	}
	\caption{\small Comparison of Preprocessing Throughput ($\TP$) of BLAZE and ABY3 for Linear Regression Inference}\label{fig:MLInf_LinReg_WAN}
\end{figure}

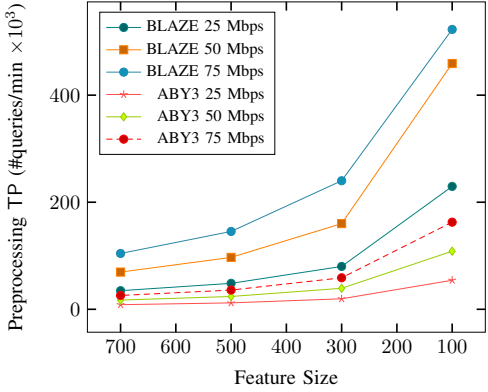
\begin{figure}[htb!]
	\centering
	\resizebox{.75\textwidth}{!}{
		\begin{tikzpicture}
		\begin{axis}[legend pos=north west, xlabel={Feature Size}, ylabel={Preprocessing $\TP$ (\#queries/min $\times10^3$)}, x dir=reverse, ,cycle list name=exotic]
		\addplot plot coordinates { (100, 229.55) (300, 80.04) (500, 48.47) (700, 34.76)};
		\addlegendentry{{\footnotesize BLAZE 25 Mbps}}
		\addplot plot coordinates { (100, 459.10) (300, 160.07) (500, 96.93) (700, 69.52)};
		\addlegendentry{{\footnotesize BLAZE 50 Mbps}}
		\addplot plot coordinates { (100, 522.80) (300, 240.11) (500, 145.40) (700, 104.27)};
		\addlegendentry{{\footnotesize BLAZE 75 Mbps}}
		\addplot plot coordinates { (100, 54.24) (300, 19.61) (500, 11.97) (700, 8.61)};
		\addlegendentry{{\footnotesize ~~ABY3 25 Mbps}}
		\addplot plot coordinates { (100, 108.49) (300, 39.23) (500, 23.94) (700, 17.23)};
		\addlegendentry{{\footnotesize ~~ABY3 50 Mbps}}
		\addplot plot coordinates { (100, 162.73) (300, 58.84) (500, 35.91) (700, 25.84)};
		\addlegendentry{{\footnotesize ~~ABY3 75 Mbps}}
		\end{axis}
		\node[align=center,font=\bfseries, xshift=2.5em, yshift=-2em] (title) at (current bounding box.north) {};
		\end{tikzpicture}
	}
	\caption{\small Comparison of Preprocessing Throughput ($\TP$) of BLAZE and ABY3 for Logistic Regression Inference}\label{fig:MLInf_LogReg_WAN}
\end{figure}

Here we provide the details of the benchmarking done on the preprocessing phase of ML Inference. The details for Linear Regression, Logistic Regression, and Neural Networks appear in \figref{MLInf_LinReg_WAN}, \figref{MLInf_LogReg_WAN}, and \figref{MLInf_NN_WAN} respectively.

\begin{figure}[htb!]
	\centering
	\resizebox{.75\textwidth}{!}{
		\begin{tikzpicture}
		\begin{axis}[legend pos=north west, xlabel={Feature Size}, ylabel={Preprocessing $\TP$ (\#queries/min)}, x dir=reverse, ,cycle list name=exotic]
		\addplot plot coordinates { (100, 754.05) (300, 422.33) (500, 293.30) (700, 224.66)};
		\addlegendentry{{\footnotesize BLAZE 25 Mbps}}
		\addplot plot coordinates { (100, 1508.10) (300, 844.65) (500, 586.60) (700, 449.32)};
		\addlegendentry{{\footnotesize BLAZE 50 Mbps}}
		\addplot plot coordinates { (100, 2262.15) (300, 1266.98) (500, 879.89) (700, 673.98)};
		\addlegendentry{{\footnotesize BLAZE 75 Mbps}}
		\addplot plot coordinates { (100, 185.11) (300, 104.51) (500, 72.80) (700, 55.86)};
		\addlegendentry{{\footnotesize ~~ABY3 25 Mbps}}
		\addplot plot coordinates { (100, 370.23) (300, 209.01) (500, 145.61) (700, 111.72)};
		\addlegendentry{{\footnotesize ~~ABY3 50 Mbps}}
		\addplot plot coordinates { (100, 555.34) (300, 313.52) (500, 218.41) (700, 167.58)};
		\addlegendentry{{\footnotesize ~~ABY3 75 Mbps}}
		\end{axis}
		\node[align=center,font=\bfseries, xshift=2.5em, yshift=-2em] (title) at (current bounding box.north) {};
		\end{tikzpicture}
	}
	\caption{\small Comparison of Preprocessing Throughput ($\TP$) of BLAZE and ABY3 for Neural Networks Inference}\label{fig:MLInf_NN_WAN}
\end{figure}

For all of the algorithms above, we observe $\approx 4\times$ gain in the preprocessing throughput over ABY3.

\section{Micro benchmarking over LAN}
\label{app:MicroBenchLAN}
In this section, we provide the benchmarking details over the LAN setting.
\subsection{Dot Product} \label{app:Bench_DotP_LAN}
For the preprocessing phase, we plot the throughput of BLAZE and ABY3 for dot product protocol (\figref{DOTPPreLAN}) over vectors of length ranging from 100 to 1000. We note at least a gain of $4\times$, which is a consequence of  $4\times$ improvement in communication, over ABY3.

\begin{figure}[htb!]
	\centering
	\resizebox{.75\textwidth}{!}{
		\begin{tikzpicture}
		\begin{axis}[legend pos=north west, xlabel={Feature Size}, ylabel={Preprocessing TP (\#it/minute)}, x dir=reverse, ,cycle list name=exotic]
		\addplot plot coordinates { (100, 9733069.31) (300, 3265913.62) (500, 1962155.69) (700, 1402339.51) (900, 1091054.38)};
		\addlegendentry{{\footnotesize ~~~This BW - 1 Gbps}}
		\addplot plot coordinates { (100, 2335011.88) (300, 805110.57) (500, 486412.67) (700, 348472.17) (900, 271483.02)};
		\addlegendentry{{\footnotesize ABY3 BW - 1 Gbps}}
		\end{axis}
		\node[align=center,font=\bfseries, xshift=2.5em, yshift=-2em] (title) at (current bounding box.north) {};
		\end{tikzpicture}
	}
	\caption{\small Comparison of Preprocessing Throughput (TP) of BLAZE and ABY3 for Dot Product Protocol for different Bandwidths (BW)}\label{fig:DOTPPreLAN}
\end{figure}

For the case of online throughput (\figref{DOTPLANOnline}), the gain over ABY3 ranges from $363\times$ to $3272\times$ across varying feature sizes. We observe a huge gain in the online throughput over ABY3 in the LAN when compared with that of WAN. This is because, in LAN, the communication time scales with the communication size unlike WAN, where the communication time remains almost same for a wide range of communication sizes. 

\begin{figure}[htb!]
	\centering
	\resizebox{.75\textwidth}{!}{
		\begin{tikzpicture}
		\begin{axis}[legend pos=north west, xlabel={Feature Size}, ylabel={Gain in Online Throughput}, cycle list name=exotic]
		\addplot plot coordinates { (100, 363.64) (300, 1090.91) (500, 1818.18) (700, 2545.45) (900, 3272.73)};
		\addlegendentry{{\footnotesize BW - 1 Gbps}}
		\end{axis}
		\node[align=center,font=\bfseries, xshift=2.5em, yshift=-2em] (title) at (current bounding box.north) {};
		\end{tikzpicture}
	}
	\caption{\small Comparison of Online Throughput (TP) of BLAZE and ABY3 for Dot Product Protocol for different Bandwidths (BW)}\label{fig:DOTPLANOnline}
\end{figure}
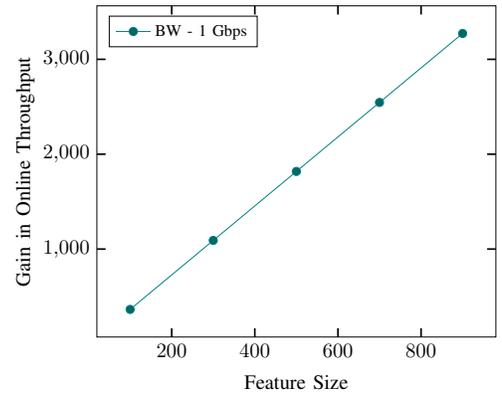

\begin{figure*}
	\centering
	\begin{subfigure}{.24\textwidth}
		\centering
		\resizebox{.98\textwidth}{!}{
			\begin{tikzpicture}
			\begin{axis}[legend pos=north west, xlabel={Feature Size}, ylabel={Offline Throughput (\#LinReg/min)}, x dir=reverse, ,cycle list name=exotic]
			\addplot plot coordinates { (100, 4890746.27) (300, 1635673.88) (500, 982057.94) (700, 701670.24) (900, 545830.09)};
			\addlegendentry{{\footnotesize ~~~This B - 128/256/512}}
			\addplot plot coordinates { (100, 1197369.09) (300, 406047.09) (500, 244476.50) (700, 174887.03) (900, 136136.27)};
			\addlegendentry{{\footnotesize ABY3 B - 128/256/512}}
			\end{axis}
			\node[align=center,font=\bfseries, xshift=2.5em, yshift=-2em] (title) at (current bounding box.north) {};
			\end{tikzpicture}
		}
		\caption{Linear Reg.: Preprocessing}\label{tab:LinRegLANa}
	\end{subfigure}
	\begin{subfigure}{.24\textwidth}
		\centering
		\resizebox{.98\textwidth}{!}{
			\begin{tikzpicture}
			\begin{axis}[legend pos=north west, xlabel={Feature Size}, ylabel={Gain in Online Throughput}, cycle list name=exotic]
			\addplot plot coordinates { (100, 138.89) (300, 342.86) (500, 572.431) (700, 799.54) (900, 886.26)};
			\addlegendentry{{\footnotesize B - 128}}
			\addplot plot coordinates { (100, 236.69) (300, 609.14) (500, 1015.23) (700, 1421.32) (900, 1594.46)};
			\addlegendentry{{\footnotesize B - 256}}
			\addplot plot coordinates { (100, 669.28) (300, 1513.3) (500, 2023.72) (700, 2365.68) (900, 2610.76)};
			\addlegendentry{{\footnotesize B - 512}}
			\end{axis}
			\node[align=center,font=\bfseries, xshift=2.5em, yshift=-2em] (title) at (current bounding box.north) {};
			\end{tikzpicture}
		}
		\caption{Linear Reg.: Online}\label{fig:LinRegLANb}
	\end{subfigure}
	\begin{subfigure}{.24\textwidth}
		\centering
		\resizebox{.98\textwidth}{!}{
			\begin{tikzpicture}
			\begin{axis}[legend pos=north west, xlabel={Feature Size}, ylabel={Offline Throughput (\#LogReg/min)}, x dir=reverse, ,cycle list name=exotic]
			\addplot plot coordinates {(300, 788314.10) (500, 596861.37) (700, 480230.82) (900, 401730.17)};
			\addlegendentry{{\footnotesize ~~~This B - 128/256/512}}
			\addplot plot coordinates {(300, 400739.89) (500, 242542.52) (700, 173895.12) (900, 135534.47)};
			\addlegendentry{{\footnotesize ABY3 B - 128/256/512}}
			\end{axis}
			\node[align=center,font=\bfseries, xshift=2.5em, yshift=-2em] (title) at (current bounding box.north) {};
			\end{tikzpicture}
		}
		\caption{Logistic Reg.: Preprocessing}\label{fig:LogRegLANa}
	\end{subfigure}
	\begin{subfigure}{.24\textwidth}
		\centering
		\resizebox{.98\textwidth}{!}{
			\begin{tikzpicture}
			\begin{axis}[legend pos=north west, xlabel={Feature Size}, ylabel={Gain in Online Throughput}, cycle list name=exotic]
			\addplot plot coordinates { (100, 6.13) (300, 17.93) (500, 29.59) (700, 41.12) (900, 52.51)};
			\addlegendentry{{\footnotesize B - 128 / 256 / 512}}
			\end{axis}
			\node[align=center,font=\bfseries, xshift=2.5em, yshift=-2em] (title) at (current bounding box.north) {};
			\end{tikzpicture}
		}
		\caption{Logistic Reg.: Online}\label{fig:LogRegLANb}
	\end{subfigure}
	\caption{Throughput~(TP) Comparison of ABY3 and BLAZE for ML Training over LAN}\label{fig:RegLAN}
\end{figure*}
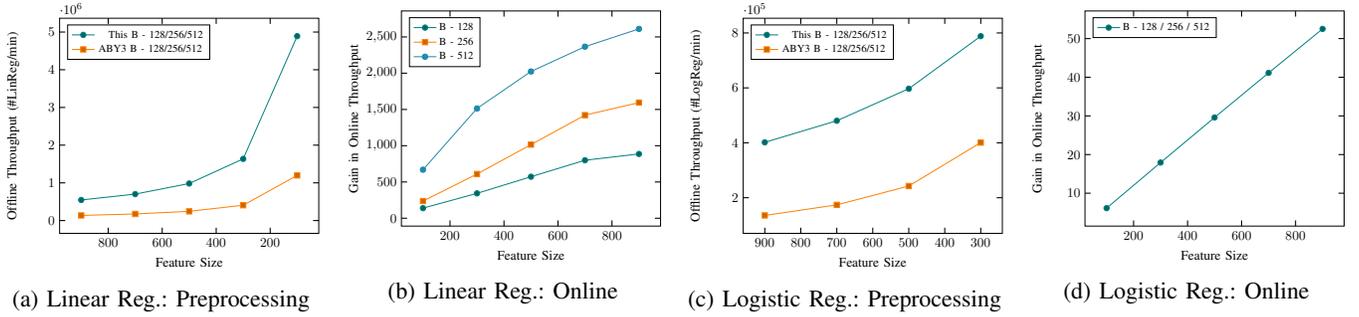

\subsection{ML Training} \label{app:Bench_MLTrain_LAN}

In this section, we explore the training phase of Linear Regression and Logistic Regression algorithms. Our performance improvement over ABY3 (\figref{RegLAN}) is reported in the terms of number of iterations over feature size varying from 100 to 1000, and batch size of $B \in\{128, 256,512\}$.

For the online phase of Linear Regression Training (\figref{LinRegLANb}), the performance gain for batch size of 128 ranges from $138 \times$ to $83.4\times$. The performance gain in the online phase increases significantly for larger batch sizes, and goes all the way up to $2610\times$ for batch size of 512. Similarly, for the case of Logistic Regression, the gain ranges from $6.11\times$ to $53.19\times$.

\tabref{MLTrainTPLAN} provides concrete details for the training phase of Linear Regression and Logistic Regression algorithms over a batch size of 128 and feature size of 784.
\begin{table}[htb!]
	\centering
	\resizebox{.98\textwidth}{!}{
		\begin{tabular}{c | c | r | c | r | c }
			\toprule
			\multirow{2}[2]{*}{Algorithm} & \multirow{2}[2]{*}{Ref.} & \multicolumn{2}{c|}{Preprocessing} & \multicolumn{2}{c}{Online}\\ 
			\cmidrule{3-6}
			& & TP ($\times 10^3$) & Gain & TP ($\times 10^6$) & Gain\\
			\midrule
			\multirow{2}{*}{\makecell{Linear\\Regression}}
			& ABY3       & 156.21 & \multirow{2}{*}{$\mathbf{4.01\times}$} & 0.16 & \multirow{2}{*}{$\mathbf{880.28\times}$}\\ 
			& {\bf This} & $\mathbf{626.54}$ &  & $\mathbf{137.97}$ & \\ 
			\midrule
			\multirow{2}{*}{\makecell{Logistic\\Regression}}
			& ABY3       & 155.42 & \multirow{2}{*}{$\mathbf{2.86\times}$} 
			& 0.16 & \multirow{2}{*}{$\mathbf{44.92\times}$}\\ 
			& {\bf This} & $\mathbf{443.81}$ &  & $\mathbf{7.19}$ & \\ 
			\bottomrule
		\end{tabular}
	}
	\caption{Throughput ($\TP$) for ML Training for a batch size B-128 and feature size $\nf$-784\label{tab:MLTrainTPLAN}}
\end{table}

\begin{table}[htb!]
	\centering
	\resizebox{.96\textwidth}{!}{
		\begin{tabular}{c|c|c|r|r|r}
			\toprule
			\multirow{2}[2]{*}{Algorithm} 
			& \multirow{2}[2]{*}{\makecell{Batch\\Size}} & \multirow{2}[2]{*}{Ref.} 
			& \multicolumn{3}{c}{Feature Size} \\ \cmidrule{4-6}
			& & & $\nf = 100$ & $\nf = 500$ & $\nf = 900$\\
			\midrule
			\multirow{6}[4]{*}{\makecell{Linear\\Regression}} 
			& \multirow{2}{*}{128}   & ABY3         
			& $4236.09$ 	 & $941.40$ 	 & $675.01$ 	 \\
			&                      & {\bf BLAZE}   
			& $8511.85$ 	 & $1961.87$ 	 & $1585.79$ 	 \\
			\cmidrule{2-6}
			& \multirow{2}{*}{256} & ABY3         
			& $2807.94$ 	 & $649.96$ 	 & $337.27$ 	 \\
			&                      & {\bf BLAZE}   
			& $6335.80$ 	 & $1587.89$ 	 & $790.92$ 	 \\
			\cmidrule{2-6}
			& \multirow{2}{*}{512} & ABY3         
			& $1516.80$ 	 & $324.34$ 	 & $172.74$ 	 \\
			&                      & {\bf BLAZE}   
			& $3564.43$ 	 & $790.82$ 	 & $410.43$ 	 \\
			\midrule
			\multirow{6}[4]{*}{\makecell{Logistic\\Regression}} 
			& \multirow{2}{*}{128}   & ABY3         
			& $2810.17$ 	 & $846.01$ 	 & $624.51$ 	 \\
			&                        & {\bf BLAZE}   
			& $2378.97$ 	 & $1230.64$ 	 & $1071.28$ 	 \\
			\cmidrule{2-6}
			& \multirow{2}{*}{256}   & ABY3         
			& $1954.21$ 	 & $590.25$ 	 & $320.45$ 	 \\
			&                        & {\bf BLAZE}   
			& $1169.23$ 	 & $753.12$ 	 & $509.58$ 	 \\
			\cmidrule{2-6}
			& \multirow{2}{*}{512} 	 & ABY3         
			& $1175.87$ 	 & $305.40$ 	 & $167.22$ 	 \\
			&                        & {\bf BLAZE}   
			& $636.94$ 	 & $391.55$ 	 & $268.39$ 	 \\
			\bottomrule
		\end{tabular}
	}
	\caption{Preprocessing Phase: Comparison of ABY3 and BLAZE for ML Training (higher = better)}\label{tab:MicroBench_TrainPreLAN}
\end{table}

In \tabref{MicroBench_TrainPreLAN}, we tabulate the performance in the preprocessing phase of the protocol of BLAZE and ABY3 for Linear Regression and Logistic Regression training, for batch size B $\in \{128, 256, 512\}$ and feature sizes $\{100, 500, 900\}$. The values in the table shows the number of iterations in the preprocessing phase that can be completed in a minute.

\begin{table}[htb!]
	\centering
	\resizebox{.96\textwidth}{!}{
		\begin{tabular}{c|c|c|r|r|r}
			\toprule
			\multirow{2}[2]{*}{Algorithm} 
			& \multirow{2}[2]{*}{\makecell{Batch\\Size}} & \multirow{2}[2]{*}{Ref.} 
			& \multicolumn{3}{c}{Feature Size} \\ \cmidrule{4-6}
			& & & $\nf = 100$ & $\nf = 500$ & $\nf = 900$\\
			\midrule
			\multirow{6}[4]{*}{\makecell{Linear\\Regression}} 
			& \multirow{2}{*}{128}   & ABY3         
			& $6622.52$  	 & $1591.93$ 	 & $810.58$  \\
			&                      & {\bf BLAZE}   
			& $41666.67$  	 & $34285.71$ 	 & $29542.10$  \\
			\cmidrule{2-6}
			& \multirow{2}{*}{256} & ABY3         
			& $3724.39$  	 & $810.59$ 	 & $412.67$  \\
			&                      & {\bf BLAZE}   
			& $35502.96$  	 & $30456.85$ 	 & $26654.82$  \\
			\cmidrule{2-6}
			& \multirow{2}{*}{512} & ABY3         
			& $1755.93$  	 & $412.65$ 	 & $205.21$  \\
			&                      & {\bf BLAZE}   
			& $34883.72$  	 & $27272.73$ 	 & $25740.03$  \\
			\midrule
			\multirow{6}[4]{*}{\makecell{Logistic\\Regression}} 
			& \multirow{2}{*}{128}   & ABY3         
			& $2623.75$  	 & $1165.43$ 	 & $683.27$  \\
			&                        & {\bf BLAZE}   
			& $7863.70$  	 & $7556.68$ 	 & $7556.68$  \\
			\cmidrule{2-6}
			& \multirow{2}{*}{256}   & ABY3         
			& $1696.98$  	 & $643.34$ 	 & $364.43$  \\
			&                        & {\bf BLAZE}   
			& $5008.35$  	 & $4893.96$ 	 & $4783.16$  \\
			\cmidrule{2-6}
			& \multirow{2}{*}{512} 	 & ABY3         
			& $982.90$  	 & $348.27$ 	 & $187.89$  \\
			&                        & {\bf BLAZE}   
			& $2542.37$  	 & $2491.69$ 	 & $2478.21$  \\
			\bottomrule
		\end{tabular}
	}
	\caption{Online Phase: Comparison of ABY3 and BLAZE for ML Training (higher = better)}\label{tab:MicroBench_TrainOnLAN}
\end{table}

Similarly, in \tabref{MicroBench_TrainOnLAN}, we tabulate the performance in the online phase of the protocol of BLAZE and ABY3 for Linear Regression and Logistic Regression training. The values in the table shows the number of online iterations that can be completed in a minute.

\subsection{ML Inference} \label{app:Bench_MLInf_LAN}

In this section, we benchmark the inference phase of Linear Regression, Logistic Regression, and Neural Networks. For inference, the benchmarking parameter is the number of queries processed per minute (\#queries/min).

\begin{figure}[htb!]
	\centering
	\begin{subfigure}{.49\textwidth}
		\centering
		\resizebox{.98\textwidth}{!}{
			\begin{tikzpicture}
			\begin{axis}[legend pos=north west, xlabel={Feature Size}, ylabel={Preprocessing TP (\#queries/min)}, x dir=reverse, ,cycle list name=exotic]
			\addplot plot coordinates { (100, 9733.07) (300, 3265.91) (500, 1962.16) (700, 1402.34) (900, 1091.05)};
			\addlegendentry{{\footnotesize ~~~This BW - 1 Gbps}}
			\addplot plot coordinates { (100, 2335.01) (300, 805.11) (500, 486.41) (700, 348.47) (900, 271.48)};
			\addlegendentry{{\footnotesize ABY3 BW - 1 Gbps}}
			\end{axis}
			\node[align=center,font=\bfseries, xshift=2.5em, yshift=-2em] (title) at (current bounding box.north) {};
			\end{tikzpicture}
		}
		\caption{Preprocessing Phase}\label{fig:LiInfLANa}
	\end{subfigure}
	\begin{subfigure}{.49\textwidth}
		\centering
		\resizebox{.98\textwidth}{!}{
			\begin{tikzpicture}
			\begin{axis}[legend pos=north west, xlabel={Feature Size}, ylabel={Gain in Online Throughput}, cycle list name=exotic]
			\addplot plot coordinates { (100, 400) (300, 1200) (500, 2000) (700, 2800) (900, 3600)};
			\addlegendentry{{\footnotesize BW - 1 Gbps}}
			\end{axis}
			\node[align=center,font=\bfseries, xshift=2.5em, yshift=-2em] (title) at (current bounding box.north) {};
			\end{tikzpicture}
		}
		\caption{Online Phase}\label{fig:LiInfLANb}
	\end{subfigure}
	\caption{Throughput (TP) Comparison of ABY3 and BLAZE for Linear Regression Inference over LAN}\label{fig:LiInfLAN}
\end{figure}
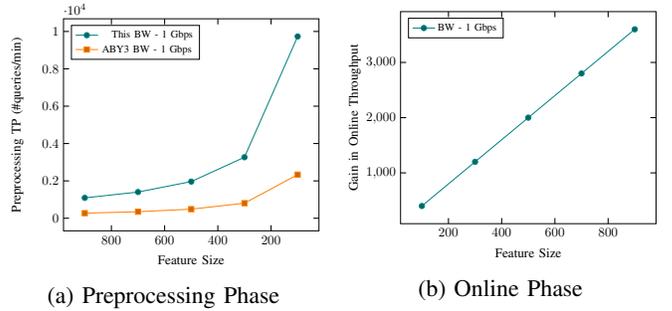

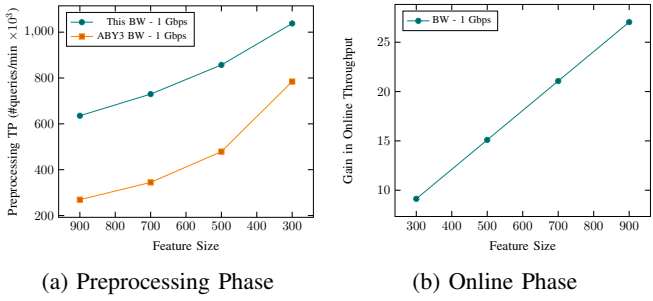
\begin{figure}[htb!]
	\centering
	\begin{subfigure}{.49\textwidth}
		\centering
		\resizebox{.98\textwidth}{!}{
			\begin{tikzpicture}
			\begin{axis}[legend pos=north west, xlabel={Feature Size}, ylabel={Preprocessing TP (\#queries/min $\times10^3$)}, x dir=reverse, ,cycle list name=exotic]
			\addplot plot coordinates { (300, 1037.99) (500, 857.01) (700, 729.77) (900, 635.42)};
			\addlegendentry{{\footnotesize ~~~This BW - 1 Gbps}}
			\addplot plot coordinates { (300, 784.51) (500, 478.82) (700, 344.56) (900, 269.10)};
			\addlegendentry{{\footnotesize ABY3 BW - 1 Gbps}}
			\end{axis}
			\node[align=center,font=\bfseries, xshift=2.5em, yshift=-2em] (title) at (current bounding box.north) {};
			\end{tikzpicture}
		}
		\caption{Preprocessing Phase}\label{fig:LoInfLANa}
	\end{subfigure}
	\begin{subfigure}{.49\textwidth}
		\centering
		\resizebox{.98\textwidth}{!}{
			\begin{tikzpicture}
			\begin{axis}[legend pos=north west, xlabel={Feature Size}, ylabel={Gain in Online Throughput}, cycle list name=exotic]
			\addplot plot coordinates { (300, 9.13) (500, 15.10) (700, 21.07) (900, 27.04)};
			\addlegendentry{{\footnotesize BW - 1 Gbps}}
			\end{axis}
			\node[align=center,font=\bfseries, xshift=2.5em, yshift=-2em] (title) at (current bounding box.north) {};
			\end{tikzpicture}
		}
		\caption{Online Phase}\label{fig:LoInfLANc}
	\end{subfigure}
	\caption{Throughput~(TP) Comparison of ABY3 and BLAZE for Logistic Regression Inference over LAN}\label{fig:LoInfLAN}
\end{figure}

\begin{figure}[htb!]
	\centering
	\begin{subfigure}{.49\textwidth}
		\centering
		\resizebox{.98\textwidth}{!}{
			\begin{tikzpicture}
			\begin{axis}[legend pos=north west, xlabel={Feature Size}, ylabel={Preprocessing TP (\#queries/min $\times10^3$)}, x dir=reverse, ,cycle list name=exotic]
			\addplot plot coordinates { (100, 29.58) (300, 16.71) (500, 11.64) (700, 8.93) (900, 7.25)};
			\addlegendentry{{\small BLAZE - 1 Gbps}}
			\addplot plot coordinates { (100, 7.40) (300, 4.18) (500, 2.91) (700, 2.23) (900, 1.81)};
			\addlegendentry{{\small ~~ABY3 - 1 Gbps}}
			\end{axis}
			\node[align=center,font=\bfseries, xshift=2.5em, yshift=-2em] (title) at (current bounding box.north) {};
			\end{tikzpicture}
		}
		\caption{Preprocessing Phase}\label{fig:NNInfLANa}
	\end{subfigure}
	\begin{subfigure}{.49\textwidth}
		\centering
		\resizebox{.98\textwidth}{!}{
			\begin{tikzpicture}
			\begin{axis}[legend pos=north west, xlabel={Feature Size}, ylabel={Gain in Online Throughput}, cycle list name=exotic]
			\addplot plot coordinates { (100, 65.46) (300, 118.18) (500, 170.89) (700, 223.60) (900, 276.31)};
			\addlegendentry{{\small 1 Gbps}}
			\end{axis}
			\node[align=center,font=\bfseries, xshift=2.5em, yshift=-2em] (title) at (current bounding box.north) {};
			\end{tikzpicture}
		}
		\caption{Online Phase}\label{fig:NNInfLANc}
	\end{subfigure}
	\caption{Throughput~(TP) Comparison of ABY3 and BLAZE for Neural Network Inference over LAN}\label{fig:NNInfLAN}
\end{figure}
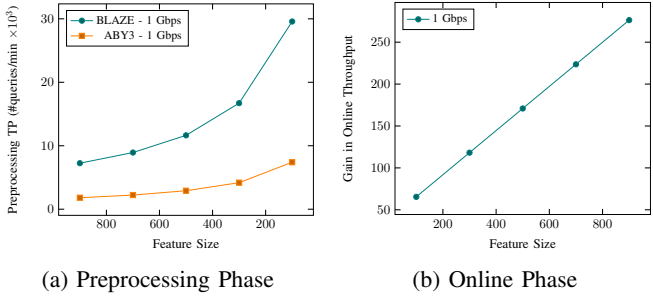

The benchmarking for the inference phase of Linear Regression, Logistic Regression, and Neural Network appears in \figref{LiInfLAN}, \figref{LoInfLAN}, and \figref{NNInfLAN} respectively. 
For the online phase of Linear Regression Inference, the performance gain over ABY3 ranges from $400 \times$ to $3600\times$. Similarly, for the case of Logistic Regression, the gain ranges from $3.16\times$ to $27.04\times$. For the case of Neural Networks, the gain range is from $65\times$ to $276\times$.

~

~

\tabref{InfTPLAN} provides concrete details for the inference phase of the aforementioned algorithms for feature size of 784.
\begin{table}[htb!]
	\centering
	\resizebox{.98\textwidth}{!}{
		\begin{tabular}{c | c | r | c | r | c }
			\toprule
			\multirow{2}[2]{*}{Algorithm} & \multirow{2}[2]{*}{Ref.} & \multicolumn{2}{c|}{Preprocessing} & \multicolumn{2}{c}{Online}\\ 
			\cmidrule{3-6}
			& & TP ($\times$$10^3$) & Gain & TP ($\times$$10^6$) & Gain\\
			\midrule
			\multirow{2}{*}{\makecell{Linear\\Regression}}
			& ABY3       & 311.384 & \multirow{2}{*}{$\mathbf{4.02\times}$} & 0.31  & \multirow{2}{*}{$\mathbf{3136\times}$}\\ 
			& {\bf This} & $\mathbf{1252.28}$ &  & $\mathbf{983.04}$ & \\ 
			\midrule
			\multirow{2}{*}{\makecell{Logistic\\Regression}}
			& ABY3       & 308.25 & \multirow{2}{*}{$\mathbf{2.23\times}$} 
			& 0.31  & \multirow{2}{*}{$\mathbf{23.57\times}$}\\ 
			& {\bf This} & $\mathbf{686.93}$ &  & $\mathbf{7.33}$ & \\ 
			\midrule
			\multirow{2}{*}{\makecell{Neural\\Networks}}
			& ABY3       & 2.04 & \multirow{2}{*}{$\mathbf{4.01\times}$} 
			& 0.003  & \multirow{2}{*}{$\mathbf{245.74\times}$}\\ 
			& {\bf This} & $\mathbf{8.14}$ &  & $\mathbf{0.65}$ & \\ 
			\bottomrule
		\end{tabular}
	}
	\caption{Throughput ($\TP$) for ML Inference for a feature size of $\nf$-784\label{tab:InfTPLAN}}
\end{table}

\section{Security of Our Constructions}
\label{app:Security}
In this section, we prove the security of our constructions using the standard real/ideal world paradigm. Our proofs work in the $\{\FSETUP, \FZKPC\}$-hybrid model, where $\FSETUP$ (\boxref{fig:FSETUP}) and $\FZKPC$ (\boxref{fig:FZKPC}) denote the ideal-world functionalities for the shared key-setup and $\piZKPC$ protocol respectively. 

Let $\Adv$ denote the real-world adversary corrupting one of the servers in $\Partyset$. We use $\Sim$ to denote an ideal-world adversary (simulator) for $\Adv$, who plays the roles of the honest servers and simulates the messages received by $\Adv$ during the protocol. The simulator initializes a boolean variable $\flag = 0$, which indicates whether an honest server $\abort$s during the protocol. For the case of $\abort$, $\flag$ is set to $1$. To distinguish the simulators for various constructions, we use the corresponding protocol name as the subscript of $\Sim$.

For a circuit $\ckt$, the simulation proceeds as follows: The simulation start with the input sharing phase, where $\Sim$ sets the input of honest parties to $0$. From the sharing protocol $\piSh$, $\Sim$ extracts the input of $\Adv$ (details are provided in the simulation for $\piSh$ protocol). This enables $\Sim$ to obtain the entire input for the $\ckt$, which in turn enables $\Sim$ to know all of the intermediate values and the output of the circuit. Looking ahead, $\Sim$ will be using these information to simulate each of the components of the $\ckt$.

We now provide the simulation details for each of the constructions in detail. For each of these, we provide the simulation for a corrupt $P_0$ and a corrupt $P_1$ separately. The case for $P_2$ is similar to that of $P_1$ and we omit details of the same.

\subsection{Sharing Protocol}

The ideal functionality realising protocol $\piSh$ is presented in \boxref{fig:FSh}.
\begin{systembox}{$\FSh$}{Functionality for protocol $\piSh$}{fig:FSh}
	\justify
	$\FSh$ interacts with the servers in $\Partyset$ and the adversary $\Sim$. $\FSh$ receives the input $\val$ from server $P_i$ while it receives $\bot$ from the other servers. If $\val = \bot$, then send $\bot$ to every server, else proceed with the computation.
	
	\begin{description}
		\item {\bf Computation of output: } Randomly select $\sqrA{\av{\val}}, \sqrB{\av{\val}}, \gv{\val}$ from $\Z{\ell}$ and set $\bv{\val} = \val + \sqrA{\av{\val}} + \sqrB{\av{\val}}$. The output shares are set as:	
		\begin{align*}
			\shr{\val}_0 &= (\sqrA{\av{\val}}, \sqrB{\av{\val}}, \bv{\val} + \gv{\val}),\\
			\shr{\val}_1 &= (\sqrA{\av{\val}}, \bv{\val}, \gv{\val}),\\ 
			\shr{\val}_2 &= (\sqrB{\av{\val}}, \bv{\val}, \gv{\val})
		\end{align*}
		
		\item {\bf Output to adversary: }  If $\Sim$ sends $\abort$, then send $(\OUTPUT, \bot)$ to all the servers. Otherwise,  send $(\OUTPUT, \shr{\val}_{\Sim})$ to the adversary $\Sim$, where $\shr{\val}_{\Sim}$ denotes the share of $\val$ corresponding to the corrupt server.
		
		\item {\bf Output to selected honest servers: } Receive $(\SELECT, \{I\})$ from adversary $\Sim$, where $\{I\}$ denotes a subset of the honest servers. If an honest server $P_i$ belongs to $I$, send $(\OUTPUT, \bot)$, else send $(\OUTPUT, \shr{\val}_{i})$,  where $\shr{\val}_{i}$ denotes the share of $\val$ corresponding to the honest server $P_i$.
	\end{description}
\end{systembox}

The simulator for the case of corrupt $P_0$ appears in \boxref{fig:Sim_piShA}.
\begin{simulatorbox}{$\Sim_{\Sh}$}{Simulator $\Sim_{\Sh}$ for the case of corrupt $P_0$}{fig:Sim_piShA}
	\justify \algoHead{Preprocessing Phase:} $\Sim_{\Sh}$ emulates $\FSETUP$ and gives the keys $(k_{01}, k_{02}$ and $\Key{\Partyset})$ to $\Adv$. By emulating $\FSETUP$, it learns the $\av{}$-values corresponding to input $\val$. If $P_i = P_0$, then $\Sim_{\Sh}$ computes $\gv{\val}$ using the key $\Key{\Partyset}$, else it samples a random $\gv{\val}$ on behalf of $\ESet$.
	
	\justify \algoHead{Online Phase:} 
	\begin{myitemize}
		\item[--] If $P_i = P_0$, $\Sim_{\Sh}$ receives $\bv{\val}$ of behalf of $\ESet$. $\Sim_{\Sh}$ sets $\flag=1$ if the received values mismatch. Else, it computes the input $\val = \bv{\val} - \av{\val}$.
		\item[--] If $P_i \ne P_0$, $\Sim_{\Sh}$ sets $\val = 0$ by assigning $\bv{\val} = \av{\val}$. $\Sim_{\Sh}$ sends $\bv{\val} + \gv{\val}$ and $\Hash(\bv{\val} + \gv{\val})$ to $\Adv$ on behalf of $P_1$ and $P_2$ respectively.
	\end{myitemize}
	%
	\noindent If $\flag = 0$ and $P_i = P_0$, $\Sim_{\Sh}$ invokes $\FSh$ with input $\val$ on behalf of $P_0$. Else it invokes $\FSh$ with input $\bot$ on behalf of $P_0$.	
\end{simulatorbox}

The simulator for the case of corrupt $P_1$ appears in \boxref{fig:Sim_piShB}.
\begin{simulatorbox}{$\Sim_{\Sh}$}{Simulator $\Sim_{\Sh}$ for the case of corrupt $P_1$}{fig:Sim_piShB}
	\justify \algoHead{Preprocessing Phase:} $\Sim_{\Sh}$ emulates $\FSETUP$ and gives the keys $(k_{01}, k_{12}$ and $\Key{\Partyset})$ to $\Adv$. By emulating $\FSETUP$, it learns the $\av{}$-values corresponding to input $\val$. If $P_i = P_0$, then $\Sim_{\Sh}$ computes $\gv{\val}$ using the key $\Key{\Partyset}$, else it computes $\gv{\val}$ using the key $k_{12}$ .
	
	\justify \algoHead{Online Phase:} 
	\begin{myitemize}
		\item[--] If $P_i = P_1$, then $\Sim_{\Sh}$ receives $\bv{\val}$ on behalf of $P_2$ and $\bv{\val} + \gv{\val}$ on behalf of $P_0$. $\Sim_{\Sh}$ sets $\flag=1$ if the received values are inconsistent. Else, it computes the input $\val = \bv{\val} - \av{\val}$.
		\item[--] If $P_i = P_j$ for $j \in \{0,2\}$, then
		\begin{itemize}
			\item[--] $\Sim_{\Sh}$ sets $\val = 0$ by assigning $\bv{\val} = \av{\val}$ and sends $\bv{\val}$ to $\Adv$ on behalf of $P_j$.
			\item[--] $\Sim_{\Sh}$ sends $\Hash(\bv{\val})$ to $\Adv$ and receives $\Hash(\bv{\val})'$ from $\Adv$ on behalf of $P_2$.
			\item[--] $\Sim_{\Sh}$ receives $\bv{\val}' + \gv{\val}'$ from $\Sim$ on behalf of $P_0$.
			\item[--] $\Sim_{\Sh}$ sets $\flag=1$ if either $\Hash(\bv{\val})' \ne \Hash(\bv{\val})$ or $\bv{\val} + \gv{\val} \ne \bv{\val}' + \gv{\val}'$. 
		\end{itemize}
	\end{myitemize}
	%
	\noindent If $\flag = 0$ and $P_i = P_1$, $\Sim_{\Sh}$ invokes $\FSh$ with input $\val$ on behalf of $P_1$. Else it invokes $\FSh$ with input $\bot$ on behalf of $P_1$.	
\end{simulatorbox}

\subsection{Joint Sharing Protocol}

The ideal functionality realising protocol $\piJSh$ is presented in \boxref{fig:FJSh}.
\begin{systembox}{$\FJSh$}{Functionality for protocol $\piJSh$}{fig:FJSh}
	\justify
	$\FJSh$ interacts with the servers in $\Partyset$ and the adversary $\Sim$. $\FJSh$ receives the input $\val$ from servers $P_i, P_j$ while it receives $\bot$ from the third server. If the values received from $P_i, P_j$ mismatch, then send $\bot$ to every server, else proceed with the computation.
	
	\begin{description}
		\item {\bf Computation of output: } Randomly select $\sqrA{\av{\val}}, \sqrB{\av{\val}}, \gv{\val}$ from $\Z{\ell}$ and set $\bv{\val} = \val + \sqrA{\av{\val}} + \sqrB{\av{\val}}$. The output shares are set as:
		\begin{align*}
		\shr{\val}_0 &= (\sqrA{\av{\val}}, \sqrB{\av{\val}}, \bv{\val} + \gv{\val}),\\
		\shr{\val}_1 &= (\sqrA{\av{\val}}, \bv{\val}, \gv{\val}),\\ 
		\shr{\val}_2 &= (\sqrB{\av{\val}}, \bv{\val}, \gv{\val})
		\end{align*}
		
		\item {\bf Output to adversary: }  If $\Sim$ sends $\abort$, then send $(\OUTPUT, \bot)$ to all the servers. Otherwise,  send $(\OUTPUT, \shr{\val}_{\Sim})$ to the adversary $\Sim$, where $\shr{\val}_{\Sim}$ denotes the share of $\val$ corresponding to the corrupt server.
		
		\item {\bf Output to selected honest servers: } Receive $(\SELECT, \{I\})$ from adversary $\Sim$, where $\{I\}$ denotes a subset of the honest servers. If an honest server $P_i$ belongs to $I$, send $(\OUTPUT, \bot)$, else send $(\OUTPUT, \shr{\val}_{i})$,  where $\shr{\val}_{i}$ denotes the share of $\val$ corresponding to the honest server $P_i$.
	\end{description}
\end{systembox}

The simulator for the case of corrupt $P_0$ appears in \boxref{fig:Sim_piJShA}.
\begin{simulatorbox}{$\Sim_{\JSh}$}{Simulator $\Sim_{\JSh}$ for the case of corrupt $P_0$}{fig:Sim_piJShA}
	\justify \algoHead{Preprocessing Phase:} $\Sim_{\JSh}$ emulates $\FSETUP$ and gives the keys $(k_{01}, k_{02}$ and $\Key{\Partyset})$ to $\Adv$. By emulating $\FSETUP$, it learns the $\av{}$-values corresponding to input $\val$. $\Sim_{\JSh}$ samples a random $\gv{\val}$ on behalf of $\ESet$.
	
	\justify \algoHead{Online Phase:} 
	\begin{myitemize}
		\item[--] If $(P_i, P_j) = (P_1, P_0)$, then
		\begin{itemize}
			\item[--] $\Sim_{\JSh}$ computes $\bv{\val} = \val + \sqrA{\av{\val}} + \sqrB{\av{\val}}$ on behalf of $P_1$.
			\item[--] $\Sim_{\JSh}$ receives $\Hash(\bv{\val}')$ from $\Adv$ on behalf of $P_2$ and sets $\flag = 1$ if $\Hash(\bv{\val})' \ne \Hash(\bv{\val})$.
			\item[--] $\Sim_{\JSh}$ then sends $\bv{\val} + \gv{\val}$ and $\Hash(\bv{\val} + \gv{\val})$ to $\Adv$ on behalf of $P_1$ and $P_2$ respectively.
		\end{itemize}
		\item[--] The case for $(P_i, P_j) = (P_2, P_0)$ follows similarly.
		\item[--] If $(P_i, P_j) = (P_1, P_2)$, then $\Sim_{\JSh}$ sets $\val = 0$ by assigning $\bv{\val} = \sqrA{\av{\val}} + \sqrB{\av{\val}}$.  $\Sim_{\JSh}$ then sends $\bv{\val} + \gv{\val}$ and $\Hash(\bv{\val} + \gv{\val})$ to $\Adv$ on behalf of $P_1$ and $P_2$ respectively.
	\end{myitemize}
	%
	\noindent If $\flag = 0$ and $P_j = P_0$, $\Sim_{\JSh}$ invokes $\FJSh$ with input $\val$ on behalf of $P_0$. Else it invokes $\FJSh$ with input $\bot$ on behalf of $P_0$.	
\end{simulatorbox}

The simulator for the case of corrupt $P_1$ appears in \boxref{fig:Sim_piJShB}.
\begin{simulatorbox}{$\Sim_{\JSh}$}{Simulator $\Sim_{\JSh}$ for the case of corrupt $P_1$}{fig:Sim_piJShB}
	\justify \algoHead{Preprocessing Phase:} $\Sim_{\JSh}$ emulates $\FSETUP$ and gives the keys $(k_{01}, k_{12}$ and $\Key{\Partyset})$ to $\Adv$. By emulating $\FSETUP$, it learns the $\av{}$-values corresponding to input $\val$. $\Sim_{\JSh}$ computes $\gv{\val}$ using the key $k_{12}$.
	
	\justify \algoHead{Online Phase:} 
	\begin{myitemize}
		\item[--] If $(P_i, P_j) = (P_1, P_0)$, then
		\begin{itemize}
			\item[--] $\Sim_{\JSh}$ computes $\bv{\val} = \val + \sqrA{\av{\val}} + \sqrB{\av{\val}}$ on behalf of $P_0$.
			\item[--] $\Sim_{\JSh}$ receives $\bv{\val}'$ from $\Adv$ on behalf of $P_2$ and sets $\flag = 1$ if $\bv{\val}' \ne \bv{\val}$.
			\item[--] $\Sim_{\JSh}$ receives $\bv{\val}' + \gv{\val}'$ from $\Adv$ on behalf of $P_0$ and sets $\flag = 1$ if $\bv{\val}' + \gv{\val}' \ne \bv{\val} + \gv{\val}$.
		\end{itemize}
		\item[--] If $(P_i, P_j) = (P_1, P_2)$, then $\Sim_{\JSh}$ receives $\bv{\val}' + \gv{\val}'$ from $\Adv$ on behalf of $P_0$ and sets $\flag = 1$ if $\bv{\val}' + \gv{\val}' \ne \bv{\val} + \gv{\val}$.
		\item[--] If $(P_i, P_j) = (P_2, P_0)$, then $\Sim_{\JSh}$ sets $\val = 0$ by assigning $\bv{\val} = \sqrA{\av{\val}} + \sqrB{\av{\val}}$.  $\Sim_{\JSh}$ then sends $\bv{\val}$ and $\Hash(\bv{\val})$ to $\Adv$ on behalf of $P_2$ and $P_0$ respectively.
	\end{myitemize}
	%
	\noindent If $\flag = 0$ and $P_i = P_1$, $\Sim_{\Sh}$ invokes $\FJSh$ with input $\val$ on behalf of $P_1$. Else it invokes $\FJSh$ with input $\bot$ on behalf of $P_1$.	
\end{simulatorbox}

\subsection{Reconstruction Protocol}

The ideal functionality realising protocol $\piRec$ is presented in \boxref{fig:FRec}.
\begin{systembox}{$\FRec$}{Functionality for protocol $\piRec$}{fig:FRec}
	\justify
	$\FRec$ interacts with the servers in $\Partyset$ and the adversary $\Sim$. $\FRec$ receives the $\sgr{\cdot}$-shares of value $\val$ from server $P_i$ for $i \in \{0,1,2\}$. The shares are
	\begin{align*}
		\sgr{\val}_0 &= (\sqrA{\av{\val}}, \sqrB{\av{\val}}),\\
		\sgr{\val}_1 &= (\sqrA{\av{\val}}', \bv{\val}'),\\ 
		\sgr{\val}_2 &= (\sqrB{\av{\val}}', \bv{\val}'')
	\end{align*}

	$\FRec$ sends $\bot$ to every server if either of the following condition is met: i) $\sqrA{\av{\val}} \ne \sqrA{\av{\val}}'$, ii) $\sqrB{\av{\val}} \ne \sqrB{\av{\val}}'$, or iii) $\bv{\val}' \ne \bv{\val}''$. Else it proceeds with the computation.
	
	\begin{description}
		\item {\bf Computation of output: } Set $\val = \bv{\val} - \sqrA{\av{\val}} - \sqrB{\av{\val}}$.
		
		\item {\bf Output to adversary: }  If $\Sim$ sends $\abort$, then send $(\OUTPUT, \bot)$ to all the servers. Otherwise,  send $(\OUTPUT, \val)$ to the adversary $\Sim$.
		
		\item {\bf Output to selected honest servers: } Receive $(\SELECT, \{I\})$ from adversary $\Sim$, where $\{I\}$ denotes a subset of the honest servers. If an honest server $P_i$ belongs to $I$, send $(\OUTPUT, \bot)$, else send $(\OUTPUT, \val)$.
	\end{description}
\end{systembox}

The simulator for the case of corrupt $P_0$ appears in \boxref{fig:Sim_piRecA}. As mentioned in the beginning of this section, $\Sim$ knows all of the intermediate values and the output of the $\ckt$. $\Sim$ uses this information to simulate the $\piRec$ protocol.
\begin{simulatorbox}{$\Sim_{\Rec}$}{Simulator $\Sim_{\Rec}$ for the case of corrupt $P_0$}{fig:Sim_piRecA}
	\justify \algoHead{Online Phase:} 
	\begin{myitemize}
		\item[--] $\Sim_{\Rec}$ sends $\bv{\val}$ and $\Hash(\bv{\val})$ to $\Adv$ on behalf of $P_1$ and $P_2$ respectively.
		\item[--] $\Sim_{\Rec}$ receives $\Hash(\sqrB{\av{\val}'})$ and $\sqrA{\av{\val}}'$ from $\Adv$ on behalf of $P_1$ and $P_2$ respectively. $\Sim_{\Rec}$ sets $\flag = 1$ if either $\Hash(\sqrA{\av{\val}'}) \ne \Hash(\sqrA{\av{\val}})$ or $\sqrB{\av{\val}}' \ne \sqrB{\av{\val}}$.
	\end{myitemize}
	%
	\noindent If $\flag = 0$, $\Sim_{\Rec}$ invokes $\FRec$ with input $(\sqrA{\av{\val}}, \sqrB{\av{\val}} )$ on behalf of $P_0$. Else it invokes $\FRec$ with input $\bot$ on behalf of $P_0$.	
\end{simulatorbox}

The simulator for the case of corrupt $P_1$ appears in \boxref{fig:Sim_piRecB}.
\begin{simulatorbox}{$\Sim_{\Rec}$}{Simulator $\Sim_{\Rec}$ for the case of corrupt $P_1$}{fig:Sim_piRecB}	
	\justify \algoHead{Online Phase:} 
	\begin{myitemize}
		\item[--] $\Sim_{\Rec}$ sends $\sqrB{\av{\val}}$ and $\Hash(\sqrB{\av{\val}})$ to $\Adv$ on behalf of $P_2$ and $P_0$ respectively.
		\item[--] $\Sim_{\Rec}$ receives $\Hash(\sqrA{\av{\val}'})$ and $\bv{\val}'$ from $\Adv$ on behalf of $P_2$ and $P_0$ respectively. $\Sim_{\Rec}$ sets $\flag = 1$ if either $\Hash(\sqrA{\av{\val}'}) \ne \Hash(\sqrA{\av{\val}})$ or $\bv{\val}' \ne \bv{\val}$.
	\end{myitemize}
	%
	\noindent If $\flag = 0$, $\Sim_{\Rec}$ invokes $\FRec$ with input $(\sqrA{\av{\val}}, \bv{\val})$ on behalf of $P_1$. Else it invokes $\FRec$ with input $\bot$ on behalf of $P_1$.
\end{simulatorbox}

\subsection{Multiplication}

The ideal functionality realising protocol $\piMult$ is presented in \boxref{fig:FMult}.
\begin{systembox}{$\FMult$}{Functionality for protocol $\piMult$}{fig:FMult}
	\justify
	$\FMult$ interacts with the servers in $\Partyset$ and the adversary $\Sim$. $\FMult$ receives $\shr{\cdot}$-shares of values $\wx$ and $\wy$ from the servers as input. If $\FMult$ receives $\bot$ from $\Sim$, then send $\bot$ to every server, else proceed with the computation.
	
	\begin{description}
		\item {\bf Computation of output: } Compute $\wx = \bv{\wx} - \sqrA{\av{\wx}} - \sqrB{\av{\wx}}, \wy = \bv{\wy} - \sqrA{\av{\wy}} - \sqrB{\av{\wy}}$ and set $\wz = \wx \wy$. Randomly select $\sqrA{\av{\wz}}, \sqrB{\av{\wz}}, \gv{\wz}$ from $\Z{\ell}$ and set $\bv{\wz} = \wz + \sqrA{\av{\wz}} + \sqrB{\av{\wz}}$. The output shares are set as:
		\begin{align*}
		\shr{\wz}_0 &= (\sqrA{\av{\wz}}, \sqrB{\av{\wz}}, \bv{\wz} + \gv{\wz}),\\
		\shr{\wz}_1 &= (\sqrA{\av{\wz}}, \bv{\wz}, \gv{\wz}),\\ 
		\shr{\wz}_2 &= (\sqrB{\av{\wz}}, \bv{\wz}, \gv{\wz})
		\end{align*}
		
		\item {\bf Output to adversary: }  If $\Sim$ sends $\abort$, then send $(\OUTPUT, \bot)$ to all the servers. Otherwise,  send $(\OUTPUT, \shr{\wz}_{\Sim})$ to the adversary $\Sim$, where $\shr{\wz}_{\Sim}$ denotes the share of $\wz$ corresponding to the corrupt server.
		
		\item {\bf Output to selected honest servers: } Receive $(\SELECT, \{I\})$ from adversary $\Sim$, where $\{I\}$ denotes a subset of the honest servers. If an honest server $P_i$ belongs to $I$, send $(\OUTPUT, \bot)$, else send $(\OUTPUT, \shr{\wz}_{i})$,  where $\shr{\wz}_{i}$ denotes the share of $\wz$ corresponding to the honest server $P_i$.
	\end{description}
\end{systembox}

The simulator for the case of corrupt $P_0$ appears in \boxref{fig:Sim_MultI}.
\begin{simulatorbox}{$\Sim_{\Mult}$}{Simulator for the case of corrupt $P_0$}{fig:Sim_MultI}
	%
	\justify \algoHead{Preprocessing Phase:}
	\begin{myitemize}
		\item[--] $\Sim_{\Mult}$ computes $\sqrA{\av{\wz}}$ and $\sqrB{\av{\wz}}$ using the keys $k_{01}$ and $k_{02}$ respectively. Also, $\Sim_{\Mult}$ samples a random $\gv{\wz}$ on behalf of $\ESet$ and prepares the $\sgr{\cdot}$-shares of $\md, \me$ honestly.
		\item[--] $\Sim_{\Mult}$ emulates $\FZKPC$ and gives $(\sqrA{\lv{\mf}}, \sqrB{\lv{\mf}})$ to $\Adv$. $\Sim_{\Mult}$ then computes the values $\psi, \chi$ followed by computing $\Gamma_{\wx \wy} = \gv{\wx}\av{\wy} + \gv{\wy}\av{\wx} + \psi - \chi$.
	\end{myitemize}
	
	\justify \algoHead{Online Phase:} 
	
	\begin{myitemize}
		\item[--] $\Sim_{\Mult}$ receives $\Hash(\starbeta{\wz})$ from $\Adv$ on behalf of $P_1$ and $P_2$. $\Sim_{\Mult}$ sets $\flag=1$ if either the received hash values are inconsistent or if $\Hash(\starbeta{\wz}) \ne \Hash(\bv{\wz} - \bv{\wx}\bv{\wy} + \psi)$.
		\item[--] Else, $\Sim_{\Mult}$ sends $\bv{\wz} + \gv{\wz}$ and $\Hash(\bv{\wz} + \gv{\wz})$ to $\Adv$ on behalf of $P_1$ and $P_2$ respectively.
	\end{myitemize}
	
	\noindent If $\flag = 0$, $\Sim_{\Mult}$ invokes $\FMult$ with input $(\shr{\wx}_0, \shr{\wy}_0)$ on behalf of $P_0$. Else it invokes $\FMult$ with input $\bot$ on behalf of $P_0$.
\end{simulatorbox}

The simulator for the case of corrupt $P_1$ appears in \boxref{fig:Sim_MultII}.
\begin{simulatorbox}{$\Sim_{\Mult}$}{Simulator for the case of corrupt $P_1$}{fig:Sim_MultII}
	%
	\justify \algoHead{Preprocessing Phase:}
	\begin{myitemize}
		\item[--] $\Sim_{\Mult}$ computes $\sqrA{\av{\wz}}$ and $\gv{\wz}$ using the keys $k_{01}$ and $k_{12}$ respectively. $\Sim_{\Mult}$ samples a random $\sqrB{\av{\wz}}$ on behalf of $P_0$ and $P_2$ and prepares the $\sgr{\cdot}$-shares of $\md, \me$ honestly.
		\item[--] $\Sim_{\Mult}$ emulates $\FZKPC$ and gives $(\sqrA{\lv{\mf}}, \mf + \lv{\mf})$ to $\Adv$. $\Sim_{\Mult}$ then computes the values $\sqr{\psi}, \sqr{\chi}$ honestly. This is followed by computing $\sqr{\Gamma_{\wx \wy}} = \gv{\wx}\sqr{\av{\wy}} + \gv{\wy}\sqr{\av{\wx}} + \sqr{\psi} - \sqr{\chi}$.
	\end{myitemize}
	
	\justify \algoHead{Online Phase:} 
	\begin{myitemize}
		\item[--] $\Sim_{\Mult}$ computes and sends $\sqr{\bv{\wz}}_{2} = \bv{\wx}\bv{\wy} - \bv{\wx}\sqrV{\av{\wy}}{2} - \bv{\wy}\sqrV{\av{\wx}}{2} + \GammaxyV{2} + \sqr{\av{\wz}}_{2}$ to $\Adv$ on behalf of $P_2$, while it receives $\sqr{\bv{\wz}}_{1}$ from $\Adv$ on behalf of $P_2$. $\Sim_{\Mult}$ computes $\bv{\wz} = \wx \wy + \av{\wz}$ and sets $\flag=1$ if $\bv{\wz} \ne \sqr{\bv{\wz}}_{1} + \sqr{\bv{\wz}}_{2}$.
		\item[--] $\Sim_{\Mult}$ sends $\Hash(\starbeta{\wz})$ to $\Adv$ on behalf of $P_0$. $\Sim_{\Mult}$ receives $\bv{\wz} + \gv{\wz}$ from $\Adv$ on behalf of $P_0$ and sets $\flag=1$ if the received value is inconsistent.
	\end{myitemize}
	
	\noindent If $\flag = 0$, $\Sim_{\Mult}$ invokes $\FMult$ with input $(\shr{\wx}_1, \shr{\wy}_1)$ on behalf of $P_1$. Else it invokes $\FMult$ with input $\bot$ on behalf of $P_1$.
\end{simulatorbox}

\subsection{Bit Extraction Protocol}
The ideal functionality realising protocol $\piBitExt$ is presented in \boxref{fig:FBitExt}.
\begin{systembox}{$\FBitExt$}{Functionality for protocol $\piBitExt$}{fig:FBitExt}
	\justify
	$\FBitExt$ interacts with the servers in $\Partyset$ and the adversary $\Sim$. $\FBitExt$ receives $\shr{\cdot}$-share of value $\val$ from the servers as input. If $\FBitExt$ receives $\bot$ from $\Sim$, then send $\bot$ to every server, else proceed with the computation.
	
	\begin{description}
		\item {\bf Computation of output: } Compute $\val = \bv{\val} + \sqrA{\av{\val}} + \sqrB{\av{\val}}$ and set $\bitb = \MSB(\val)$ where $\MSB$ denotes the most significant bit. Randomly select $\sqrA{\av{\bitb}}, \sqrB{\av{\bitb}}, \gv{\bitb}$ from $\Z{1}$ and set $\bv{\bitb} = \bitb \xor \sqrA{\av{\bitb}} \xor \sqrB{\av{\bitb}}$. The output shares are set as:
		\begin{align*}
		\shr{\bitb}_0 &= (\sqrA{\av{\bitb}}, \sqrB{\av{\bitb}}, \bv{\bitb} \xor \gv{\bitb}),\\
		\shr{\bitb}_1 &= (\sqrA{\av{\bitb}}, \bv{\bitb}, \gv{\bitb}),\\ 
		\shr{\bitb}_2 &= (\sqrB{\av{\bitb}}, \bv{\bitb}, \gv{\bitb})
		\end{align*}
		
		\item {\bf Output to adversary: }  If $\Sim$ sends $\abort$, then send $(\OUTPUT, \bot)$ to all the servers. Otherwise,  send $(\OUTPUT, \shr{\bitb}_{\Sim})$ to the adversary $\Sim$, where $\shr{\bitb}_{\Sim}$ denotes the share of $\bitb$ corresponding to the corrupt server.
		
		\item {\bf Output to selected honest servers: } Receive $(\SELECT, \{I\})$ from adversary $\Sim$, where $\{I\}$ denotes a subset of the honest servers. If an honest server $P_i$ belongs to $I$, send $(\OUTPUT, \bot)$, else send $(\OUTPUT, \shr{\bitb}_{i})$,  where $\shr{\bitb}_{i}$ denotes the share of $\bitb$ corresponding to the honest server $P_i$.
	\end{description}
\end{systembox}

The simulator for the case of corrupt $P_0$ appears in \boxref{fig:Sim_BitExtI}.
\begin{simulatorbox}{$\Sim_{\BitExt}$}{Simulator for the case of corrupt $P_0$}{fig:Sim_BitExtI}
	%
	\justify 
	\algoHead{Preprocessing Phase:}
	\begin{myitemize}
		\item[--] $\Sim_{\BitExt}$ computes $\vr_1$ and $\vr_2$ using the keys $k_{01}$ and $k_{02}$ on behalf of $P_1$ and $P_2$ respectively. 
		\item[--] The steps corresponding to $\piJSh$ protocol are simulated similar to $\Sim_{\JSh}$ (\boxref{fig:Sim_piJShA}) for the case of corrupt $P_0$.
		\item[--] The steps corresponding to the garbling are simulated according to the underlying garbling scheme.
	\end{myitemize}
	%
	\justify 
	\algoHead{Online Phase:} 
	\begin{myitemize}
		\item[--] The steps corresponding to $\piJSh$ are simulated similar to $\Sim_{\JSh}$ (\boxref{fig:Sim_piJShA}), for the case of corrupt $P_0$.
		\item[--] The steps corresponding to the garbling are simulated according to the underlying garbling scheme.
	\end{myitemize}
	
	\noindent If $\flag = 0$, $\Sim_{\BitExt}$ invokes $\FBitExt$ with input $(\sqrA{\av{\val}}, \sqrB{\av{\val}}, \bv{\val} + \gv{\val})$ on behalf of $P_0$. Else it invokes $\FBitExt$ with input $\bot$ on behalf of $P_0$.
\end{simulatorbox}

The simulator for the case of corrupt $P_1$ appears in \boxref{fig:Sim_BitExtII}.
\begin{simulatorbox}{$\Sim_{\BitExt}$}{Simulator for the case of corrupt $P_1$}{fig:Sim_BitExtII}
	%
	\justify 
	\algoHead{Preprocessing Phase:}
	\begin{myitemize}
		\item[--] $\Sim_{\BitExt}$ computes $\vr_1$ and $\vr_2$ using the keys $k_{01}$ and $k_{02}$ on behalf of $P_0$ and $P_2$ respectively. 
		\item[--] The steps corresponding to $\piJSh$ protocol are simulated similar to $\Sim_{\JSh}$ (\boxref{fig:Sim_piJShA}) for the case of corrupt $P_1$.
		\item[--] The steps corresponding to the garbling are simulated according to the underlying garbling scheme.
	\end{myitemize}
	%
	\justify 
	\algoHead{Online Phase:} 
	\begin{myitemize}
		\item[--] $\Sim_{\BitExt}$ receives the actual keys for $\vu_1$ from $P_1$ on behalf of $P_2$. $\Sim_{\BitExt}$ sets $\flag=0$ if the received value mismatches with the value computed by himself on behalf of $P_0$. 
		\item[--] The steps corresponding to the garbling are simulated according to the underlying garbling scheme.
		\item[--] $\Sim_{\BitExt}$ sends bit $\vv$ and hash of the corresponding key to $P_1$ on behalf of $P_2$.
		\item[--] The steps corresponding to $\piJSh$ are simulated similar to $\Sim_{\JSh}$ (\boxref{fig:Sim_piJShA}), for the case of corrupt $P_1$.
	\end{myitemize}
	
	\noindent If $\flag = 0$, $\Sim_{\BitExt}$ invokes $\FBitExt$ with input $(\sqrA{\av{\val}}, \bv{\val}, \gv{\val})$ on behalf of $P_1$. Else it invokes $\FBitExt$ with input $\bot$ on behalf of $P_1$.
\end{simulatorbox}

\subsection{Bit2A Protocol}

The ideal functionality realising protocol $\PiBitA$ is presented in \boxref{fig:FBitA}.
\begin{systembox}{$\FBitA$}{Functionality for protocol $\PiBitA$}{fig:FBitA}
	\justify
	$\FBitA$ interacts with the servers in $\Partyset$ and the adversary $\Sim$. $\FBitA$ receives $\shrB{\cdot}$-share of a bit $\bitb$ from the servers as input. If $\FBitA$ receives $\bot$ from $\Sim$, then send $\bot$ to every server, else proceed with the computation.
	
	\begin{description}
		\item {\bf Computation of output: } Compute $\bitb = \bv{\bitb} \xor \sqrA{\av{\bitb}} \xor \sqrB{\av{\bitb}}$. Let $\arval{\bitb}$ denotes the value of bit $\bitb$ over an arithmetic ring $\Z{\ell}$. Randomly select $\sqrA{\av{\arval{\bitb}}}, \sqrB{\av{\arval{\bitb}}}, \gv{\arval{\bitb}}$ from $\Z{\ell}$ and set $\bv{\arval{\bitb}} = \arval{\bitb} + \sqrA{\av{\arval{\bitb}}} + \sqrB{\av{\arval{\bitb}}}$. The output shares are set as:
		\begin{align*}
		\shr{\arval{\bitb}}_0 &= (\sqrA{\av{\arval{\bitb}}}, \sqrB{\av{\arval{\bitb}}}, \bv{\arval{\bitb}} + \gv{\arval{\bitb}}),\\
		\shr{\arval{\bitb}}_1 &= (\sqrA{\av{\arval{\bitb}}}, \bv{\arval{\bitb}}, \gv{\arval{\bitb}}),\\ 
		\shr{\arval{\bitb}}_2 &= (\sqrB{\av{\arval{\bitb}}}, \bv{\arval{\bitb}}, \gv{\arval{\bitb}})
		\end{align*}
		
		\item {\bf Output to adversary: }  If $\Sim$ sends $\abort$, then send $(\OUTPUT, \bot)$ to all the servers. Otherwise,  send $(\OUTPUT, \shr{\arval{\bitb}}_{\Sim})$ to the adversary $\Sim$, where $\shr{\arval{\bitb}}_{\Sim}$ denotes the share of $\arval{\bitb}$ corresponding to the corrupt server.
		
		\item {\bf Output to selected honest servers: } Receive $(\SELECT, \{I\})$ from adversary $\Sim$, where $\{I\}$ denotes a subset of the honest servers. If an honest server $P_i$ belongs to $I$, send $(\OUTPUT, \bot)$, else send $(\OUTPUT, \shr{\arval{\bitb}}_{i})$,  where $\shr{\arval{\bitb}}_{i}$ denotes the share of $\arval{\bitb}$ corresponding to the honest server $P_i$.
	\end{description}
\end{systembox}

The simulator for the case of corrupt $P_0$ appears in \boxref{fig:Sim_BitAI}.
\begin{simulatorbox}{$\Sim_{\BitA}$}{Simulator for the case of corrupt $P_0$}{fig:Sim_BitAI}
	%
	\justify 
	\algoHead{Preprocessing Phase:}
	\begin{myitemize}
		\item[--] The steps corresponding to $\piJSh$ and $\piMult$ protocols are simulated similar to $\Sim_{\JSh}$ (\boxref{fig:Sim_piJShA}) and $\Sim_{\Mult}$ (\boxref{fig:Sim_MultI}) respectively, for the case of corrupt $P_0$.
	\end{myitemize}
	%
	\justify 
	\algoHead{Online Phase:} 
	\begin{myitemize}
		\item[--] The steps corresponding to $\piJSh$ and $\piMult$ protocols are simulated similar to $\Sim_{\JSh}$ (\boxref{fig:Sim_piJShA}) and $\Sim_{\Mult}$ (\boxref{fig:Sim_MultI}) respectively, for the case of corrupt $P_0$.
	\end{myitemize}
	
	\noindent If $\flag = 0$, $\Sim_{\BitA}$ invokes $\FBitA$ with input $\shrB{\bitb}_0$ on behalf of $P_0$. Else it invokes $\FBitA$ with input $\bot$ on behalf of $P_0$.
\end{simulatorbox}

The simulator for the case of corrupt $P_1$ appears in \boxref{fig:Sim_BitAII}.
\begin{simulatorbox}{$\Sim_{\BitA}$}{Simulator for the case of corrupt $P_1$}{fig:Sim_BitAII}
	%
	\justify 
	\algoHead{Preprocessing Phase:}
	\begin{myitemize}
		\item[--] The steps corresponding to $\piJSh$ and $\piMult$ protocols are simulated similar to $\Sim_{\JSh}$ (\boxref{fig:Sim_piJShB}) and $\Sim_{\Mult}$ (\boxref{fig:Sim_MultII}) respectively, for the case of corrupt $P_1$.
	\end{myitemize}
	%
	\justify 
	\algoHead{Online Phase:} 
	\begin{myitemize}
		\item[--] The steps corresponding to $\piJSh$ and $\piMult$ protocols are simulated similar to $\Sim_{\JSh}$ (\boxref{fig:Sim_piJShB}) and $\Sim_{\Mult}$ (\boxref{fig:Sim_MultII}) respectively, for the case of corrupt $P_1$.
	\end{myitemize}
	
	\noindent If $\flag = 0$, $\Sim_{\BitA}$ invokes $\FBitA$ with input $\shrB{\bitb}_1$ on behalf of $P_1$. Else it invokes $\FBitA$ with input $\bot$ on behalf of $P_1$.
\end{simulatorbox}

\subsection{Dot Product Protocol}

The ideal functionality realising protocol $\piDotP$ is presented in \boxref{fig:FDotP}.
\begin{systembox}{$\FDotP$}{Functionality for protocol $\piSh$}{fig:FDotP}
	\justify
	$\FDotP$ interacts with the servers in $\Partyset$ and the adversary $\Sim$. $\FDotP$ receives $\shr{\cdot}$-shares of vectors $\vecX$ and $\vecY$ from the servers as input. Here $\vecX$ and $\vecY$ are $\nf$-length vectors.	If $\FDotP$ receives $\bot$ from $\Sim$, then send $\bot$ to every server, else proceed with the computation.
	
	\begin{description}
		\item {\bf Computation of output: } Compute $\wx_i = \bv{\wx_i} - \sqrA{\av{\wx_i}} - \sqrB{\av{\wx_i}}, \wy_i = \bv{\wy_i} - \sqrA{\av{\wy_i}} - \sqrB{\av{\wy_i}}$ for $i \in [\nf]$ and set $\wz = \sum_{i = 1}^{\nf} \wx_i \wy_i$. Randomly select $\sqrA{\av{\wz}}, \sqrB{\av{\wz}}, \gv{\wz}$ from $\Z{\ell}$ and set $\bv{\wz} = \wz + \sqrA{\av{\wz}} + \sqrB{\av{\wz}}$. The output shares are set as:
		\begin{align*}
		\shr{\wz}_0 &= (\sqrA{\av{\wz}}, \sqrB{\av{\wz}}, \bv{\wz} + \gv{\wz}),\\
		\shr{\wz}_1 &= (\sqrA{\av{\wz}}, \bv{\wz}, \gv{\wz}),\\ 
		\shr{\wz}_2 &= (\sqrB{\av{\wz}}, \bv{\wz}, \gv{\wz})
		\end{align*}
		
		\item {\bf Output to adversary: }  If $\Sim$ sends $\abort$, then send $(\OUTPUT, \bot)$ to all the servers. Otherwise,  send $(\OUTPUT, \shr{\wz}_{\Sim})$ to the adversary $\Sim$, where $\shr{\wz}_{\Sim}$ denotes the share of $\wz$ corresponding to the corrupt server.
		
		\item {\bf Output to selected honest servers: } Receive $(\SELECT, \{I\})$ from adversary $\Sim$, where $\{I\}$ denotes a subset of the honest servers. If an honest server $P_i$ belongs to $I$, send $(\OUTPUT, \bot)$, else send $(\OUTPUT, \shr{\wz}_{i})$,  where $\shr{\wz}_{i}$ denotes the share of $\wz$ corresponding to the honest server $P_i$.
	\end{description}
\end{systembox}

The simulator for the case of corrupt $P_0$ appears in \boxref{fig:Sim_DotPI}.
\begin{simulatorbox}{$\Sim_{\DotP}$}{Simulator for the case of corrupt $P_0$}{fig:Sim_DotPI}
	%
	\justify 
	\algoHead{Preprocessing Phase:}
	\begin{myitemize}
		\item[--] For the preprocessing corresponding to each of the $\nf$ multiplications, $\Sim_{\DotP}$ simulates similar to the simulation for the preprocessing phase of $\piMult$ protocol given in $\Sim_{\Mult}$ (\boxref{fig:Sim_MultI}), for the case of corrupt $P_0$.
	\end{myitemize}
	%
	\justify 
	\algoHead{Online Phase:} 
	\begin{myitemize}
		\item[--] $\Sim_{\DotP}$ receives $\Hash(\starbeta{\wz})$ from $\Adv$ on behalf of $P_1$ and $P_2$. $\Sim_{\DotP}$ sets $\flag=1$ if either the received hash values are inconsistent or if $\Hash(\starbeta{\wz}) \ne \Hash(\bv{\wz}-\sum_{i=1}^{\nf}\bv{\wx_i}\bv{\wy_i} + \psi)$.
		\item[--] Else, $\Sim_{\DotP}$ sends $\bv{\wz} + \gv{\wz}$ and $\Hash(\bv{\wz} + \gv{\wz})$ to $\Adv$ on behalf of $P_1$ and $P_2$ respectively.
	\end{myitemize}
	
	\noindent If $\flag = 0$, $\Sim_{\DotP}$ invokes $\FDotP$ with input $(\shr{\vecX}_0, \shr{\vecY}_0)$ on behalf of $P_0$. Else it invokes $\FDotP$ with input $\bot$ on behalf of $P_0$.
\end{simulatorbox}

The simulator for the case of corrupt $P_1$ appears in \boxref{fig:Sim_DotPII}.
\begin{simulatorbox}{$\Sim_{\DotP}$}{Simulator for the case of corrupt $P_1$}{fig:Sim_DotPII}
	%
	\justify 
	\algoHead{Preprocessing Phase:}
	\begin{myitemize}
		\item[--] For the preprocessing corresponding to each of the $\nf$ multiplications, $\Sim_{\DotP}$ simulates similar to the simulation for the preprocessing phase of $\piMult$ protocol given in $\Sim_{\Mult}$ (\boxref{fig:Sim_MultII}), for the case of corrupt $P_1$.
	\end{myitemize}
	%
	\justify 
	\algoHead{Online Phase:} 
	\begin{myitemize}
		\item[--] $\Sim_{\DotP}$ computes and sends $\sqr{\bv{\wz}}_{2}$ to $\Adv$ on behalf of $P_2$, while it receives $\sqr{\bv{\wz}}_{1}$ from $\Adv$ on behalf of $P_2$. $\Sim_{\DotP}$ computes $\bv{\wz} = \vecX \band \vecY + \av{\wz}$ and sets $\flag=1$ if $\bv{\wz} \ne \sqr{\bv{\wz}}_{1} + \sqr{\bv{\wz}}_{2}$.
		\item[--] $\Sim_{\DotP}$ sends $\Hash(\starbeta{\wz})$ to $\Adv$ on behalf of $P_0$. $\Sim_{\DotP}$ receives $\bv{\wz} + \gv{\wz}$ from $\Adv$ on behalf of $P_0$ and sets $\flag=1$ if the received value is inconsistent.
	\end{myitemize}
	
	\noindent If $\flag = 0$, $\Sim_{\DotP}$ invokes $\FDotP$ with input $(\shr{\vecX}_1, \shr{\vecY}_1)$ on behalf of $P_1$. Else it invokes $\FDotP$ with input $\bot$ on behalf of $P_1$.
\end{simulatorbox}

\subsection{Truncation Protocol}

The ideal functionality realising protocol $\piTrunc$ is presented in \boxref{fig:FTrunc}.
\begin{systembox}{$\FTrunc$}{Functionality for protocol $\piTrunc$}{fig:FTrunc}
	\justify
	$\FTrunc$ interacts with the servers in $\Partyset$ and the adversary $\Sim$. 
	
	\begin{description}
		\item {\bf Computation of output: } Randomly select $\vr \in \Z{\ell}$ and set $\trunc{\vr} = \vr/2^d$. Here $\trunc{\vr}$ denotes the truncated value of $\vr$. 
		Randomly select $\sqrA{\av{\vr}}, \sqrB{\av{\vr}}, \gv{\vr}$ from $\Z{\ell}$ and set $\bv{\vr} = \vr + \sqrA{\av{\vr}} + \sqrB{\av{\vr}}$. The output shares of $\vr$ are set as:
		\begin{align*}
		\shr{\vr}_0 &= (\sqrA{\av{\vr}}, \sqrB{\av{\vr}}, \bv{\vr} + \gv{\vr}),\\
		\shr{\vr}_1 &= (\sqrA{\av{\vr}}, \bv{\vr}, \gv{\vr}),\\ 
		\shr{\vr}_2 &= (\sqrB{\av{\vr}}, \bv{\vr}, \gv{\vr})
		\end{align*}
		
		Randomly select $\sqrA{\av{\trunc{\vr}}}, \sqrB{\av{\trunc{\vr}}}, \gv{\trunc{\vr}}$ from $\Z{\ell}$ and set $\bv{\trunc{\vr}} = \trunc{\vr} + \sqrA{\av{\trunc{\vr}}} + \sqrB{\av{\trunc{\vr}}}$. The output shares of $\trunc{\vr}$ are set as:
		\begin{align*}
		\shr{\trunc{\vr}}_0 &= (\sqrA{\av{\trunc{\vr}}}, \sqrB{\av{\trunc{\vr}}}, \bv{\trunc{\vr}} + \gv{\trunc{\vr}}),\\
		\shr{\trunc{\vr}}_1 &= (\sqrA{\av{\trunc{\vr}}}, \bv{\trunc{\vr}}, \gv{\trunc{\vr}}),\\ 
		\shr{\trunc{\vr}}_2 &= (\sqrB{\av{\trunc{\vr}}}, \bv{\trunc{\vr}}, \gv{\trunc{\vr}})
		\end{align*}
		
		\item {\bf Output to adversary: }  If $\Sim$ sends $\abort$, then send $(\OUTPUT, \bot)$ to all the servers. Otherwise,  send $(\OUTPUT, (\shr{\vr}_{\Sim}, \shr{\trunc{\vr}}_{\Sim}))$ to the adversary $\Sim$, where $(\shr{\vr}_{\Sim}, \shr{\trunc{\vr}}_{\Sim})$ denotes the share of $(\vr, \trunc{\vr})$ corresponding to the corrupt server.
		
		\item {\bf Output to selected honest servers: } Receive $(\SELECT, \{I\})$ from adversary $\Sim$, where $\{I\}$ denotes a subset of the honest servers. If an honest server $P_i$ belongs to $I$, send $(\OUTPUT, \bot)$, else send $(\OUTPUT, (\shr{\vr}_{i}, \shr{\trunc{\vr}}_{i}))$,  where $(\shr{\vr}_{i}, \shr{\trunc{\vr}}_{i})$ denotes the share of $(\vr, \trunc{\vr})$ corresponding to the honest server $P_i$.
	\end{description}
\end{systembox}

The simulator for the case of corrupt $P_0$ appears in \boxref{fig:Sim_TruncI}.
\begin{simulatorbox}{$\Sim_{\Trunc}$}{Simulator for the case of corrupt $P_0$}{fig:Sim_TruncI}
	\justify 
	~
	\begin{myitemize}
		\item[--] $\Sim_{\Trunc}$ computes the values $R_1$ and $R_2$ using the keys $k_{01}$ and $k_{02}$ respectively.
		\item[--] The steps corresponding to $\piJSh$ are simulated similar to $\Sim_{\JSh}$ (\boxref{fig:Sim_piJShA}), for the case of corrupt $P_0$.
	\end{myitemize}
	
	\noindent $\Sim_{\Trunc}$ invokes $\FTrunc$ on behalf of $P_0$.
\end{simulatorbox}

The simulator for the case of corrupt $P_1$ appears in \boxref{fig:Sim_TruncII}.
\begin{simulatorbox}{$\Sim_{\Trunc}$}{Simulator for the case of corrupt $P_1$}{fig:Sim_TruncII}
	\justify
	~
	\begin{myitemize}
		\item[--] $\Sim_{\Trunc}$ compute the value $R_1$ using the key $k_{01}$, while it samples random $R_2$.
		\item[--] The steps corresponding to $\piJSh$ are simulated similar to $\Sim_{\JSh}$ (\boxref{fig:Sim_piJShB}), for the case of corrupt $P_1$.
		\item[--] $\Sim_{\Trunc}$ receives $\Hash(\vu)$ from $\Adv$ on behalf of $P_2$ and sets $\flag=1$ if the received value is inconsistent.
	\end{myitemize}
	
	\noindent If $\flag = 0$, $\Sim_{\Trunc}$ invokes $\FTrunc$ on behalf of $P_1$. Else it invokes $\FTrunc$ with input $\bot$ on behalf of $P_1$.
\end{simulatorbox}

\end{document}